\documentclass[a4paper,11pt]{article}
\pdfoutput=1
\usepackage{jheppub}

\title{\boldmath Spin-independent Interferences and Spin-dependent Interactions with Scalar Dark Matter}
\author{R. Martinez}
\author{and F. Ochoa}

\affiliation{Departamento de F\'{\i}sica, Universidad Nacional de Colombia, \\Ciudad Universitaria, K. 45 No. 26-85, Bogot\'{a} D.C., Colombia}

\emailAdd{remartinezm@unal.edu.co}
\emailAdd{faochoap@unal.edu.co}
 
\abstract{We explore mechanisms of interferences under which the spin-independent interaction in the scattering of scalar dark matter with nucleus is suppressed in relation to the spin-dependent one. We offer a detailed derivation of the nuclear amplitudes based on the interactions with quarks in the framework of a nonuniversal $U(1)'$ extension of the standard model. By assuming a range of parameters compatible with collider searches, electroweak observables and dark matter abundance, we find scenarios for destructive interferences with and without isospin symmetry. The model reveals solutions with mutually interfering scalar particles, canceling the effective spin-independent coupling with only scalar interactions, which requires an extra Higgs boson with mass $M_{H}>125$ GeV. The model also possesses scenarios with only vector interactions through two neutral gauge bosons, $Z$ and $Z'$, which do not exhibit interference effects. Due to the nonuniversality of the $U(1)'$ symmetry, we distinguish two family structures of the quark sector with different numerical predictions. In one case, we obtain cross sections that pass all the Xenon-based detector experiments. In the other case, limits from LUX experiment enclose an exclusion region for dark matter between $9$ and $800$ GeV. We examine a third scenario with isospin-violating couplings where interferences between scalar and vector boson exchanges cancel the scattering. We provide solutions where interactions with Xenon-based detectors is suppressed for light dark matter, below $6$ GeV, while interactions with Germanium- and Silicon-based detectors exhibit solutions up to the regions of interest for positive signals reported by CoGeNT and CDMS-Si experiments, and compatible with the observed DM relic density for DM mass in the range $8.3-10$ GeV. Spin-dependent interactions become the dominant source of scattering around the interference regions, where Maxwellian speed distribution is considered.}

\begin{document}
\maketitle
\flushbottom

\section{Introduction}

The observation of the scalar particle at the CERN Large Hadron Collider by the ATLAS and CMS collaborations \cite{Phys.Lett.B716.1A, Phys.Lett.B716.1B} with the properties of the single Higgs boson of the standard model (SM)\cite{SMA, SMB, SMC, HiggsA, HiggsB}, has confirmed the success of this model to explain most of the observations in particle physics. Now, the mass of the Higgs boson at $125$ GeV, is a known parameter that we can use to obtain new constraints of the multiple extensions of the SM. These extensions are still motivated by unanswered theoretical questions and experimental observations that the SM does not explain. For example, the particle content of the SM accounts for the visible luminous matter of the universe. However, the astrophysical evidence of non luminous matter, i.e. dark matter (DM), is a mistery with a compelling explanation in particle physics if the SM is extended to models with some kind of weakly interacting massive particles (WIMPs) as candidates for microscopic constituents of the DM sector of the universe \cite{lee-weinber-77, wimpsA, wimpsB, wimpsC, wimpsD, wimpsE, wimpsF}. The present experimental evidence of DM are based mostly on its gravitational effects coming from observations as the measurements of the rotational speed of stars in galaxies \cite{galaxyA, galaxyB} or observations of gravitational lensing by galaxy clusters \cite{lensingA, lensingB}. Also, its existence is supported by cosmological facts as the formation of the large-scale structures during the evolution of the universe \cite{cosmoA, cosmoB, cosmoC}. These observations are not in contradiction with the hypothesis of a stable fundamental WIMP particle with a mass in the $1$ to $1000$ GeV range that feels interactions with the strength of the weak nuclear force. Thus, there are chances to obtain information of the true nature of the DM by means other than just the gravitational interaction. These searches have focused mostly on three different mechanisms of detection. First, by detecting particles resulting from WIMP annihilation somewhere in the universe, as for example, the detection of positron and electron pairs carried out at PAMELA, ATIC and Fermi LAT experiments \cite{pam, ati, femlat}. Second, by searching for WIMP pair production at the LHC collider \cite{DMproductionA, DMproductionB}. Finally, through measurements of the nuclear recoil energy coming from elastic scattering with WIMP particles, as in CDMS \cite{cdms-ge}, CoGeNT \cite{cogent}, Xenon100 \cite{xe100} and LUX \cite{lux} experiments. On the other hand, the analysis of the experimental results, must be accompanied by precise theoretical assumptions, which will allow us to understand the experimental data if a positive signal is confirmed, or to guess where and how to continue the research in case of negative results. Many theoretical models have been proposed in the literature with scalar, fermionic or vector WIMP candidates. In the references from \cite{DM-susyA} to \cite{hep-ph/1511.08432}, we list some examples in supersymmetry, extra dimension, little Higgs models and in abelian and non-abelian extensions of the SM.

Experiments for WIMP direct detection through elastic scattering of the nucleus have made great progress by developing new detection techniques sensitive at different WIMP mass regions. In fact, positive signals of a WIMP particle have been claimed independently by the CoGeNT \cite{cogent2}, DAMA/LIBRA \cite{dama} and CDMS-II \cite{cdms2} collaborations. However, these results exhibit failure of internal consistency and/or compatibility with other similar experiments that have not shown positive results. Thus, to solve the question if the DM sector of the universe is made of fundamental particles with direct detection techniques requires a careful understanding of the possible interactions involved in the WIMP-nucleus scattering process. In general terms, these interactions presumedly must exhibit the following features. First, the scattering with atomic nucleus is a consequence of some microscopic coupling with the quark content of the nucleons (protons and neutrons). These couplings falls into one or more of five categories: scalar, pseudoscalar, vector, pseudovector and tensor interactions. Second, the interactions can or can not depend on the spin content of the nucleus. The first one is called spin-dependent interactions (SD) while the latter is spin-independent (SI) \cite{wimpsA, wimpsB, wimpsC, wimpsD}. The scalar-, vector- and tensor-like interactions add coherently over all the nucleons, leading to SI interactions. In contrast, interactions with axial currents, which is the case for pseudoscalar and pseudovector interactions, couple mostly through the spin of the nucleus. Qualitatively, the SD interaction is suppressed by mutual cancellations between nucleons with opposite spins. Only in the case of nucleus with unpaired nucleons, a net coupling with the spin arises. Third, in general, it is assumed that the WIMP does not couple directly with the ordinary matter, otherwise, it would exhibit large annihilation ratios, in contradiction with the evidence of stable DM from estimations for the relic DM density in the universe today. Thus, the couplings with quarks must be carried out by other intermediary particles. It can occur via the exchange of known SM particles, for example, Higgs bosons, or by new extra particles from extensive models, for example, heavy scalar bosons in two Higgs doublet models \cite{gunion} or squarks exchange in supersymetric models \cite{wimpsA, wimpsB, wimpsC, wimpsD}. 

Scalar WIMP candidates exhibit SD elastic cross section proportional to the factor $\beta ^2=(v/c)^2$, as shown in Eq. (\ref{SI-SD-cross-sections}), where $v$ is the speed at which the WIMP particle collides with the nucleus. As a first approximation, this speed corresponds to the circular speed of the Sun around the galactic center, i.e, about $220$ km s$^{-1}$. Thus, the SD interaction is suppressed by an additional factor $\beta ^2 \sim 5 \times 10^{-7}$, while the SI interaction adds to the over all nucleons without a similar suppression. Then, the SD coupling is typically ignored in scalar WIMPs compared with its SI part. However, if for some reason, the SI contribution is suppressed to values below the SD ones, then the main source for scattering of a scalar WIMP is through the nuclear spin. Obviously, we could just make all the SI coupling constants zero or fit small values to obtain negligible SI cross sections. However, this assumption could spoil the observed relic density of DM (small couplings lead us to unobservable excesses of DM densities), and it requires a fine-tuning of the parameters without a natural explanation for their smallness.

The main goal of this work is to explore mechanisms that naturally lead us to large suppression of the SI cross section in direct detection with scalar WIMP, below the SD contributions, and still to have microscopic non-zero couplings with the ordinary matter. To achieve this goal, we make use of quantum interference between different intermediary particles in the WIMP-quark interactions. The mechanism of interference has previously been invoked to produce isospin-violating effects to explain some experimental results. For instance, interferences between Higgs and photon exchanges in the context of asymmetric composite DM \cite{nobile1} leads to asymmetries in the couplings with neutrons and protons. Also, scenarios with fermionic WIMPs and interferences with extra neutral weak bosons $Z'$, have been proposed to accommodate experimental data in the framework of general low-energy effective theories \cite{nobile2}, specific models with extended sectors \cite{hidenA, hidenB, hidenC} and abelian extensions \cite{PRD91-12-123541}. We have the intention of obtaining destructive interference in the SI sector. For that, we perform our analysis in the framework of an $U(1)'$ abelian extension of the SM, which includes an extra neutral $Z'$ gauge boson; specifically, the nonuniversal family extension introduced in Refs. \cite{martinezochoa1A, martinezochoa1B, martinezochoa2} give us a natural background %where, in addition to provide the minima elements to derive our results, other types of problems can be addressed, as for example the SM flavor puzzle, the family structure of fermions and the physics of neutrinos. Also, this model contains an specific two Higgs doublet model (2HDM), which implies many phenomenological and theoretical options. Finally, the model exhibits nonuniversal couplings in the quark sector, providing an scenary to explore family nonuniversal effects in nuclear interactions. We emphasize however, that although our numerical results are particular solutions in this model, the same type of analysis can be extended into other models with extra $Z'$ gauge bosons. 
which provide elements to derive new results. In addition to a $Z'$ gauge boson, the extra $U(1)'$ symmetry is nonuniversal in the quark sector, which implies the necessity of at least two scalar doublets in order to generate all the Yukawa couplings and obtain a complete massive spectrum in the quark sector. Thus, the model is a natural combination of a specialized two Higgs doublet model (2HDM) and a model with extra $Z'$ gauge bosons. Both sectors will have important implications in the WIMP-nucleus scattering. 

This paper is organized as follows. Section 2 is devoted to describe the theoretical model. Since this model has been discussed in previous works \cite{martinezochoa1A, martinezochoa1B, martinezochoa2}, we just describe some general properties and show the basic couplings. In section 3, based on the fundamental couplings at the quark level, we will obtain the nuclear SI and SD effective couplings and cross sections at zero momentum transfer. In section 4, we explore solutions for destructive interferences that will nullify the SI cross sections. We also evaluate isospin violating scenarios. In section 5, we will compare the SD and SI cross sections, first for a discrete value of the speed of collision, and later, more accurately, by comparing differential event rates with a Maxwellian distribution of speeds. Finally, we summarize our conclusions in section 6.  

\section{The Model}

\subsection{Overview}

The particle content of the model, shown in Tables \ref{tab:SM-espectro} and \ref{tab:exotic-espectro}, is composed of ordinary SM particles and new extra non-SM particles, where column $G_{sm}$ indicates the transformation rules under the SM gauge group $(SU(3)_c,SU(2)_L,U(1)_Y)$, and column $U(1)_X$ contains the values of the new quantum number $X$. Below, based on fundamental facts we describe some general properties of the model.

\begin{table}[tbp]
%\caption{Ordinary SM particle content, with $i=$1,2,3}
\begin{center}
%\caption{\small Ordinary SM particle content, with $i=$1,2,3} \vspace{-0.5cm}
%\label{tab:SM-espectro}
\begin{equation*}
\begin{tabular}{c c c}
\hline\hline
$Spectrum$ & $G_{sm}$ & $U(1)_{X }$ \\ \hline \\
$\ 
\begin{tabular}{c}
%\\
$q^i_{L}=\left( 
\begin{array}{c}
U^i \\ 
D^i 
\end{array}%
\right) _{L}$ %\\
%\\ 
\end{tabular}%
\ $ & 
$(3,2,1/3)$
&
$\ 
\begin{tabular}{c}
$1/3$ for $i=1$ \\ 
$0$ for $i=2,3$%
\end{tabular}
\ $

\\ \\%\hline
$U_R^i$
& 
$(3^*,1,4/3)$
&
\begin{tabular}{c}
%\\ 
$2/3$ %\\
%\\
\end{tabular}

\\ \\%\hline
$D_R^i$
&
$(3^*,1,-2/3)$
&
\begin{tabular}{c}
%\\
$-1/3$ %\\
%\\
\end{tabular}

\\ \\%\hline
$\ 
\begin{tabular}{c}
%\\
$\ell ^i_{L}=\left( 
\begin{array}{c}
\nu ^i \\ 
e^i 
\end{array}%
\right) _{L}$ \\
%\\ 
\end{tabular}
\ $
& 
$(1,2,-1)$
&
$-1/3$ 

\\  \\%\hline
$e_R^i$
&
$(1,1,-2)$
&
\begin{tabular}{c}
%\\
$-1$ %\\
%\\
\end{tabular}

\\  \\%\hline
$\ 
\begin{tabular}{c}
%\\
$\phi _{1}=\left( 
\begin{array}{c}
\phi _1^{+} \\ 
\frac{1}{\sqrt{2}}(\upsilon _1+\xi _1+i\zeta _1) 
\end{array}%
\right) $% \\ 
%\\
\end{tabular}
\ $
&
$(1,2,1)$
&
$2/3$

\\  \\%\hline
$\ 
\begin{tabular}{c}
%\\
$W _{\mu }=\left( 
\begin{array}{cc}
W _{\mu }^{3}&\sqrt{2}W_{\mu }^+ \\ 
\sqrt{2}W_{\mu }^-& -W _{\mu }^{3}
\end{array}%
\right) $% \\ 
%\\
\end{tabular}
\ $
&
$(1, 2 \times 2^*, 0)$
&
$0$

\\  \\%\hline
$\ 
B_{\mu }
\ $
&
$(1, 1, 0)$
&
$0$ \\
\hline
\end{tabular}%
\end{equation*}%
\end{center}
\vspace{-0.5cm}
\caption{\small Ordinary SM particle content, with $i=$1,2,3} 
\label{tab:SM-espectro}
\end{table}

\begin{table}[tbp]
\begin{center}
%\caption{\small Extra non-SM particle content, with $n=$1,2}\vspace{-0.5cm}
%\label{tab:exotic-espectro}
\begin{equation*}
\begin{tabular}{c c c}
\hline\hline
$Spectrum$ & $G_{sm}$ & $U(1)_{X }$ \\ \hline \\
$\ 
\begin{tabular}{c}
$T_L$ 
\end{tabular}%
\ $ & 
$(3,1,4/3)$
&
$1/3$

\\ \\%\hline
\begin{tabular}{c}
$T_R$
\end{tabular}
& 
$(3^*,1,4/3)$
&
$2/3$

\\ \\%\hline
\begin{tabular}{c}
$J_L^n$ 
\end{tabular}
&
$(3,1,-2/3)$
&
$0$

\\ \\ %\hline
$\ 
\begin{tabular}{c}
$J_R^n$  
\end{tabular}
\ $
& 
$(3^*,1,-2/3)$
&
$-1/3$ 

\\ \\%\hline
\begin{tabular}{c}
$(\nu _R^i)^c$ 
\end{tabular}
&
$(1,1,0)$
&
$-1/3$

\\ \\ %\hline
\begin{tabular}{c}
$N _R^i$ 
\end{tabular}
&
$(1,1,0)$
&
$0$

\\ \\%\hline
$\ 
\begin{tabular}{c}
$\phi _{2}=\left( 
\begin{array}{c}
\phi _2^{+} \\ 
\frac{1}{\sqrt{2}}(\upsilon _2+\xi _2+i\zeta _2) 
\end{array}%
\right) $ 
\end{tabular}
\ $
&
$(1,2,1)$
&
$1/3$

\\ \\%\hline 
\begin{tabular}{c}
$\chi  = \frac{1}{\sqrt{2}}(\upsilon _{\chi}+\xi _{\chi }+i\zeta _{\chi})$ 
\end{tabular}
&
$(1,1,0)$
&
$-1/3$

\\ \\%\hline 
\begin{tabular}{c}
$\sigma  = \frac{1}{\sqrt{2}}(\xi _{\sigma}+i\zeta _{\sigma})$ 
\end{tabular}
&
$(1,1,0)$
&
$-1/3$

\\  \\ %\hline
%\begin{tabular}{c}
%\\
$Z'_{\mu}$ %\\
%\\
%\end{tabular}
&
$(1,1,0)$
&
$0$
\\ \hline
\end{tabular}%
\end{equation*}%
\end{center}
\vspace{-0.5cm}
\caption{\small Extra non-SM particle content, with $n=$1,2}\vspace{-0.5cm}
\label{tab:exotic-espectro}
\end{table}

\begin{list}{}{}

\item[-] The equations that cancel the chiral anomalies are obtained in the reference \cite{martinezochoa1A}. These equations lead us to a set of non-trivial solutions for $U(1)_X$ that requires a structure of three families, where the left-handed quarks $q^i_L$ have nonuniversal charges: family with $i=1$ has $X_1=1/3$, while $X_{2,3}=0$ for $i=2,3$. The match with the physical quarks gives rise to different options. We choose the two structures, A and B, shown in table \ref{tab:family-matching}. In addition, the cancellation of anomalies require the existence of an extended quark sector. A simple possibility is introducing quasichiral singlets ($T$ and $J^{n}$, where $n=1,2$), i.e. singlets that are chiral under $U(1)_X$ and vector-like under $G_{sm}$. Due to the global symmetry in Eq. (\ref{global-symm}) below, this sector will not participate in the WIMP-nucleus scattering. We emphasize however, that by introducing appropriate discrete symmetries, it is possible to obtain scenarios where these quarks can mediate the scattering.

\item[-] It is desirable to obtain a realistic model compatible with the oscillation of neutrinos. For this purpose, the model introduces new neutrinos  $(\nu ^i_R)^c$ and $N ^i_R$ which may generate seesaw neutrino masses. This sector will be irrelevant in the present analysis. However, the option to study direct detection with fermionic DM exists if we arrange conditions for $N_R$ to be a WIMP candidate.

\item[-] An extra neutral gauge boson, $Z'_{\mu}$, is required to make the $U(1)_X$ transformation a local symmetry.

\item[-] Due to the nonuniversal structure of the quark doublets, an additional scalar doublet, $\phi _2$, identical to $\phi _1$ under  $G_{sm}$ but with different $U(1)_X$ charges is required in order to obtain massive fermions after the spontaneus symmetry breaking, where the electroweak vacuum expectation value (VEV) is $\upsilon = \sqrt{\upsilon  _1 ^2+\upsilon _2^2}$. 

\item[-] An extra scalar singlet, $\chi $, with VEV $\upsilon _{\chi}$ is required to produce the symmetry breaking of the $U(1)_X$ symmetry. We assume that it happens on a large scale $\upsilon _{\chi} \gg \upsilon$. Since this field is a singlet under the $G_{sm}$ symmetry, there are no couplings between $\chi$ and the SM left-handed doublets $q^{i}_{L}$ in the Yukawa Lagrangian. Its coupling with the ordinary matter is possible only through mixing with the quasiquiral quarks $T_R$ and $J^{n}_R$. As an example, the real part of $\chi $ may explain the diphoton excess recently announced by the ATLAS and CMS collaborations \cite{CMS750,ATLAS750} at $750$ GeV, as studied in \cite{1512.05617}.    

\item[-] Another scalar singlet, $\sigma $, is introduced, which will be our WIMP candidate. In order to reproduce the observed DM relic density, this particle must accomplish the following minima conditions

\begin{enumerate}
\item[(i)] Since $\sigma $ acquires a nontrivial charge $U(1)_X$, it must be complex in order to be a massive candidate.  

\item[(ii)] To avoid odd powers terms in the scalar Lagrangian, which leads to unstable DM, we impose the following global continuos symmetry

\begin{eqnarray}
\sigma  \rightarrow e^{i\theta }\sigma .
\label{global-symm}
\end{eqnarray}

\item[(iii)] In spite of the above symmetry, the model still can generate odd power terms via spontaneous symmetry breaking. To avoid this, $\sigma $ must not generate VEV during the lifetime of our universe.

\end{enumerate}
 
Even though the field $\chi$ is defined with the same quantum numbers as $\sigma$, and may exhibit the same couplings with nucleus, the former does not accomplish conditions (ii) and (iii). As a consequence, field $\chi$ is too unstable to survive at the current energies of our universe. Thus, we consider that the full scattering of the nucleus with scalar singlets is due only to $\sigma$. 

\end{list}{}{}

\subsection{Lagrangians}

The Lagrangians that describe all interactions of the above particles are constructed from the symmetries of the model. First, the most general renormalizable and $G_{sm} \times U(1)_X$ invariant scalar potential is

\begin{eqnarray}
V&=&\mu _1^2 \left| \phi _1 \right|^2 +\mu _2^2 \left| \phi _2 \right|^2 + \mu _3^2 \left| \chi \right|^2 +\mu _4^2 \left| \sigma \right|^2+\mu _5^2\left( \chi ^*\sigma  +h.c \right)  \nonumber \\
&+&f_1\left(\phi _2^{\dagger} \phi _1 \sigma +h.c.\right)+f_2\left(\phi _2^{\dagger} \phi _1 \chi +h.c.\right) \nonumber \\
&+& \lambda _1 \left| \phi _1 \right|^4+\lambda _2 \left| \phi _2 \right|^4+\lambda _3 \left| \chi \right|^4+\lambda _4 \left| \sigma \right|^4 \nonumber \\
&+& \left| \phi _1 \right|^2\left[ \lambda _6 \left| \chi \right| ^2+ \lambda '_6 \left| \sigma \right|^2+ \lambda ''_6 \left( \chi ^*\sigma  +h.c. \right) \right] \nonumber \\
&+& \left| \phi _2 \right|^2\left[ \lambda _7 \left| \chi \right| ^2+ \lambda '_7 \left| \sigma \right|^2+ \lambda ''_7 \left( \chi ^*\sigma +h.c. \right) \right] \nonumber \\
&+&\lambda _5 \left| \phi _1 \right|^2\left| \phi _2 \right|^2+\lambda '_5 \left| \phi _1^{\dagger} \phi _2 \right|^2 +\lambda _8  \left| \chi \right|^2\left| \sigma \right|^2+\lambda '_8  \left[\left(\chi ^* \sigma \right)^2+h.c.\right].
\label{higgs-pot-1}
\end{eqnarray} 
In addition, if we impose the global symmetry from equation (\ref{global-symm}), terms where only appears $\sigma$ or $\sigma ^*$ are not allowed, which lead us to the constraints $\mu _5=f_1=\lambda ''_{6,7}=\lambda '_8=0$.

Second, the kinetic sector of the Higgs Lagrangian is:

\begin{eqnarray}
\mathcal{L}_{kin} &=&\sum_S (D_{\mu}S)^{\dagger}(D^{\mu}S),
\label{higgs-kinetic}
\end{eqnarray}
where the covariant derivative is defined as

\begin{eqnarray}
D^{\mu}=\partial ^{\mu }-igW^{\mu}_{\alpha} T^{\alpha }_S-ig'\frac{Y_S}{2}B^{\mu}-ig_XX_SZ'^{\mu}.
\label{covariant}
\end{eqnarray}
The parameters $2T^{\alpha }_S$ correspond to the Pauli matrices when $S=\phi _{1,2}$ and $T^{\alpha }_S=0$ when  $S=\chi , \sigma $, while $Y_S$ and $X_S$ correspond to the hypercharge and $U(1)_X$ charge according to the values in Tables. \ref{tab:SM-espectro} and \ref{tab:exotic-espectro}. $g_X$ is the new coupling constant from the extra $U(1)_X$ gauge symmetry, while $g$ and $g'$ are the same as in the SM, which accomplish the constraint $g'=g\tan{\theta _W}=S_W/C_W$, with $\theta _W$ the Weinberg angle that rotate the neutral gauge bosons into SM-like gauge bosons:

\begin{eqnarray}
\begin{pmatrix}
A_{\mu} \\
Z_{\mu } \\ 
\end{pmatrix} &= &\begin{pmatrix}
C_{W} & S_{W} \\
-S_{W} & C_{W}
\end{pmatrix}
\begin{pmatrix}
W_{\mu}^3 \\
B_{\mu } \\ 
\end{pmatrix},
\label{sm-gauge-eigenvec}
\end{eqnarray}  
where $A_{\mu}$ is identified with the photon, while $Z_{\mu}$ is a weak neutral boson. However, as we will see in equation (\ref{mixing-zzp}), due to a mixing with the extra boson $Z'_{\mu }$, this state is not a mass eigenstate.

With regard to  the interactions with fermions, the Dirac Lagrangian reads:

\begin{eqnarray}
\mathcal{L}_{D}=i\sum_{i}\left(\overline{f^i_L}\gamma ^\mu D_{\mu} f^i_L+\overline{f^i_R}\gamma ^\mu D_{\mu}f^i_R\right), 
\end{eqnarray}
where $f$ represents any of the SM or non-SM weak eigenstates, and the index $i=1,2,3$ runs over the three families; while the Yukawa Lagrangian for the quark sector is:

\begin{eqnarray}
-\mathcal{L}_Q &=& \overline{q_L^{a}}(\widetilde{\phi }_1 h^{U}_{1})_{aj}U_R^{j}+ \overline{q_L^{3}}\left(\widetilde{\phi} _2h^{U}_2 \right)_{3j}U_R^{j}+\overline{q_L^{a}}\left(\phi _2 h^{D}_{2} \right)_{aj}D_R^{j}+\overline{q_L^{3}}\left(\phi _1 h^{D}_1\right)_{3j}D_R^{j} \notag \\
&+&\overline{q_L^{a}}\left(\phi  _2 h^{J}_{2} \right)_{am} J^{m}_R+ \overline{q_L^{3}} (\phi _1 h^{J}_{1})_{3m} J^{m}_R+\overline{q_L^{a}} (\widetilde{\phi } _1 h^{T}_{1})_aT_R+\overline{q_L^{3}}\left(\widetilde{\phi} _2 h^{T}_{2} \right)_3T_R  \notag \\
&+&\overline{T_{L}}\left( \sigma h_{\sigma }^{U}+\chi h_{\chi }^{U}\right)_{j}{U}_{R}^{j}+\overline{T_{L}}\left( \sigma h_{\sigma}^{T}+\chi h_{\chi }^{T}\right){T}_{R}
\nonumber \\
&+&\overline{J_{L}^n}\left( \sigma ^*h_{\sigma }^{D}+\chi ^*h_{\chi }^{D}\right)_{nj}{D}_{R}^{j}+\overline{J_{L}^n}\left( \sigma ^*h_{\sigma }^{J}+\chi ^*h_{\chi }^{J}\right)_{nm}{J}_{R}^{m}+h.c.,
 \label{quark-yukawa-1}
\end{eqnarray}
where $\widetilde{\phi}_{1,2}=i\sigma_2 \phi_{1,2}^*$ are conjugate scalar doublets, and $a=1,2$. %is the index that label the second and third quark doublets and $n(m)=1,2$ is the index of the exotic $J^{n(m)}$ quarks. %and $h^{Q}_{r}$ are non-diagonal matrices in the flavor space associated with each quark $Q$ and scalar field $r=\phi _{1,2}, \chi _0$ and $\sigma _0$.
For the leptonic sector we obtain:

\begin{eqnarray}
-\mathcal{L}_{\ell}&=&\overline{\ell ^{i}_{L}}\left( \widetilde{\phi}_1h_{1 }^{\nu } \right)_{ij}{\nu }_{R}^{j}+ \overline{\ell ^{i}_{L}}\left( \widetilde{\phi}_2h_{2 }^{N}\right)_{ij}{N}_{R}^{j}\nonumber \\
&+&\overline{(\nu ^{i}_{R})^c}\left( \sigma ^*h_{\sigma }^{N}+\chi ^* h_{\chi }^{N}\right)_{ij}{N}_{R}^j +\frac{1}{2}M_{Ni}\overline{(N^{i}_{R})^c}{N}_{R}^{i}\nonumber \\
&+&\overline{\ell ^{i}_{L}}\left( \phi _1h_{1}^{e } \right)_{ij}{e}_{R}^{j}+h.c.
\label{yukawa-leptons-1}
\end{eqnarray}
In particular, we can see in the quark Lagrangian in equation (\ref{quark-yukawa-1}) that due to the nonuniversality of the $U(1)_X$ symmetry, not all couplings between quarks and scalars are allowed by the gauge symmetry, which leads us to specific zero-texture Yukawa matrices as studied in ref. \cite{martinezochoa1A}. Also, if we consider again the global symmetry in (\ref{global-symm}), without any other symmetries on the fermionic singlets, then the terms with only $\sigma$ or $\sigma ^*$ must disapears, which leads us to the constraints $h_{\sigma} ^f=0$ for the Yukawa couplings of $\sigma$ with any fermion $f$. Thus, we do not have point-like interactions of the WIMP with matter.

Finally, analogous to general 2HDM, we can obtain different realizations for the couplings. In order to avoid large flavor changing neutral currents, we explore two limits equivalents to the 2HDM types I and II. In type I, only the scalar doublet $\phi _1$ provides masses to both the up- and down-type quarks, while in type II, the doublets $\phi _1$ and $\phi _2$ give masses to the up- and down-type quarks, respectively.

\subsection{Mass eigenstates}

To identify the free parameters of the model, we must rotate the fields into mass eigenstates. For the scalar sector, %after applying the minimum conditions, the mass matrices are obtained from the derivatives of the expectation value of the scalar potential $M_{ij}^{2}=\partial ^2 \langle V \rangle/\partial S_i \partial S_j |_{S_j=0}$ for each scalar component of the model. 
after diagonalization of the mass matrices, we obtain the following mass eigenstates \cite{martinezochoa1B}:

\begin{eqnarray}
\begin{pmatrix}
G^{\pm}\\
H^{\pm}
\end{pmatrix}&=&R_{\beta}\begin{pmatrix}
\phi _1^{\pm}\\
\phi _2^{\pm}
\end{pmatrix}, \ \ \  
\begin{pmatrix}
G_{0}\\
A_{0}
\end{pmatrix}=R_{\beta}\begin{pmatrix}
\zeta _1\\
\zeta _2
\end{pmatrix}, \nonumber \\ 
\begin{pmatrix}
H\\
h
\end{pmatrix}&=&R_{\alpha}\begin{pmatrix}
\xi _1\\
\xi _2
\end{pmatrix}, \ \ \
\begin{pmatrix}
H_{\chi }\\
G_{\chi }
\end{pmatrix}\sim I\begin{pmatrix}
\xi _{\chi }\\
\zeta _{\chi }
\end{pmatrix}, 
\label{scalar-eigenvectors}
\end{eqnarray}      
where $I$ is the identity, and the rotation matrices are defined according to

\begin{eqnarray}
R_{\beta, \alpha}&=&\begin{pmatrix}
C_{\beta, \alpha} & S_{\beta, \alpha} \\
-S_{\beta, \alpha} & C_{\beta, \alpha}
\end{pmatrix}.
\end{eqnarray}
The mixing angle $\beta $ is defined through the ratio of the electroweak VEV as $\tan{\beta }=T_{\beta}=\upsilon _2/\upsilon _1$, 
while $\alpha $ is related to $\beta$ as
 \begin{eqnarray}
\sin{2\alpha }&\approx &\sin{2\beta}\left[1-\frac{\sqrt{2}C_{2\beta}S_{2\beta}\upsilon ^2}{f_2\upsilon _{\chi }}\left(\lambda _{1}C_{\beta }^2-\frac{\lambda _5+\lambda '_5}{2}C_{2\beta }-\lambda _{2}S_{\beta }^2\right) \right],
\label{alf-beta-const}
\end{eqnarray} 
where we have taken the dominant contribution assuming that $\upsilon ^2 \ll \left|f_2 \upsilon _{\chi}\right|$. The parameters $\lambda _{1,2,5}$ and $\lambda '_{5}$ are coupling constants from interactions between Higgs doublets $\phi_{1,2}$, and $f_2$ is the coupling of the cubic term between doublets and the singlet $\chi $. In order to reduce the parameter space, we neglect the second term and take:

\begin{eqnarray}
\sin 2{\alpha }\approx \sin 2{\beta } &\Rightarrow &\alpha \approx \beta.
\label{alf-beta-const2}
\end{eqnarray}
In particular, we identify the field $h$ as the observed $125$ GeV Higgs boson, and $H$ is an extra CP-even neutral Higgs boson.
%Then, we will take $T_{\beta}$ as the free parameter. %As for the eigenvalues, we obtain for the physical particles the following squared masses at dominant order:

%\begin{eqnarray}
%M_{H^{\pm}}^2&\approx&M_{H}^2\approx M_{A_{0}}^2\approx-\frac{f_2\upsilon _{\chi}}{\sqrt{2}}\left(\frac{1+T_{\beta}^2}{T_{\beta}}\right) \nonumber \\
%M_{H_{\chi }}^2 &\approx &2\hat{\lambda }_{33} \upsilon _{\chi} ^2, \nonumber \\
%M_{h}^2 &\approx& \frac{2\upsilon ^2}{\left(1+T_{\beta }^2\right)^2}\left(\hat{\lambda }_{11}+2\hat{\lambda }_{12}T_{\beta }^2+\hat{\lambda }_{22}T_{\beta }^4\right).
%\label{scalar-mass}
%\end{eqnarray}

As for the neutral gauge sector, after the symmetry breaking and using the basis in (\ref{sm-gauge-eigenvec}), we obtain from the kinetic Lagrangian in equation (\ref{higgs-kinetic}) the following mass Lagrangian

\begin{eqnarray}
\mathcal{L}_{Z-Z'}=\frac{1}{2}M_Z^{2}Z^{\mu}Z_{\mu}+\frac{1}{2}M_{Z'}^2Z'^{\mu}Z'_{\mu }-(1+C_{\beta}^2)\frac{2g_XC_W}{3g}M_Z^2Z^{\mu}Z'_{\mu },
\label{mixing-zzp}
\end{eqnarray}
where

\begin{eqnarray}
M_Z&\approx& \frac{g\upsilon }{2C_W} ,\ \ \  \text{and} \  \ M_{Z'}\approx  \frac{g_X\upsilon _{\chi }}{3}.
\label{neutralgauge-masses}
\end{eqnarray}
Since the Lagrangian (\ref{mixing-zzp}) exhibits a $Z-Z'$ mixing term, we must rotate the neutral fields to obtain mass eigenstates. By defining the mixing angle as

\begin{eqnarray}
S_{\theta} \approx  (1+C_{\beta}^2)\frac{2g_XC_W}{3g}\left(\frac{M_Z}{M_{Z'}}\right)^2,
\label{mixing-angle}
\end{eqnarray}
we obtain the total rotation from weak to mass eigenstates as
%the charged eigenstates $W_{\mu }^{\pm}=(W_{\mu }^{1}\mp W_{\mu }^{2})/\sqrt{2}$ with mass $M_{W}=g\upsilon /2$, while for the neutral sector, we obtain the following squared mass matrix in the neutral gauge basis  $({W_{\mu }^3,B_{\mu},Z'_{\mu}})$:

\begin{eqnarray}
\begin{pmatrix}
A_{\mu} \\
Z_{1\mu } \\
Z_{2\mu } 
\end{pmatrix} &= &R_0
\begin{pmatrix}
W_{\mu}^3 \\
B_{\mu } \\
Z'_{\mu } 
\end{pmatrix},
\ \ \ \ \text{with} \ \ R_0=\begin{pmatrix}
S_W & C_W  & 0 \\ 
&&\\
C_WC_{\theta} &  -S_WC_{\theta}  &  S_{\theta}   \\
 &&\\
-C_WS_{\theta}  & S_WS_{\theta}   & C_{\theta}  \\
\end{pmatrix}.
\label{gauge-eigenvec}
\end{eqnarray}
We see that in the limit $S_{\theta}=0$, we obtain $Z_1=Z$ and $Z_2=Z'$.  

\section{Constraints}

We will find that the WIMP-nucleon elastic cross section depends on 9 free fundamental parameters of the model, which we classify into three categories: parameters of coupling, parameters of mass and parameters of mixing. Into the parameters of couplings we identify 3 coupling constants: the coupling constant $g_X$ defined in equation (\ref{covariant}), and the two coupling constants, $\lambda '_6$ and $\lambda '_7$ that couple the scalar WIMP $\sigma $ particle with the two Higgs doublets, as shown in equation  (\ref{higgs-pot-1}). We parameterize these couplings in terms of $\lambda '_6$ and the ratio $\lambda _r=\lambda '_7/\lambda '_6$. Into the parameters of mass, we have three unknown masses: the mass of the $Z_2$ gauge boson, that we will approximate to $M_{Z_2}\approx M_{Z'}$, as defined in (\ref{neutralgauge-masses}), the mass of the scalar WIMP ($M_{\sigma }$) and the mass of the CP-even Higgs boson $H$ ($M_{H}$). Finally, the mixing parameters correspond to three mixing angles from the diagonalization into mass eigenstates. They are: the two mixing angles from the scalar sector ($\beta $ and $\alpha$) and one angle from the $Z-Z'$ mixing term ($\theta $). However, these angles are not independent from each other. The angle $\beta $ is equal to $\alpha $, according to the constraint in (\ref{alf-beta-const2}), while $\theta$ is related to both $\beta$ and $M_{Z'}\approx M_{Z_{2}}$ through (\ref{mixing-angle}). Thus, our space of parameters is reduced to 7 free parameters: $(g_X, \lambda '_6, \lambda _r, M_{\sigma}, M_{H}, M_{Z_{2}}, T_{\beta})$.

On the other hand, the above parameters can be constrained from theoretical conditions and/or phenomenological observables. We will include some limits into our present analysis to obtain results compatible with other observations.

\begin{table}[tbp]
\begin{center}
%\caption{\small Match between left-handed quark states and phenomenological quarks with their $U(1)_X$ charges} \vspace{-0.5cm}
%\label{tab:family-matching}
\begin{equation*}
\begin{tabular}{c|c c c|c c c}
\hline\hline
 & $U^1_L$ & $U^2_L$ & $U^3_L$ & $D^1_L$ & $D^2_L$ & $D^3_L$ \\ \hline
\text{Model A} & $u$  & $c$  & $t$  & $d$  & $s$ & $b$ \\
\text{Model B} & $t$  & $u$  & $c$  & $b$  & $d$  & $s$ \\ \hline
$U(1)_X$ \text{charges}& $1/3$ & $0$ & $0$ & $1/3$ & $0$ & $0$ \\ \hline
\end{tabular}
\end{equation*}%
\end{center} 
\vspace{-0.5cm}
\caption{\small Match between left-handed quark states and phenomenological quarks with their $U(1)_X$ charges} 
\label{tab:family-matching}
\end{table}

\subsection{$U(1)_X$ gauge coupling constant}

The coupling constant $g_X$ can be constrained from observables at high and low energies, as shown in \cite{martinezochoa1A, martinezochoa1B}. First, by measurements of dilepton events, limits on $pp\rightarrow Z_2 \rightarrow e^+e^-(\mu ^+\mu^-)$ cross sections at LHC are reported, obtaining values as large as $g_X \approx 0.4$ at $M_{Z_{2}}\approx 3$ TeV. Also, deviations of electroweak parameters due to a small $Z$-$Z'$ mixing leads to important constraints on the gauge coupling. From Z pole observables measured at CERN-LEP and SLAC colliders, limits up to $g_X \approx 0.3-0.4$ in the range $M_{Z_{2}}=3-4$ TeV were obtained. Thus, the limit $g_X=0.4$ is an appropriate superior bound.

\begin{figure}[tb] 
\centering
\includegraphics[width=12cm,height=10cm,angle=0]{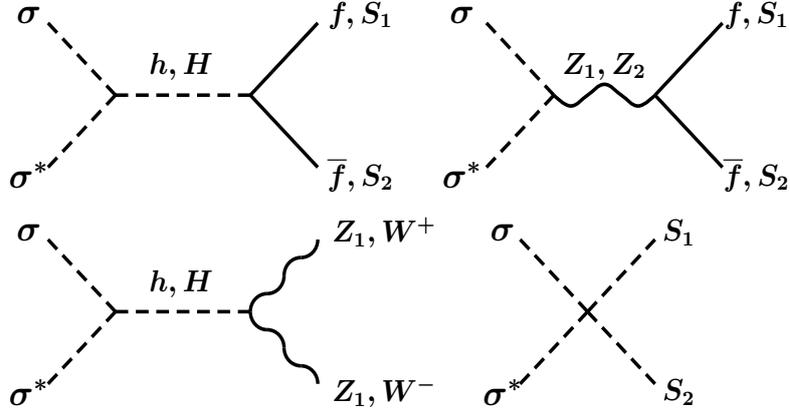}\vspace{-3cm}
\caption{\small Diagrams contributing to DM annihilation to fermions $f$, scalar particle pairs $S_{1,2}$ and vector boson pairs.}
\label{fig1}
\end{figure}

\subsection{Scalar coupling constants}

\begin{figure}[tb] 
\centering
\includegraphics[width=7.3cm,height=6cm,angle=0]{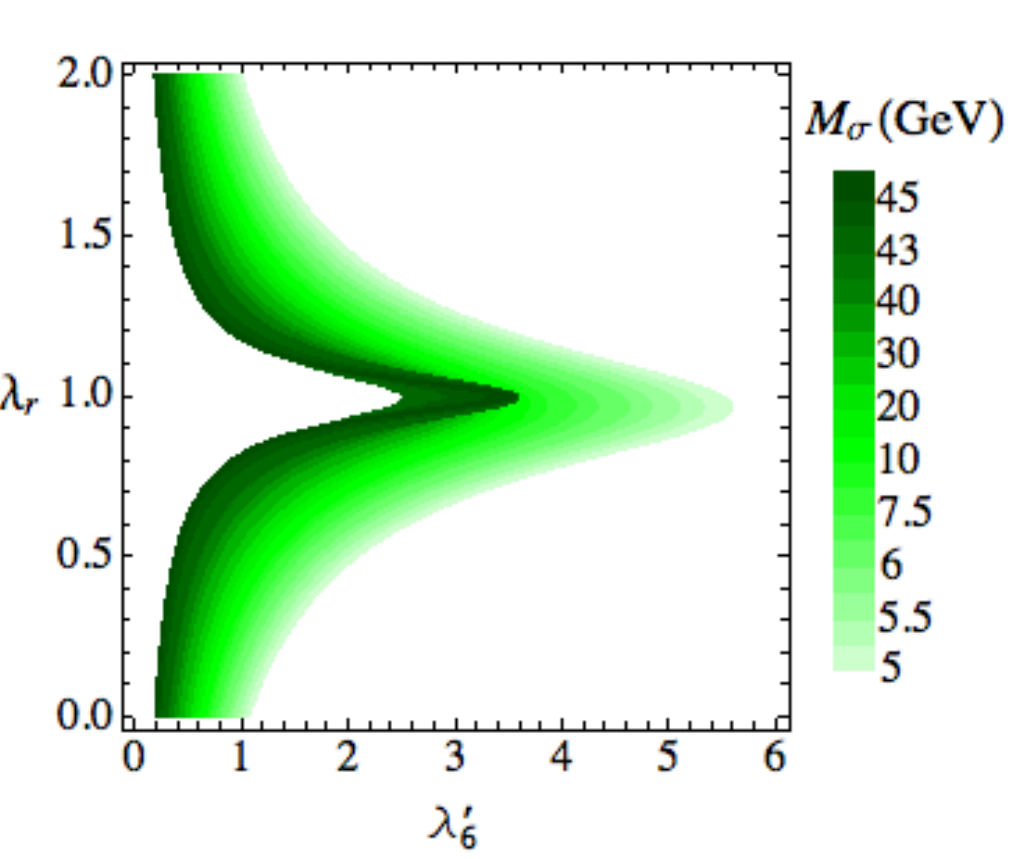}\hspace{0.2cm}
\includegraphics[width=7.3cm,height=6cm,angle=0]{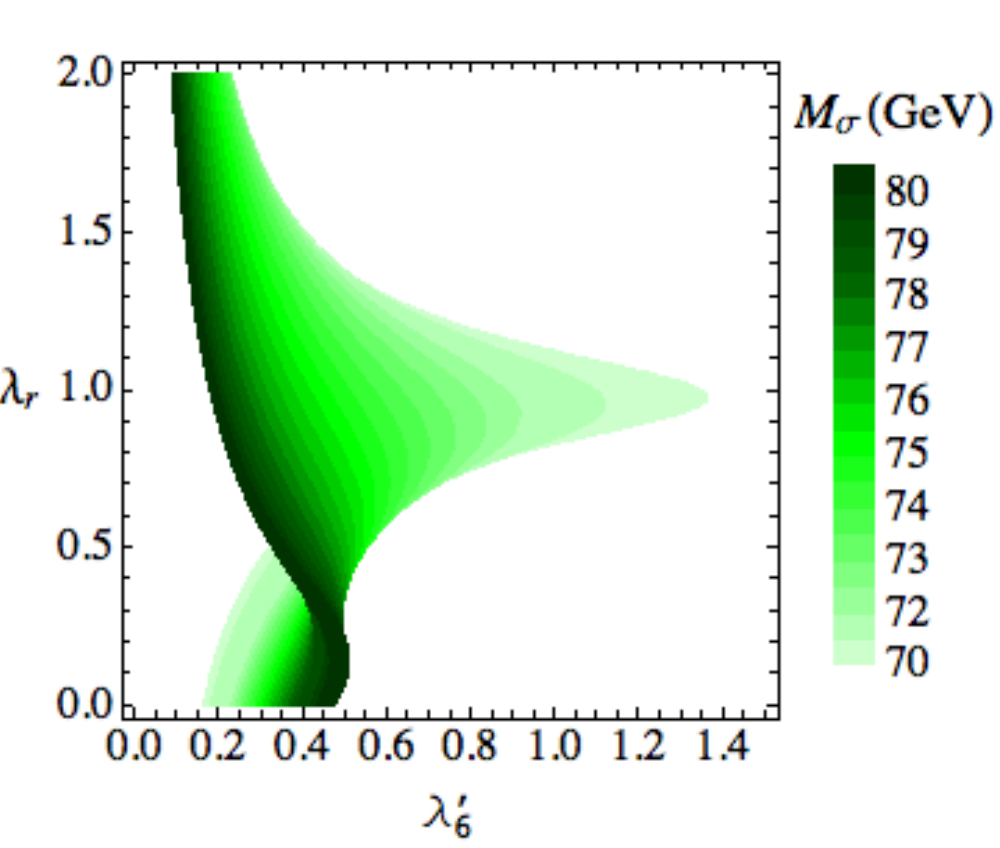}\vspace{-0.5cm}
\caption{\small Limits from DM relic density for the scalar couplings $\lambda'_{6}$ and $\lambda_r$ with DM masses in the $5$ to $45$ GeV (left) and $70$ to $80$ GeV (right) ranges, and $T_{\beta}=10$. }
\label{fig2}
\end{figure}

From stability conditions of the Higgs potential, the scalar coupling constants $\lambda '_6$ and $\lambda '_7$, must take positive values, as shown in \cite{martinezochoa2}. Also, they can not take arbitrarily large values, otherwise, we would obtain excess of WIMPs annihilation, spoiling the observed relic DM density. In order to connect the regions from WIMP scattering with the limits from relic abundance, we evaluate the allowed points in the space of parameters $(\lambda' _{6}, \lambda _r, M_{\sigma})$ compatible with the observed abundance $\Omega h^2=0.1198 \pm 0.0051$ at $95\%$ C.L. In figure \ref{fig1}, we show the most important WIMP annihilation processes, where $f$ denotes fermions with masses above $1$ GeV ($\tau, c, b, t$), and $S_{1,2}$ are Higgs boson pairs ($h,H,H^{\pm},A_0$). As an example, in figure \ref{fig2} we perform the scan of the WIMP mass in the plane $\lambda' _6 - \lambda _r$ with $T_{\beta}=10$ and for type II model. We set $M_{Z_{2}}=3$ TeV and $g_X=0.4$. We show the ranges $M_{\sigma}=5-45$ GeV and $70-80$ GeV in the left and right plots, respectively. Between $45$ GeV and $70$ GeV we have the resonance associated with production of the 125 GeV Higgs boson at $M_{\sigma}=M_h/2\approx 63$ GeV, corresponding to the process $\sigma \sigma ^{*} \rightarrow h \rightarrow f\overline{f}$. In this range, we obtain an excess of WIMP annihilation, and the relic density drops below the experimental limits. We see that the more massive the WIMP, the smaller the limits for $\lambda'_{6}$. A superior limit $\lambda' _{6}\approx 5.8$ is obtained for $M_{\sigma} \sim 5$ GeV and $\lambda _r=1$. Similar limits are obtained in the framework of the type I model.  
% as studied in \cite{martinezochoa1B, martinezochoa2}. By combining results from both references, the ranges $0.02 \leq \lambda '_6 \leq 1.5$ and $0 \leq \lambda _r \leq 2$ are feasible values.

\subsection{$H$ and $Z_2$ masses}

The masses of the neutral Higgs boson $H$ and the neutral gauge boson $Z_2$ have some direct limits from colliders. For $M_H$ there are many decay channels that impose different limits \cite{data-particle} from searches for light neutral Higgs bosons (with masses below $125$ GeV) to very heavy Higgs bosons (at the TeV scale).  Since we will consider that the SM-like Higgs boson at 125 GeV is the lightest one, we set larger values for $M_H$. For $M_{Z_{2}}$, we take the experimental limit near $3$ TeV \cite{zprimemassA, zprimemassB}. For the WIMP candidate $\sigma $ we adopt the typical range $1-1000$ GeV. 

Thus, our space of parameters is reduced to 7 parameters: $(g_X, \lambda '_6, \lambda _r, M_{\sigma}, M_{H}, M_{Z_{2}}, T_{\beta})$, where $g_X,  \lambda '_6$, $\lambda _r$ and $M_{\sigma}$ obtain indirect constraints from phenomenological facts, while $M_{H}$ and $M_{Z_{2}}$ have the lowest bounds from direct searches in colliders. Table \ref{tab:parameters-constraints} summarizes these conclusions.

\section{Elastic Cross Section}

%\subsection{The intermediary particles}
As stated before, the WIMP particle does not interact directly with the ordinary matter. The mechanism to produce scattering is through intermediary particles that couple simultaneusly with the WIMP and the quark content of the nucleus. In the model, these particles are of two types: scalars and vector bosons. For the scalar couplings, due to the symmetries of the model, the only source of scattering is through the $125$ GeV Higgs boson ($h$) and the extra CP-even neutral Higgs boson ($H$), %coming from the real part of the scalar doublets $\phi _1$ and $\phi _2$, written in Tabs. \ref{tab:SM-espectro} and \ref{tab:exotic-espectro}. Specifically, the mass and weak eigenvectors rotate as \cite{martinezochoa1B}
where the mixing angle obeys the constraint from Eq. (\ref{alf-beta-const}) or (\ref{alf-beta-const2}). The vector couplings, on the other hand, occur only with the neutral gauge sector of the model, i.e., through the $Z_1$ and $Z_2$ eigenstates which obey the the constraint from Eq. (\ref{mixing-angle}). %Below, we determine the interactions of the intermediary particles with the WIMP and the quarks. 
%The rotation into mass eigenvectors are:
%The mixing angle obey the constraint from Eq. (\ref{mixing-angle}). Although the photon $A_{ \mu}$ appears in the diagonalization, it will not be relevant because the WIMP is electrically neutral.

\begin{table}[tbp]
\begin{center}
%\caption{\small Constraints for free parameters. There is not a defined limit for $T_{\beta}$ } \vspace{-0.5cm}
%\label{tab:parameters-constraints}
\begin{equation*}
\begin{tabular}{c|ccc}
\hline\hline
\text{Type} & Parameter & Constraint & Source \\ \hline
\text{Coupling} 
& 
\begin{tabular}{c}
$g_X$ \\
$\lambda '_6$ \\
$\lambda '_r$
\end{tabular}
&
\begin{tabular}{c}
$0-0.4$ \\
$0-6$ \\
$0-2$
\end{tabular}
& 
\begin{tabular}{c}
$pp\rightarrow Z_2 \rightarrow \ell^+\ell^-$ \text{and electroweak}\\
\text{Stability and relic density} \\
\text{Stability and relic density} 
\end{tabular}
  \\ \hline
\text{Mass} &
\begin{tabular}{c}
$M_{Z_{2}}$ \\
$M_{H}$ \\
$M_{\sigma }$
\end{tabular}  
& 
\begin{tabular}{c}
$>3$ TeV \\
$>125$ GeV \\
$1 - 1000$ GeV
\end{tabular}  
& 
\begin{tabular}{c}
\text{Colliders} \\
\text{Colliders} \\
\text{Relic density}
\end{tabular}
 \\ \hline
\text{Mixing}& 
$T_{\beta}$  
&
0-$\infty$
& 
-  \\ \hline
\end{tabular}
\end{equation*}%
\end{center}
\vspace{-0.5cm} 
\caption{\small Constraints for free parameters. Although we do not specify a limit for $T_{\beta}$, non appreciable change is observed for $T_{\beta}>10$.} \vspace{-0.5cm}
\label{tab:parameters-constraints}
\end{table}

\subsection{Fundamental couplings}

After identifying the intermediary particles of the interaction, we will need to know how they couple with the WIMP and the quarks. First, from the Higgs potential compatible with the symmetries of the model, the couplings of WIMP with both Higgs bosons are \cite{martinezochoa1B}:

\begin{eqnarray}
V_{\sigma -higgs}&=&\upsilon \lambda_h h\left|\sigma \right|^2+\upsilon \lambda_HH\left|\sigma \right|^2
\label{scalar-potential}
\end{eqnarray}    
where:

\begin{eqnarray}
\lambda_h&=&\left(-\lambda '_6S_{\alpha }C_{\beta }+\lambda '_7C_{\alpha }S_{\beta }\right) \nonumber \\
\lambda_H&=&\left(\lambda '_6C_{\alpha }C_{\beta }+\lambda '_7S_{\alpha }S_{\beta }\right).
\label{higgs-coupling}
\end{eqnarray}
By applying the constraint from (\ref{alf-beta-const2}), and defining the ratio $\lambda _r=\lambda '_7/\lambda '_6$, the above coupling constants become:

\begin{eqnarray}
\lambda_h&=&\lambda '_6\left(\lambda _r-1\right)C_{\beta }S_{\beta } \nonumber \\
\lambda_H&=&\lambda '_6\left(1+\lambda _rT^2_{\beta }\right)C^2_{\beta }.
\label{higgs-coupling-2}
\end{eqnarray}

Second, from the kinetic part of the Higgs Lagrangian, the couplings between $\sigma $ and the gauge neutral bosons are obtained. For the trilineal terms, it is found that \cite{martinezochoa1B}

\begin{eqnarray}
\mathcal{L}_{\sigma -vector } &=& i\overline{g}_1\left(\sigma \overleftrightarrow{\partial _{\mu }}\sigma ^*\right) Z^{\mu }_1+i\overline{g}_2\left(\sigma  \overleftrightarrow{\partial _{\mu }}\sigma ^*\right) Z^{\mu }_2,
\label{gauge-interaction}
\end{eqnarray}
where

\begin{eqnarray}
\overline{g}_1=\frac{g_X}{3}S_{\theta}, \ \ \ \ \overline{g}_2=\frac{g_X}{3}C_{\theta},
\end{eqnarray}
with the definition $a\overleftrightarrow{\partial _{\mu }}b=a\partial _{\mu }b-(\partial _{\mu }a)b$. The mixing angle $S_{\theta }$ is given in (\ref{mixing-angle}). Now, we proceed to write the couplings with the quarks. For the interaction through the scalar bosons, we use the Yukawa Lagrangian in (\ref{quark-yukawa-1}). In particular, the couplings with $h$ and $H$ have the form:

\begin{eqnarray}
-\mathcal{L}_{q-higgs } &=&\frac{1}{\sqrt{2}}\overline{U^{(0)}_L}\left[\left(-S_{\alpha}\eta _1^{U,0}+C_{\alpha}\eta _2^{U,0} \right)h+\left(C_{\alpha}\eta _1^{U,0}+S_{\alpha}\eta _2^{U,0} \right)H\right]U^{(0)}_R \nonumber \\
&+&\frac{1}{\sqrt{2}}\overline{D^{(0)}_L}\left[\left(-S_{\alpha}\eta _1^{D,0}+C_{\alpha}\eta _2^{D,0} \right)h+\left(C_{\alpha}\eta _1^{D,0}+S_{\alpha}\eta _2^{D,0} \right)H\right]D^{(0)}_R \nonumber \\
&+& H.c.,
\label{quark-higgs}
\end{eqnarray}
where $\eta _a^{Q,0}$ are the Yukawa matrices in weak eigenstates. As usual in general 2HDM \cite{prd63095007}, we may impose restrictions to avoid large flavor changing neutral currents, obtaining the equivalents of type I and type II 2HDM. After rotation to mass eigenstates, the Lagrangian can be generically written as:

\begin{eqnarray}
-\mathcal{L}_{q-higgs } &=&\frac{m_U}{\upsilon }\overline{U_L}\left[c_U^{h}h+c_U^{H}H\right]U_R+\frac{m_D}{\upsilon }\overline{D_L}\left[c_D^{h}h+c_D^{H}H\right]D_R + H.c., \nonumber \\
\label{quark-higgs-2}
\end{eqnarray}
where the coefficients $c_Q^{\mathcal{H}}$ are given in table \ref{tab:yukawa-coef} for type I and type II Yukawa couplings, before and after applying the constraint (\ref{alf-beta-const2}). 

\begin{table}[tbp]
\begin{center}
%\caption{\small Coefficients for Type I and II Yukawa couplings, before and after applying the constraint from Eq.  (\ref{alf-beta-const2}) } \vspace{-0.5cm}
%\label{tab:yukawa-coef}
\begin{equation*}
\begin{tabular}{c|cc|cc}
\hline\hline
& \multicolumn{2}{c|}{Type I} & \multicolumn{2}{c}{Type II}  \\ \hline
 & \small{Before (\ref{alf-beta-const2})} &  \small{After (\ref{alf-beta-const2})} & \small{Before (\ref{alf-beta-const2})} &  \small{After (\ref{alf-beta-const2})} \\ \hline
$c_U^h$& $C_{\alpha}/S_{\beta}$  & $1/T_{\beta}$  & $C_{\alpha}/S_{\beta}$ &  $1/T_{\beta}$  \\
$c_D^h$ & $C_{\alpha}/S_{\beta}$ & $1/T_{\beta}$  & $-S_{\alpha}/C_{\beta}$  &  $-T_{\beta}$  \\ 
$c_U^H$ & $S_{\alpha}/S_{\beta}$ & $1$ & $S_{\alpha}/S_{\beta}$ & $1$  \\
$c_D^H$ & $S_{\alpha}/S_{\beta}$ & $1$ & $C_{\alpha}/C_{\beta}$ & $1$ \\ \hline
\end{tabular}
\end{equation*}%
\end{center}
\vspace{-0.5cm}
\caption{\small Coefficients for Type I and II Yukawa couplings, before and after applying the constraint from Eq.  (\ref{alf-beta-const2}) } \vspace{-0.5cm}
\label{tab:yukawa-coef}
\end{table}

Finally, the interactions between the quarks and the neutral gauge bosons arise from the Dirac Lagrangian. For the neutral weak sector, it is parameterized as \cite{martinezochoa1B}:

\begin{eqnarray}
\mathcal{L}_{NW}&=&Z_{1\mu }\overline{Q}\gamma ^{\mu}\left[\frac{g}{2C_W}(v_Q^{SM}-\gamma _5a_Q^{SM})C_{\theta}-\frac{g_X}{2}(v_Q^{NSM}-\gamma _5a_Q^{NSM})S_{\theta}\right]Q \nonumber \\
&-&Z_{2\mu }\overline{Q}\gamma ^{\mu}\left[\frac{g}{2C_W}(v_Q^{SM}-\gamma _5a_Q^{SM})S_{\theta}+\frac{g_X}{2}(v_Q^{NSM}-\gamma _5a_Q^{NSM})C_{\theta}\right]Q, 
\label{weak current}
\end{eqnarray} 
where the vector and axial couplings are defined according to table \ref{tab:vector-axial-couplings}. The above Lagrangian can be written in a simple form if we define rotations into the modified vector and axial couplings as

\begin{eqnarray}
\begin{pmatrix}
\overline{v}_{Q}^{(1)} \\
\\
\overline{v}_{Q}^{(2)}
\end{pmatrix}&=&
\mathcal{R}\begin{pmatrix}
v_Q^{SM} \\
\\
v_Q^{NSM}
\end{pmatrix}, \ \ \ \ \begin{pmatrix}
\overline{a}_{Q}^{(1)} \\
\\
\overline{a}_{Q}^{(2)}
\end{pmatrix}=
\mathcal{R}\begin{pmatrix}
a_Q^{SM} \\
\\
a_Q^{NSM}
\end{pmatrix}
\label{modified-coup}
\end{eqnarray}
where:

\begin{eqnarray}
\mathcal{R}=-\frac{g_X}{g}\begin{pmatrix}
-\frac{g}{g_X}C_{\theta}& C_WS_{\theta} \\
& \\
\frac{g}{g_X}S_{\theta} & C_WC_{\theta}
\end{pmatrix},
\end{eqnarray}
obtaining

\begin{eqnarray}
\mathcal{L}_{NW}&=&\frac{g}{2C_W}\left[\overline{Q}\gamma ^{\mu}\left(\overline{v}_{Q}^{(1)}-\gamma _5\overline{a}_{Q}^{(1)} \right)QZ_{1\mu }+\overline{Q}\gamma ^{\mu}\left(\overline{v}_{Q}^{(2)}-\gamma _5\overline{a}_{Q}^{(2)} \right)QZ_{2\mu } \right].
\label{weak-neutral-lag}
\end{eqnarray}

\begin{table}[tbp]
\begin{center}
%\caption{\small Vector and Axial couplings for the weak neutral currents $Z$ (SM-type) and $Z'$ (non-SM type) and for each quarks, with $n=2,3$ } \vspace{-0.5cm}
%\label{tab:vector-axial-couplings}
\begin{equation*}
\begin{tabular}{c c c c c c}
 \hline \hline \vspace{0.1cm} 
$Quark$&$v_Q^{SM}$&$a_Q^{SM}$&&$v_Q^{NSM}$&$a_Q^{NSM}$   \\  \hline \vspace{0.2cm}
$U^1$&$1/2-4S_W^2/3$&$1/2$&&$-1$&$1/3$\\ \vspace{0.2cm}
$U^n$&$1/2-4S_W^2/3$&$1/2$&&$-2/3$&$2/3$\\ \vspace{0.2cm}
$D^1$&$-1/2+2S_W^2/3$&$-1/2$&&$0$&$-2/3$\\ \vspace{0.2cm}
$D^n$&$-1/2+2S_W^2/3$&$-1/2$&&$1/3$&$-1/3$ \\ \hline
\end{tabular}
\end{equation*}
\end{center}
\vspace{-0.5cm}
\caption{\small Vector and Axial couplings for the weak neutral currents $Z$ (SM-type) and $Z'$ (non-SM type) and for each quarks, with $n=2,3$ } 
\label{tab:vector-axial-couplings}
\end{table}
%It is interesting to note that if the $Z-Z'$ mixing angle is zero, $S_{\theta }=0$ ,the $Z_1=Z$ boson decouple from the WIMP as shown in (\ref{gauge-interaction}), while the $Z_2=Z'$ boson decouple from

\begin{figure}[tb] 
\centering
\includegraphics[width=8cm,height=4.5cm,angle=0]{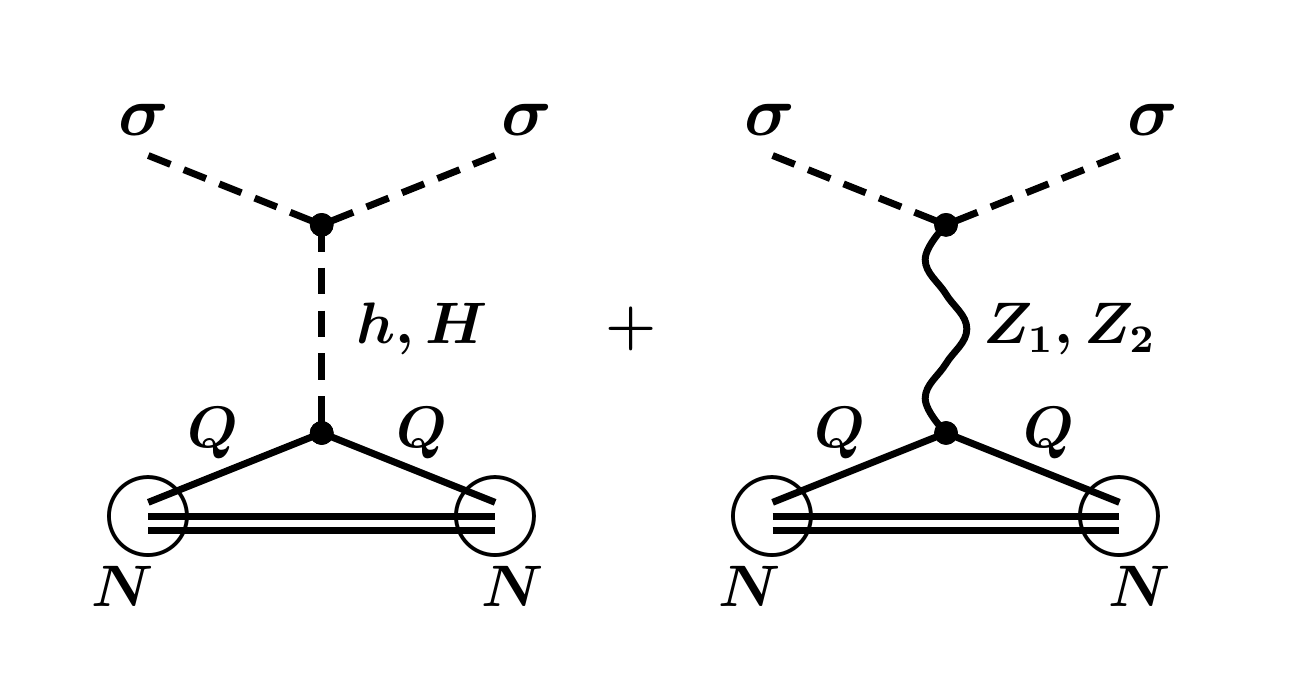}%\vspace{-15cm}
\caption{\small Elastic scattering between the scalar WIMP $\sigma $ and quarks from a nucleon $N$}
\label{fig3}
\end{figure}

In conclusion, the scattering of the WIMP with a nucleus in the model can be mediated by four particles: two scalar bosons (the known Higgs boson $h$ and the extra CP-even Higgs boson $H$) and two gauge bosons (the known neutral $Z_1$ boson and the extra $Z_2$ boson). Thus, the elastic scattering with one nucleon (proton or neutron) at the microscopic level is described by figure \ref{fig3}. As is standard \cite{wimpsA, wimpsB, wimpsC, wimpsD}, the calculation of the nuclear matrix elements starts by obtaining effective couplings for the interaction of the WIMP with the quarks, which is later traslated into effective couplings with nucleons, and finally at nuclear level. These amplitudes will be evaluated at the zero momentum transfer limit.

\subsection{Effective couplings with quarks}

From the two Higgs Lagrangians in (\ref{scalar-potential}) and (\ref{quark-higgs-2}), we obtain the matrix element for the $\sigma Q \rightarrow \sigma Q$ scattering through only scalar particles\footnote{We perform our calculation with the relativistic normalization, defined such that $\langle {\bf p}| {\bf p'}\rangle=2E_{{\bf p}}V\delta _{{\bf p},{\bf p'}}$ for each particle state}:

\begin{eqnarray}
i\mathcal{M}_{S}=\sum _{\mathcal{H}=h,H}\frac{-i\lambda _{\mathcal{H}}m_Qc_Q^{\mathcal{H}}}{q^2-M_{\mathcal{H}}^{2}}\overline{u^{s'}}(k')u^{s}(k),
\label{scalar-element}
\end{eqnarray}
while the gauge contribution from (\ref{gauge-interaction}) and (\ref{weak-neutral-lag}) gives:

\begin{eqnarray}
i\mathcal{M}_{G}=\sum _{\mathcal{Z}=Z_1,Z_2}\frac{\overline{g}_{\mathcal{Z}}g}{2C_W}(p+p')^{\mu }D_{\mu \nu}\overline{u^{s'}}(k')\gamma ^{\nu }\left(\overline{v}_{Q}^{(\mathcal{Z})}-\gamma _5\overline{a}_{Q}^{(\mathcal{Z})}\right)u^{s}(k).
\label{gauge-element}
\end{eqnarray}

In the above expressions, we call $(p, k)$ the momentum of the initial $\sigma$ and $Q$ respectively, $(p', k')$ the corresponding final states and $q$ the momentum of the intermediary particle. $u^s(k)$ is the wave function of a quark with spin $s$ and momentum $k$, while $D_{\mu \nu}$ is the propagator of the intermediary gauge bosons, defined in the  Feynman gauge as:

\begin{eqnarray}
D_{\mu \nu}=\frac{-ig_{\mu \nu}}{q^2-M_{\mathcal{Z}}^{2}}.
\label{gauge-propagator}
\end{eqnarray} 

Then, the total matrix element between final and initial states is the superposition

\begin{eqnarray}
i\mathcal{M}_{fi}=i\mathcal{M}_{S}+i\mathcal{M}_{G}.
\end{eqnarray}

Since the galactic WIMP moves at non-relativistic speeds, the momentum transfer, $q$, through the intermediary particles is negligible in relation to their masses. Thus, the above matrix element at low energies become: 

\begin{eqnarray}
i\mathcal{M}_{fi}^{Low}&=&i\sum _{\mathcal{H}=h,H}S_{\mathcal{H}}m_Qc_Q^{\mathcal{H}}\overline{u^{s'}}(k')u^{s}(k) \nonumber \\
&+&i\sum _{\mathcal{Z}=Z_1,Z_2}G_{\mathcal{Z}}(p+p')^{\mu }\overline{u^{s'}}(k')\gamma _{\mu }\left(\overline{v}_{Q}^{(\mathcal{Z})}-\gamma _5\overline{a}_{Q}^{(\mathcal{Z})}\right)u^{s}(k),
\label{low-element}
\end{eqnarray}
where we have defined the dimensionally inverse mass squared effective couplings:

\begin{eqnarray}
S_{\mathcal{H}}=\frac{\lambda _{\mathcal{H}}}{M_{\mathcal{H}}^{2}}, \ \ \ \ \  G_{\mathcal{Z}}=\frac{\overline{g}_{\mathcal{Z}}g}{2C_WM_{\mathcal{Z}}^{2}}.
\label{effective-couplings}
\end{eqnarray}

The matrix element in (\ref{low-element}) can be derived from the following effective Lagrangian:

\begin{eqnarray}
\mathcal{L}_{Q}&=&\sum _{\mathcal{H}=h,H}S_{\mathcal{H}}c_Q^{\mathcal{H}}m_Q\overline{Q}Q|\sigma |^2
-i\sum _{\mathcal{Z}=Z_1,Z_2}G_{\mathcal{Z}}\left(\sigma \overleftrightarrow{\partial ^{\mu }}\sigma ^*\right) \left(\overline{Q}\gamma _{\mu }\overline{v}_{Q}^{(\mathcal{Z})}Q\right) \nonumber \\
&+&i\sum _{\mathcal{Z}=Z_1,Z_2}G_{\mathcal{Z}}\left(\sigma \overleftrightarrow{\partial ^{\mu }}\sigma ^*\right) \left(\overline{Q}\gamma _{\mu }\gamma _5\overline{a}_{Q}^{(\mathcal{Z})}Q\right),
\label{efective-lagr}
\end{eqnarray}
where we have separated the vector interactions (the $\gamma _{\mu}$ term) from the vector-axial one ($\gamma _{\mu }\gamma _{5}$ term).

\subsection{Effective couplings with cucleons\label{Secc:nucleon-couplings}}

Now,  based on the effective interactions with quarks from (\ref{efective-lagr}), we can calculate the matrix elements for nucleons. For this calculation, we will describe the nucleon as a bound state composed by valence quarks, virtual sea of quark-antiquark pairs and gluons. This bound state can be described as a spinor with momentum $k_N=(E_N, \mathbf{k}_N)$ and the following free-particle wave function:

\begin{eqnarray}
u^{s}(k_N)=\sqrt{E_N+M_N}\begin{pmatrix}
\xi ^s  \\
\frac{\mathbf{k}_N \cdot \pmb{\sigma }}{E_N+M_N}\xi ^s
\end{pmatrix},
\label{spinor}
\end{eqnarray}    
with $M_N$ the mass of the nucleon, $s$ the internal spin state, $1/2$ or $-1/2$. and $\pmb{\sigma }$ the Pauli matrices. Calling $| \sigma \rangle =|p, 0\rangle$ the states of WIMP with momentum $p$ and spin $0$, $| N \rangle =|k_N, s\rangle$ the states of the nucleon before the interaction, and $| \sigma ' \rangle =|p', 0\rangle$, $| N' \rangle =|k'_N, s'\rangle$ the corresponding states after the interaction, then the matrix element for the $\sigma N \rightarrow \sigma N$ scattering is:

\begin{eqnarray}
\mathcal{M}_{fi}&=&\sum _Q \langle \sigma ', N' | \mathcal{L}_{Q} | N, \sigma \rangle,
\label{nucleon-element}
\end{eqnarray}
with $\mathcal{L}_{Q}$ defined in Eq. (\ref{efective-lagr}). For scalar elements, we obtained:

\begin{eqnarray}
\langle \sigma ' ||\sigma |^2|\sigma \rangle &=& 1\nonumber \\
\langle \sigma ' |\sigma \overleftrightarrow{\partial ^{\mu }}\sigma ^* |\sigma \rangle&=&i(p'+p)^{\mu}.
\label{scalar-amplituds}
\end{eqnarray}
For the nucleon elements, as detailed in the appendix \ref{App:Form factors}, we obtained:

\begin{eqnarray}
\sum _Q\langle N' |c_Q^{\mathcal{H}}m_Q\overline{Q}Q| N \rangle &=& \overline{u^{s'}}(k'_N)M_NF_N^{\mathcal{H}}u^{s}(k_N) \nonumber \\
\sum _Q\langle N' | \overline{Q}\gamma _{\mu }\overline{v}_{Q}^{(\mathcal{Z})}Q| N \rangle &=&\overline{u^{s'}}(k'_N)\gamma _{\mu }V_N^{\mathcal{Z}}u^{s}(k_N) \nonumber \\
\sum _Q \langle N' | \overline{Q}\gamma _{\mu }\gamma _5\overline{a}_{Q}^{(\mathcal{Z})}Q| N \rangle &=&\overline{u^{s'}}(k'_N)\Gamma _{\mu ,5}^{(\mathcal{Z})}(q^2)u^{s}(k_N),
\label{nucleon-amplitudes}
\end{eqnarray}
where the form factors $F_N^{\mathcal{H}}$ and $V_N^{\mathcal{Z}}$ are defined in Eqs. (\ref{scalar-formfactor}) and (\ref{vector-formfactor}) of the appendix \ref{App:Form factors}. The other form factor, $\Gamma _{\mu ,5}^{(\mathcal{Z})}(q^2)$, is defined in Eqs. (\ref{axial-formfactor}) or (\ref{axial-formfactor-2}), which depends on the function $f_N^{\mathcal{Z}}(q^2)$ written in Eq. (\ref{spin-formfactor-zero}) in the limit of zero momentum transfer. The matrix element from (\ref{nucleon-element}) is solved by applying the amplitudes (\ref{scalar-amplituds}) and (\ref{nucleon-amplitudes}), with the explicit form of the wave function (\ref{spinor}). It is simpler if we choose the inertial system where the initial nucleus is at rest, i.e., where the initial momentum of the nucleon is $k_N=(M_N,\mathbf{0})$. After making all the above replacements, we obtained the following amplitude:

\begin{eqnarray}
\mathcal{M}_{fi}&=&\sum _{\mathcal{H}}S_{\mathcal{H}}M_NF_N^{\mathcal{H}} \sqrt{2M_N\left(E'_N+M_N\right)}(\xi ^{s'})^{\dagger }\xi ^s \nonumber \\
&+&\sum _{\mathcal{Z}}G_{\mathcal{Z}}V_N^{\mathcal{Z}}(p'+p)^{\mu}(\xi ^{s'})^{\dagger }\left[a_{\mu }\sqrt{2M_N\left(E'_N+M_N\right)}+\sigma _{\mu }\left(\mathbf{k'}_N \cdot \pmb{\sigma}^{\dagger} \right) \right]\xi ^s  \nonumber \\
&-&\sum _{\mathcal{Z}}\frac{1}{2}G_{\mathcal{Z}}f_N^{\mathcal{Z}}(q^2)(p'+p)^{\mu}(\xi ^{s'})^{\dagger }\left[\sigma _{\mu }\sqrt{2M_N\left(E'_N+M_N\right)}+a_{\mu } \left(\mathbf{k'}_N \cdot \pmb{\sigma}^{\dagger}\right)  \right]\xi ^s, \nonumber \\
\label{nucleon-element-2}
\end{eqnarray}  
where the vectors $a_{\mu }$ and  $\sigma _{\mu }$ are defined in (\ref{c-pauli-matrices}). In terms of the momentum that the WIMP transfer to each nucleon, $q=p-p'=k'_N-k_N$, and the invariants

\begin{eqnarray}
(p'+p)^{\mu}a_{\mu }&=&2E_{\sigma }-q^0, \nonumber \\
(p'+p)^{\mu}\sigma _{\mu }&=&\left(\mathbf{q}-2\mathbf{p}\right) \cdot \pmb{\sigma },
\label{invariants}
\end{eqnarray}
we can obtain the complete amplitude at finite momentum transfer. However,
 %\begin{eqnarray}
%\mathcal{M}_{fi}&=&\sum _{\mathcal{H}}S_{\mathcal{H}}M_NF_N^{\mathcal{H}} \sqrt{2M_N\left(2M_N+q^0\right)}(\xi ^{s'})^{\dagger }\xi ^s \nonumber \\
%&+&\sum _{\mathcal{Z}}G_{\mathcal{Z}}V_N^{\mathcal{Z}}(\xi ^{s'})^{\dagger }\left[\left(2E_{\sigma }-q^0\right)\sqrt{2M_N\left(2M_N+q^0\right)}+\left[\left(\mathbf{q}-2\mathbf{p}\right) \cdot \pmb{\sigma }\right]\left[\mathbf{q} \cdot \pmb{\sigma}^{\dagger}\right]  \right]\xi ^s  \nonumber \\
%&-&\sum _{\mathcal{Z}}\frac{1}{2}G_{\mathcal{Z}}f_N^{\mathcal{Z}}(q^2)(\xi ^{s'})^{\dagger }\left[\left(\mathbf{q}-2\mathbf{p}\right) \cdot \pmb{\sigma }\sqrt{2M_N\left(2M_N+q^0\right)}+\left(2E_{\sigma }-q^0\right)\left( \mathbf{q} \cdot \pmb{\sigma}^{\dagger} \right) \right]\xi ^s,\nonumber \\
%\label{nucleon-element-3}
%\end{eqnarray} 
at this stage it is convenient to define the amplitude in the limit of zero momentum transfer, i.e. the limit of  (\ref{nucleon-element-2}) when $(q^0, \mathbf{q})=(0,\mathbf{0})$, from where the amplitude becomes:

\begin{eqnarray}
\mathcal{M}_{0}&=&\sum _{\mathcal{H}}2S_{\mathcal{H}}M_N^{2}F_N^{\mathcal{H}}(\xi ^{s'})^{\dagger }\xi ^s \nonumber \\
&+&\sum _{\mathcal{Z}}4G_{\mathcal{Z}}V_N^{\mathcal{Z}}\left[E_{\sigma}M_N  \right](\xi ^{s'})^{\dagger }\xi ^s  \nonumber \\
&+&\sum _{\mathcal{Z}}2G_{\mathcal{Z}}f_N^{\mathcal{Z}}(0)M_N(\xi ^{s'})^{\dagger }\left(\mathbf{p} \cdot \pmb{\sigma }\right)\xi ^s.\nonumber \\
\label{nucleon-element-zerotransfer}
\end{eqnarray} 
If we describe the spin part of the nucleon through the two-component operator $\chi _N$, the above element can be obtained from the following Lagrangian at zero momentum transfer:

\begin{eqnarray}
\mathcal{L}_{0}^{N}&=&\sum _{\mathcal{H}}2S_{\mathcal{H}}M_N^{2}F_N^{\mathcal{H}}|\sigma |^2\overline{\chi _N}\chi _N \nonumber \\
&+&\sum _{\mathcal{Z}}4G_{\mathcal{Z}}V_N^{\mathcal{Z}}\left[E_{\sigma}M_N  \right]|\sigma |^2\overline{\chi _N}\chi _N\nonumber \\
&+&\sum _{\mathcal{Z}}2G_{\mathcal{Z}}f_N^{\mathcal{Z}}(0)M_N|\sigma |^2\overline{\chi _N}\left(\mathbf{p} \cdot \pmb{\sigma }\right)\chi _N.
\label{nucleon-lagrangian}
\end{eqnarray}

\subsection{Nuclear amplitude}

The final stage is to calculate the nuclear matrix elements. At zero momentum transfer and in the rest system of the nucleus, the ground state of the nucleus is determined by the ket $|A, m_j\rangle$, where $A$ is the number of nucleons and $j$ is the angular momentum with internal states $m_j=-j, -j+1,..., j-1, j$ (before interactions). The WIMP-nucleus scattering amplitude is:

\begin{eqnarray}
\mathcal{M}_{0}^{A}&=&\sum _N\langle \sigma ' ; A, m'_j | \mathcal{L}_{0}^{N} | A, m_j; \sigma \rangle \nonumber \\
&=&\sum _N\left[\sum _{\mathcal{H}}2S_{\mathcal{H}}\langle A, m'_j |\overline{\chi _N}M_N^{2}F_N^{\mathcal{H}}\chi _N| A, m_j \rangle \right. \nonumber \\
&&\ \ \ \ +\sum _{\mathcal{Z}}4G_{\mathcal{Z}}E_{\sigma}  \langle A, m'_j |\overline{\chi _N}V_N^{\mathcal{Z}}M_N\chi _N| A, m_j \rangle \nonumber \\
&&\ \ \ \ \left.+\sum _{\mathcal{Z}}2G_{\mathcal{Z}}\langle A, m'_j |\overline{\chi _N}f_N^{\mathcal{Z}}(0)M_N\left(\mathbf{p} \cdot \pmb{\sigma }\right)\chi _N| A, m_j \rangle \right].
\label{nuclear-amplitude-zero}
\end{eqnarray}
where the first equation from (\ref{scalar-amplituds}) was applied for the scalar elements, and the first sum is over all the nucleons. Each nuclear amplitude is obtained by coherently adding the nucleon factors through the nuclear wave function \cite{wimpsA, wimpsB, wimpsC}. At zero momentum transfer, as shown in appendix \ref{App:nuclear-amplit}, each amplitude gives:

\begin{eqnarray}
\langle A, m'_j |\overline{\chi _N}M_N^{2}F_N^{\mathcal{H}}\chi _N| A, m_j \rangle&=& AM_N^{2}F_N^{\mathcal{H}}\delta _{m'_{j}m_{j}} \nonumber \\
 \langle A, m'_j |\overline{\chi _N}V_N^{\mathcal{Z}}M_N\chi _N| A, m_j \rangle&=&AV_N^{\mathcal{Z}}M_N\delta _{m'_{j}m_{j}} \nonumber \\
\langle A, m'_j |\overline{\chi _N}f_N^{\mathcal{Z}}(0)M_N\left(\mathbf{p} \cdot \pmb{\sigma }\right)\chi _N| A, m_j \rangle &=& \frac{1}{j}Af_N^{\mathcal{Z}}(0)M_N\langle S_A^{N} \rangle\mathbf{p} \cdot\langle m'_j|\mathbf{J}|m_j \rangle,
\label{nuclear-amplitudes}
\end{eqnarray}
where $\langle S_A^{N} \rangle$ is the expectation value for a nucleon $N$ to have spin in the direction of the total angular momentum of the nucleus $A$, and $\mathbf{J}$ is the angular momentum operator of the nucleus. With (\ref{nuclear-amplitudes}) the matrix element in (\ref{nuclear-amplitude-zero}) becomes:

\begin{eqnarray}
\mathcal{M}_{0}^{A}&=&\sum _N\left[\sum _{\mathcal{H}}2AS_{\mathcal{H}}M_N^{2}F_N^{\mathcal{H}}\delta _{m'_{j}m_{j}} \right. \nonumber \\
&&\ \ \ \ +\sum _{\mathcal{Z}}4AG_{\mathcal{Z}}E_{\sigma}V_N^{\mathcal{Z}}M_N\delta _{m'_{j}m_{j}}\nonumber \\
&&\ \ \ \ \left.+\sum _{\mathcal{Z}}2AG_{\mathcal{Z}}f_N^{\mathcal{Z}}(0)M_N\frac{\langle S_A^{N} \rangle}{j}\mathbf{p} \cdot \langle m'_j|\mathbf{J}|m_j \rangle \right].
\label{nuclear-amplitude-zero2}
\end{eqnarray}
We can see that the first two terms of the above element do not depend on the spin variables, while the last one does. Thus, it is convenient to separate the amplitude into spin-independent (SI) and spin-dependent (SD) interactions:

 \begin{eqnarray}
 \mathcal{M}_{0}^{SI}&=&\sum _N\left[\sum _{\mathcal{H}}2AS_{\mathcal{H}}M_N^{2}F_N^{\mathcal{H}}\delta _{m'_{j}m_{j}}+\sum _{\mathcal{Z}}4AG_{\mathcal{Z}}E_{\sigma}V_N^{\mathcal{Z}}M_N\delta _{m'_{j}m_{j}}\right] \nonumber \\
\mathcal{M}_{0}^{SD}&=&\sum _N\sum _{\mathcal{Z}}2AG_{\mathcal{Z}}f_N^{\mathcal{Z}}(0)M_N\frac{\langle S_A^{N} \rangle}{j}\mathbf{p} \cdot \langle m'_j|\mathbf{J}|m_j \rangle.
\label{SI-SD-amplitudes}
 \end{eqnarray} 
For the SI amplitude, we can factorize terms as follows:

 \begin{eqnarray}
 \mathcal{M}_{0}^{SI}&=&4AM_NE_{\sigma}\delta _{m'_{j}m_{j}}\sum _N\left[\sum _{\mathcal{H}}\frac{M_N}{2E_{\sigma}}S_{\mathcal{H}}F_N^{\mathcal{H}}+\sum _{\mathcal{Z}}G_{\mathcal{Z}}V_N^{\mathcal{Z}}\right].
 \label{SI-amplitude}
 \end{eqnarray}
Since the speed of the WIMP is non-relativistic, and the mass of protons and neutrons are almost the same, we will take $E_{\sigma}=M_{\sigma}$ and $M_N=M_p=M_n$. It is usual to parameterize the nucleon mass in terms of the proton mass. Thus, if we define the effective WIMP-nucleon coupling as:

\begin{eqnarray}
f_N=\frac{M_p}{2M_{\sigma}}\sum _{\mathcal{H}=h,H}S_{\mathcal{H}}F_N^{\mathcal{H}}+\sum _{\mathcal{Z}=Z_1,Z_2}G_{\mathcal{Z}}V_N^{\mathcal{Z}},
\label{effective-coup}
\end{eqnarray}
and taking $AM_N=M_A$ as the total mass of the nucleus, then the SI amplitude can be written in a short form as:

 \begin{eqnarray}
 \mathcal{M}_{0}^{SI}&=&4M_AM_{\sigma}\delta _{m'_{j}m_{j}}\sum _Nf_N.
 \label{SI-amplitude}
 \end{eqnarray}
It is interesting to note that, in general, the coupling $f_N$ in (\ref{effective-coup}) is not the same for protons and neutrons. In fact, two sources of isospin asymmetry arise. First, through the form factor $F_N^{\mathcal{H}}$ defined in Eq. (\ref{scalar-formfactor}). Analogous as obtained by authors in reference \cite{gunion} in the framework of a generic 2HDM, the coefficients $c_Q^{\mathcal{H}}$ are different for quarks $u$ and $d$ in Type II models, which leads us to different interactions between protons and neutrons, since the $(u,d)$ content are different. Second, through the vector coupling $V_N^{\mathcal{Z}}$, which also is different for protons and neutrons, as shown in Eq. (\ref{vector-formfactor}). It is convenient to separate the sum over nucleons into protons ($N=p$) and neutrons ($N=n$). If there are $Z$ protons and $A-Z$ neutrons, the Eq. (\ref{SI-amplitude}) is written as:

 \begin{eqnarray}
 \mathcal{M}_{0}^{SI}&=&4M_AM_{\sigma}\delta _{m'_{j}m_{j}}\left[f_pZ+f_n(A-Z)\right]\nonumber \\
&=&4M_AM_{\sigma}\delta _{m'_{j}m_{j}}f_p\left[Z+\frac{f_n}{f_p}(A-Z)\right].
 \label{SI-amplitude2}
 \end{eqnarray}

On the other hand, the SD amplitude in (\ref{SI-SD-amplitudes}) can be written as:

 \begin{eqnarray}
\mathcal{M}_{0}^{SD}&=&2M_A\mathbf{p} \cdot \langle m'_j|\mathbf{J}|m_j \rangle\sum _{\mathcal{Z}}G_{\mathcal{Z}}\left[\frac{1}{j}\sum _Nf_N^{\mathcal{Z}}(0)\langle S_A^{N} \rangle\right]
\label{SD-amplitude}
 \end{eqnarray}
where, again, we define the nuclear mass as $AM_N=M_A$. However, not all the nucleons contribute to the total angular momentum. Only those spins with the expectation of pointing in the same direction, and that does not cancel with an opposite spin, will contribute to the nuclear spin. As usual \cite{wimpsA, wimpsB, wimpsC, wimpsD}, we generalize the individual expectation value $\langle S_A^{N} \rangle$ to the expectation value for a group of protons (neutrons) to contribute to the nuclear spin, $\langle S_{p(n)} \rangle$, such that:

\begin{eqnarray}
\sum _Nf_N^{\mathcal{Z}}(0)\langle S_A^{N} \rangle=f_p^{\mathcal{Z}}(0)\langle S_p \rangle+f_n^{\mathcal{Z}}(0)\langle S_n \rangle.
\label{average-spin}
\end{eqnarray} 
If the spin coupling parameter is defined as:

\begin{eqnarray}
\Lambda _{\mathcal{Z}}=\frac{1}{j}\left[f_p^{\mathcal{Z}}(0)\langle S_p \rangle+f_n^{\mathcal{Z}}(0)\langle S_n \rangle\right],
\label{spin-coup}
\end{eqnarray}
and taking into account the known relation $\mathbf{p}=E_{\sigma }\pmb{\beta }\approx M_{\sigma }\pmb{\beta }$ with $\pmb{\beta }=\mathbf{v}/c$, the SD amplitude in (\ref{SD-amplitude}) becomes:

\begin{eqnarray}
\mathcal{M}_{0}^{SD}&=&2M_AM_{\sigma }\pmb{\beta } \cdot \langle m'_j|\mathbf{J}|m_j \rangle\sum _{\mathcal{Z}=Z_1,Z_2}G_{\mathcal{Z}}\Lambda _{\mathcal{Z}}.
\label{SD-amplitude2}
 \end{eqnarray}
The spin expectation values $\langle S_N \rangle$ can be calculated through different nuclear models, for example, the odd-group model \cite{wimpsC}. In table \ref{tab:nuclear-parameters} we list values of the spin parameters for some isotopes, including other nuclear parameters.

In conclusion, the nuclear matrix element for WIMP-nucleus scattering at zero momentum transfer is:

\begin{eqnarray}
\mathcal{M}_{0}^{A}=\mathcal{M}_{0}^{SI}+\mathcal{M}_{0}^{SD},
\label{nuclear-amplitude-final}
\end{eqnarray}
where the SI and SD amplitudes are given by (\ref{SI-amplitude2}) and (\ref{SD-amplitude2}).

\begin{table}[tbp]
\begin{center}
%\caption{\small Nuclear parameters for 4 isotopes, incluing mass ($M_A$), number of protons ($Z$), spin expectation per nucleon ($\langle S_N \rangle$) and total spin $j$ \cite{wimpsA, wimpsB, wimpsC, wimpsD}} \vspace{-0.5cm}
%\label{tab:nuclear-parameters}
\begin{equation*}
\begin{tabular}{cccccc}
 \hline \hline \vspace{0.1cm} 
Nucleus & $M_A$ (GeV) & $Z$ & $\langle S_p \rangle$ &  $\langle S_n \rangle$ & $j$ \\  \hline \vspace{0.2cm}
$^{19}F$ & $18$ & $9$ & $0.46$ & $0$ & $1/2$\\ \vspace{0.2cm}
$^{29}Si$ & $27$ & $14$ & $0$ & $0.15$ & $1/2$\\ \vspace{0.2cm}
$^{73}Ge$ & $68.5$ & $32$ & $0$ & $0.491$ & $9/2$\\ \vspace{0.2cm}
$^{131}Xe$ & $123$ & $54$ & $0$ & $-0.166$ & $3/2$ \\ \hline
\end{tabular}
\end{equation*}
\end{center}
\vspace{-0.5cm}
\caption{\small Nuclear parameters for 4 isotopes, incluing mass ($M_A$), number of protons ($Z$), spin expectation per nucleon ($\langle S_N \rangle$) and total spin $j$ \cite{wimpsA, wimpsB, wimpsC, wimpsD}} \vspace{-0.5cm}
\label{tab:nuclear-parameters}
\end{table}

\subsection{Cross section}

For any polarized $2\rightarrow 2$ elastic process, the differential cross section at finite momentum transfer is (in the relativistic normalization):

\begin{eqnarray}
\frac{d\sigma }{d|\mathbf{q}|^2}=\frac{|\mathcal{M}_{fi}|^2}{16\pi E_1^{2}E_2^{2}v^2},
\end{eqnarray}
where $E_{1,2}$ are the energies of the incoming particles, and $v$ is its relative speed. In the non-relativistic limit, $E_{1,2}\approx M_{1,2}$. In the case of the WIMP-nucleus cross section, we are interested in obtaining the limit of zero momentum transfer, where $\mathcal{M}_{fi}=\mathcal{M}_{0}^{A}$ does not depend on $|\mathbf{q}|^2$. Thus, the cross section at zero momentum transfer gives \cite{wimpsA, wimpsB, wimpsC}:

\begin{eqnarray}
\sigma _0= \int _0^{4m_r^{2}v^2}\frac{d\sigma (|\mathbf{q}|=0)}{d|\mathbf{q}|^2}d|\mathbf{q}|^2=\int _0^{4m_r^{2}v^2} \frac{|\mathcal{M}_{0}^{A}|^2}{16\pi M_{\sigma }^{2}M_A^{2}v^2}d|\mathbf{q}|^2=\frac{m_r^{2}}{4\pi M_{\sigma }^{2}M_A^{2}}|\mathcal{M}_{0}^{A}|^2,
\label{cross-section-zerotransfer}
\end{eqnarray}
where $m_r=M_{\sigma }M_A/(M_{\sigma }+M_A)$ is the reduced mass of the WIMP-nucleus system\footnote{In the non-relativistic normalization, the amplitude is defined as $M_{fi}=\mathcal{M}_{fi}/4M_1M_2$, such that the cross section in (\ref{cross-section-zerotransfer}) is equivalent to the usual expression $4m_r^{2}|M_{0}^{A}|^2/\pi $}. If the experiment does not measure the polarization of the particles, we must average the initial spin states and add all the possible final states. Thus, we obtain the unpolarized cross section:

\begin{eqnarray}
\sigma _0^{unpol}=\frac{m_r^{2}}{4\pi M_{\sigma }^{2}M_A^{2}}\frac{\overline{|\mathcal{M}_{0}^{A}|^2}}{(2s+1)(2j+1)},
\label{unpol-cross-section}
\end{eqnarray}
where $s$ is the spin of the incoming WIMP and $j$ the spin of the incoming nucleus, while the average amplitude is:

\begin{eqnarray}
\overline{|\mathcal{M}_{0}^{A}|^2}=\sum _{m_{s}}\sum _{m'_{s}}\sum _{m_{j}}\sum _{m'_{j}}|\mathcal{M}_{0}^{A}|^2,
\label{average-amplitude}
\end{eqnarray}
where $m_{s}, m'_{s}$, $m_{j}$ and $m'_j$ are the internal spin projections of the incoming and outcoming particles. In our case, since $s=s'=0$ (scalar WIMP), there is neither the sum over $m_{s}$ nor over $m'_s$. For the nucleus, both $m_{j}$ and $m'_{j}$ add over $-j, -j+1,..., j-1, j$. Thus, after applying the amplitude from (\ref{nuclear-amplitude-final}) and adding the spin states, as shown in appendix \ref{App.cross section}, we obtain the unpolarized cross section:

\begin{eqnarray}
\sigma _0^{unpol}=\sigma _0^{SI}+\sigma _0^{SD},
\label{unpol-cross-section2}
\end{eqnarray}
where the SI and SD cross sections are:

\begin{eqnarray}
\sigma _0^{SI}&=&\frac{4m_r^{2}}{\pi }f_p^{2}\left|Z+\frac{f_n}{f_p}(A-Z)\right|^2, \nonumber \\
\sigma _0^{SD}&=&\frac{4m_r^{2}}{3\pi }\left|\pmb{\beta }\right|^2j(j+1)\left|\sum _{\mathcal{Z}=Z_1,Z_2}G_{\mathcal{Z}}\Lambda _{\mathcal{Z}}\right|^2,
\label{SI-SD-cross-sections}
\end{eqnarray}
where $|\pmb{\beta }|=v/c$ and $f_N=f_p$ or $f_n$ is given by (\ref{effective-coup}).

\section{Interferences in Spin-independent Interactions}

Before exploring some phenomenological consequences, it is important to normalize our theoretical equations in terms of the parameters provided by the different experimental collaborations, which report the cross sections normalized to a single proton and by assuming that the interactions with protons and neutrons are the same. For the SI interactions, it is equivalent to make $f_p=f_n$ in Eq. (\ref{SI-SD-cross-sections}), which we will call the ``measured'' coupling $\overline{f}$ and the measured cross section $\overline{\sigma }_0^{SI}$:

\begin{eqnarray}
\overline{\sigma }_0^{SI}=\frac{4m_{r}^{2}}{\pi }A^2\overline{f}^{2}.
\label{measured-cross}
\end{eqnarray}
Thus, the reported one-proton cross section (for $A=Z=1$) is:

\begin{eqnarray}
\overline{\sigma }_p^{SI}=\frac{4m_{p\sigma}^{2}}{\pi }\overline{f}^{2},
\label{measured-proton-cross}
\end{eqnarray}      
where $m_{p\sigma}=M_{\sigma }M_p/(M_{\sigma }+M_p)$ is the WIMP-proton reduced mass. On the other hand, our theoretical couplings $f_p$ and $f_n$ do not match, in general, with the experimental one $\overline{f}$. Thus, the predicted one-proton cross section

\begin{eqnarray}
\sigma _p^{SI}=\frac{4m_{p\sigma}^{2}}{\pi }f_p^{2},
\label{predicted-proton-cross}
\end{eqnarray} 
does not in general coincide with (\ref{measured-proton-cross}). In order to do comparisons, we must introduce a normalization factor into our theoretical cross sections. For that we will match the predicted nuclear SI cross section from (\ref{SI-SD-cross-sections}) with the measured one from (\ref{measured-cross}), obtaining:

\begin{eqnarray}
\text{if}\ \ \overline{\sigma }_0^{SI}=\sigma _0^{SI} \ \ \ \ \Rightarrow \ \ \ \ \ f_p^{2}=\overline{f}^2\frac{A^2}{\left|Z+\frac{f_n}{f_p}(A-Z)\right|^2},
\end{eqnarray}    
which lead us for the one-proton cross sections:

\begin{eqnarray}
\sigma _p^{SI}=\frac{4m_{p\sigma}^{2}}{\pi }\overline{f}^2\frac{A^2}{\left|Z+\frac{f_n}{f_p}(A-Z)\right|^2}=\overline{\sigma }_p^{SI}\frac{A^2}{\left|Z+\frac{f_n}{f_p}(A-Z)\right|^2}. 
\label{normalized-proton-cross}
\end{eqnarray} 
By calling

\begin{eqnarray}
N=\frac{\left|Z+\frac{f_n}{f_p}(A-Z)\right|^2}{A^2},
\end{eqnarray}
we find that our predicted cross section for a single proton must be normalized as

\begin{eqnarray}
\sigma _p^{SI} \longrightarrow N\sigma _p^{SI},
\label{normalization}
\end{eqnarray}
in order to compare with the experimental limits.

\subsection{Interactions with Higgs exchange\label{sec:higgs-interaction}}

From (\ref{SI-SD-cross-sections}), we can see that the SI cross section can exhibit different scenarios for destructive interference. For the total interference, we obtain that

\begin{eqnarray}
\sigma _0^{SI}=0 \ \ \ \text{if:} \ \ \begin{cases}
\frac{f_n}{f_p}=-\frac{Z}{A-Z}, \ \ \ \text{or}  \\ 
f_p=f_n=0.
\end{cases} 
\label{total-interference}
\end{eqnarray}
The former requires isospin violation in the WIMP-nucleon interaction, while the latter can arise from a quantum interference between different intermediary channels of the interaction, and must be symmetric between protons and neutrons. In addition, with isospin violation we find two extreme scenarios with partial interference, where the cross section cancels for only protons or only neutrons:

\begin{eqnarray}
\text{for }f_p\neq f_n\ \ : \begin{cases}
f_p=0\ \ \ \text{proton-phobic interactions}  \\ 
f_n=0\ \ \ \text{neutron-phobic interactions}.
\end{cases} 
\label{isospin-interference}
\end{eqnarray}

In particular, as suggested by some authors \cite{isospinviolationA, isospinviolationB, isospinviolationC, isospinviolationD}, an interesting option arises when interference by isospin asymmetry cancels the interaction for Xenon-based detectors, which require a ratio $f_n/f_p \approx -0.7$, giving, as far as we know, the most popular approximation to understand the tensions of the experimental data between Xenon- and other-based detectors.

On the other hand, taking into account the definition in Eq. (\ref{effective-coup}), the effective proton and neutron couplings occur from exchanges between Higgs and gauge bosons. To explore the effects from each channel, in this section we first ``turn off'' the gauge interactions by making $g_X=0$, obtaining:

\begin{figure}[t]
\centering
\includegraphics[width=5cm,height=4cm,angle=0]{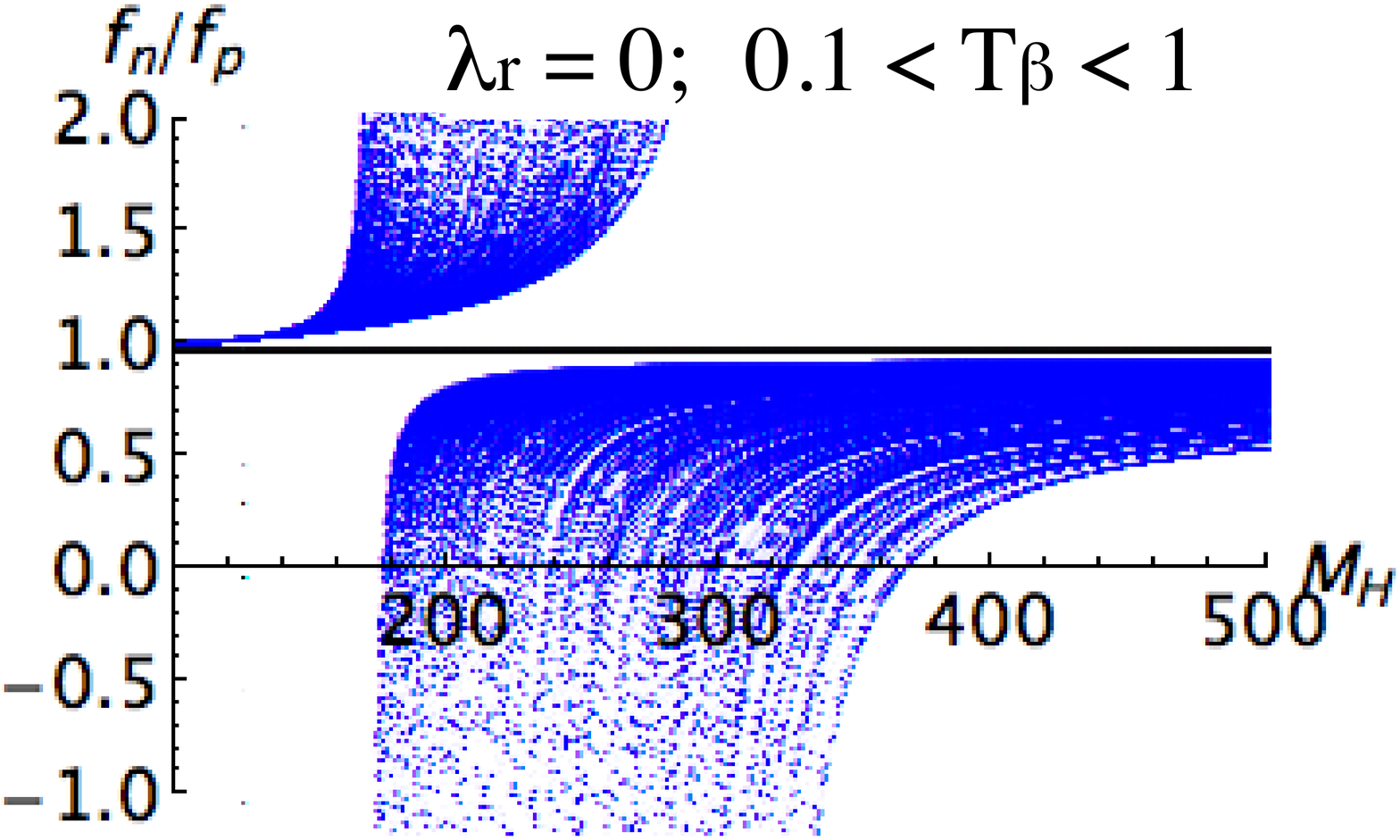}
\includegraphics[width=5cm,height=4cm,angle=0]{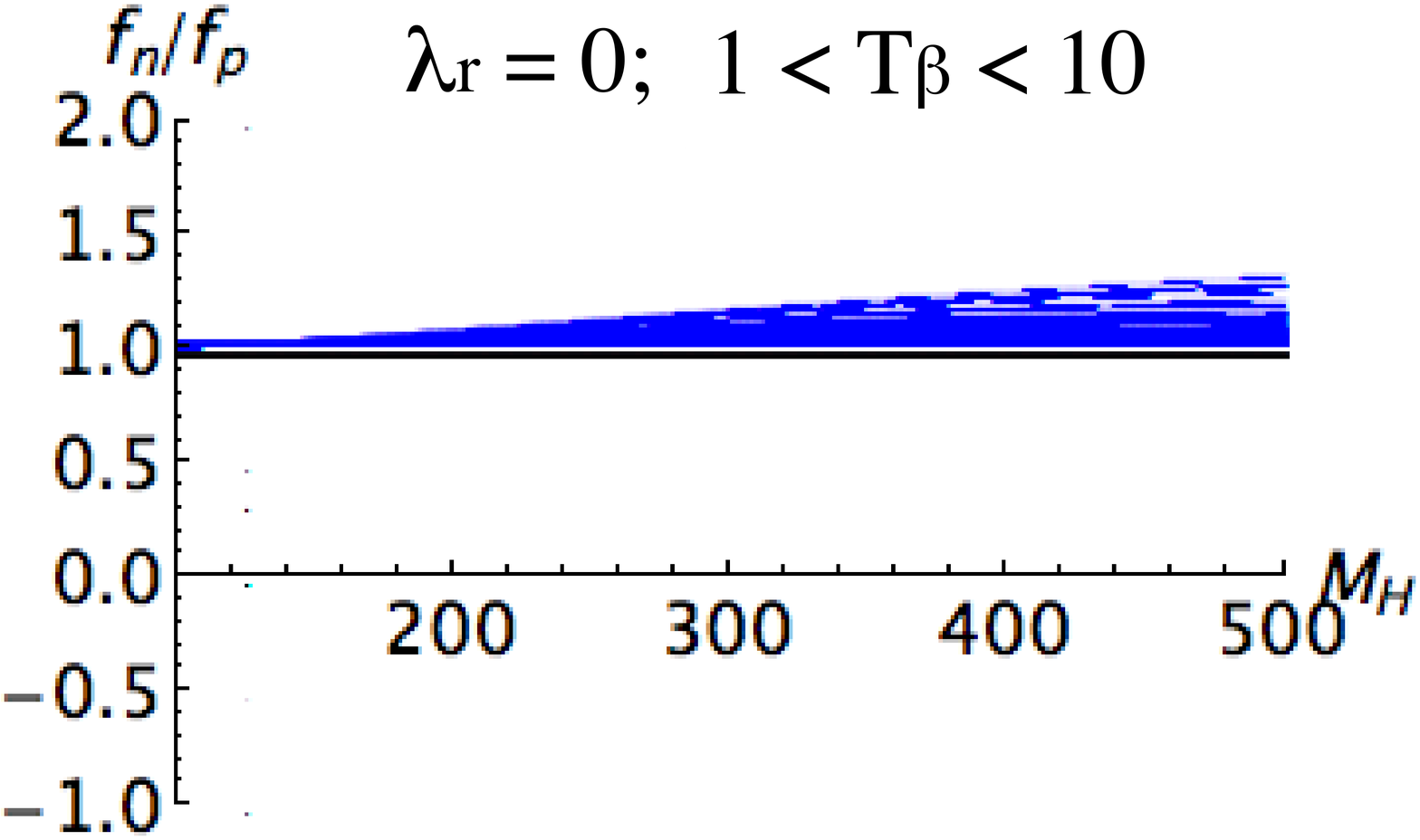}
\includegraphics[width=5cm,height=4cm,angle=0]{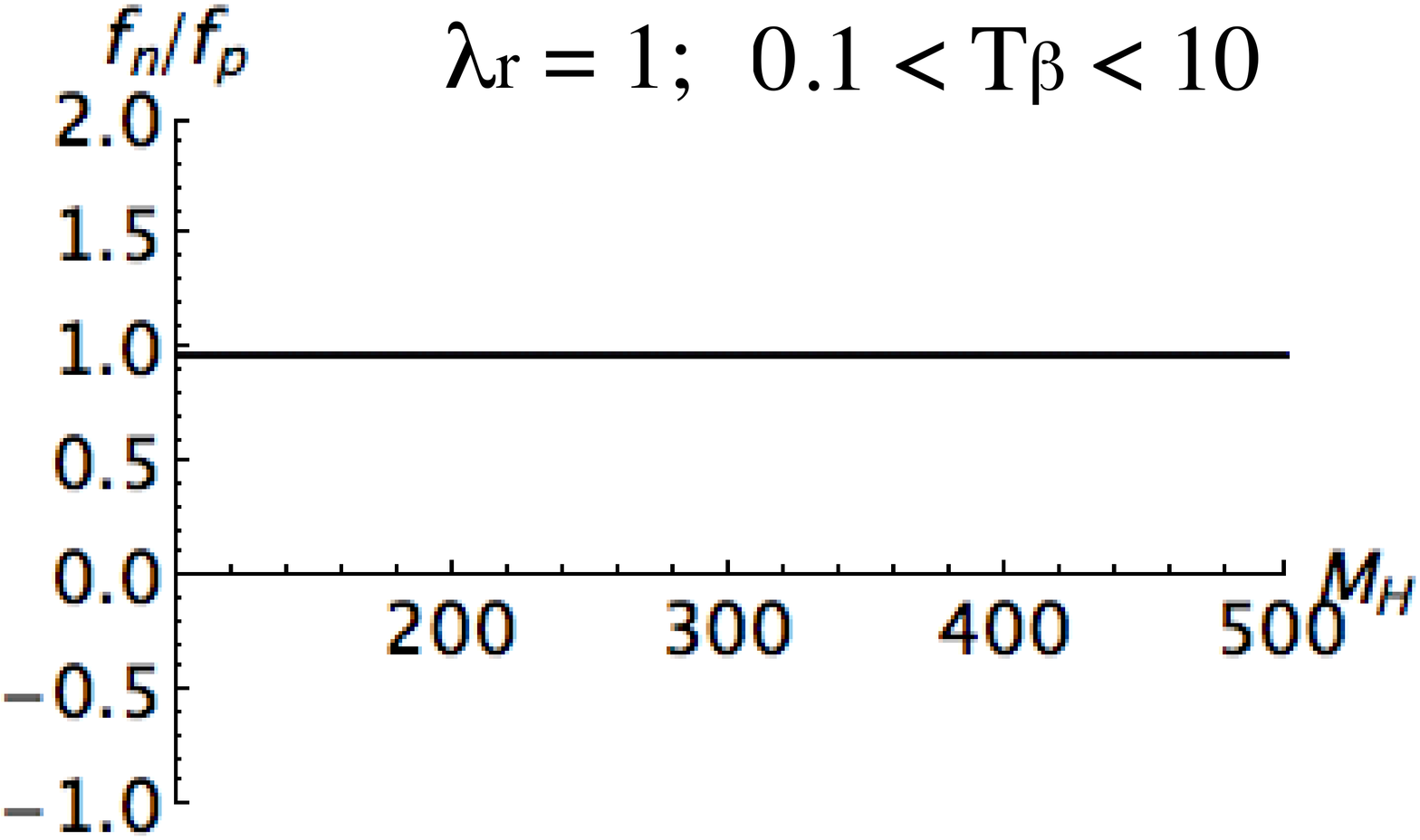}
\includegraphics[width=5cm,height=4cm,angle=0]{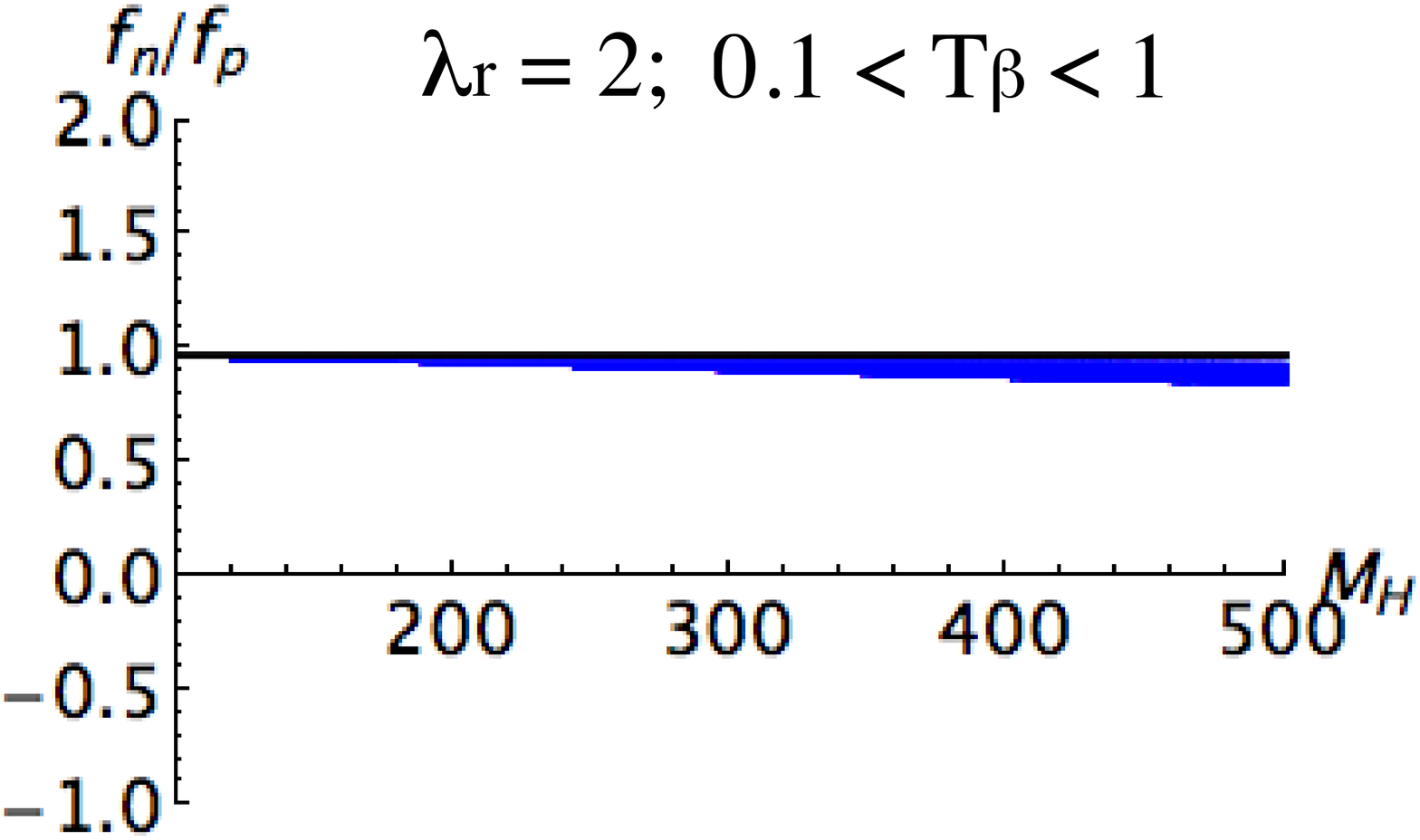}
\includegraphics[width=5cm,height=4cm,angle=0]{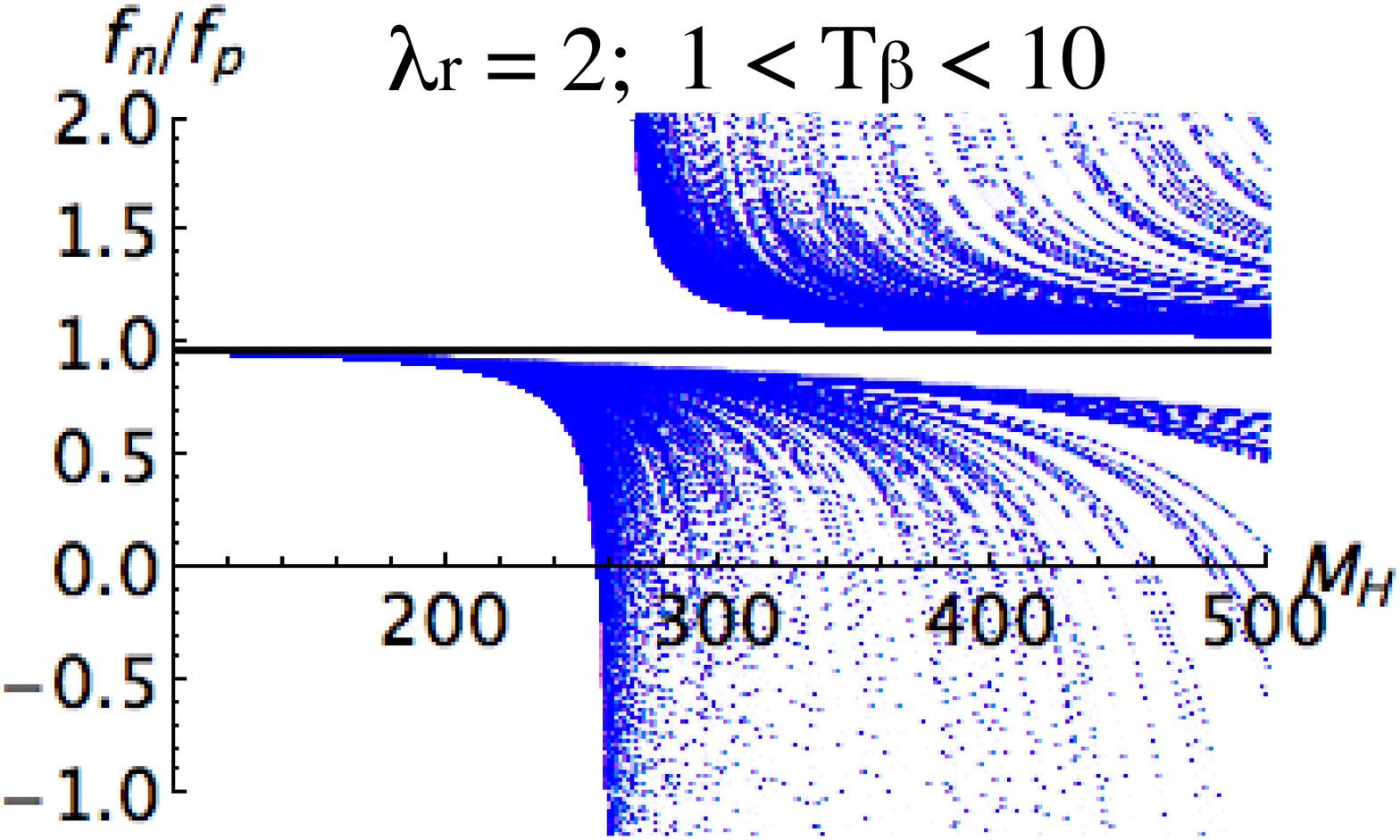}
\caption{\small Ratio between neutron- and proton-WIMP effective coupling by Higgs exchange in Type I (horizontal line) and Type II (extended regions) models and for three values of $\lambda_{r}$ and different regions of $T_{\beta}$}
\label{fig4}
\end{figure}

\begin{eqnarray}
f_N=\frac{M_p}{2M_{\sigma}}\sum _{\mathcal{H}=h,H}S_{\mathcal{H}}F_N^{\mathcal{H}},
\label{effective-scalar-coup}
\end{eqnarray}
with $S_{\mathcal{H}}$ defined in (\ref{effective-couplings}) (where the couplings $\lambda _{\mathcal{H}}$ are given by (\ref{higgs-coupling-2}) for each Higgs boson) and $F_N^{\mathcal{H}}$ in (\ref{scalar-formfactor}). In this case, the space of parameters is $(\lambda '_6,$ $\lambda _r,$ $T_{\beta},$ $M_H,$ $M_{\sigma })$. Figure \ref{fig4}, shows the neutron to proton ratio $f_n/f_p$ as function of the mass of the second Higgs boson $H$, for both Type I and II models, where we scan regions in the ranges of $T_{\beta}$ shown in the labels of each graphic and  $\lambda _r$ takes three values: $0$ ($\lambda '_7=0$), $1$ ($\lambda '_7=\lambda '_6$) and $2$ ($\lambda '_7=2\lambda '_6$), while the parameters $M_{\sigma }$ and $\lambda '_6$ are removed by the ratio. First, as to be expected, the Type I model (black horizontal line) is near $1$ along all the mass range, since the Yukawa coefficients do not distiguish between $u$ and $d$ quarks, which leads us to the same effective coupling with protons and neutrons (a small asymmetry arises due to the different values of the form factors $f_{TQ}^{N}$ for protons and neutrons that participate in (\ref{scalar-formfactor}), but are not appreciable). By contrast, the Type II model exhibits points with isospin violation in different regions of $M_H$ according to the parameter values. For example, for $\lambda _r=0$ and small $T_{\beta}$ (below 1), the ratio $f_n/f_p$ takes positive values (larger than 1) below $300$ GeV, while for values above $200$ GeV, there are abrupt drops in negative regions. The behaviour is very different for larger $T_{\beta}$ (above 1), where the allowed region collapse into a small band above $1$, increasing with $M_H$. The difference between regions below and above $T_{\beta}=1$ can be understood from the relative magnitudes between the Yukawa coefficients $c_U^{h}=1/T_{\beta}$ and $c_D^{h}=-T_{\beta}$ (see table \ref{tab:yukawa-coef}), which is the source for the isospin asymmetry in type II model. If $T_{\beta}<1$, we see that $c_U^{h} > \left|c_D^{h}\right|$, from where the couplings of protons results dominant for a large range of the space of parameters, i.e. solutions with $f_n/f_p<1$ are favored. The opposite situation occurs for $T_{\beta}>1$, where the regions reveal solutions only for  $f_n/f_p>1$. At $\lambda _r=1$, however, the situation changes radically. At this limit, according to Eq. (\ref{higgs-coupling-2}), the SM-like Higgs boson, $h$, decouples from the WIMP. Since this coupling is the only source of isospin violation in type II model for scalar interactions, the asymmetry disappears, and we obtain the same effective coupling with protons and neutrons in both models. Finally, for $\lambda _r=2$, the regions appear inverted in relation to the first case. This inversion comes from the change of sign of the coupling with the Higgs boson $h$ in Eq. (\ref{higgs-coupling-2}) when $\lambda _r>1$.

On the other hand, it is interesting to see that there are solutions where $f_n/f_p=0$ and $\infty $. The former leads us to neutron-phobic scenarios, while the latter corresponds to proton-phobic interactions. Furthermore, we observe some values of $M_H$ where both $f_p$ and $f_n$ cancel, similar to the second scenario of interference in (\ref{total-interference}). In this case, the scalar interaction does not contribute to the WIMP-nucleus scattering for all types of nucleus regardless their atomic number.     

%to the fast falls shown by the points, both couplings cancel for nearly the same $M_H$, similar as the second scenario of interference in (\ref{total-interference}).
%
 %Thus, we can find solutions with interfering Higgs bosons, such that the scalar interaction does not contribute to the WIMP-nucleus scattering in spite of the existence of couplings between scalar and ordinary matter. 
%Also, we observe that solutions cross all the negative region, including the claimed value of $f_n/f_p \approx -0.7$. However, due to the proximity of masses $M_H$ that cancels neutron and proton interactions, the interference would happen for all type of nucleus regardless their atomic number. 

\begin{figure}[t]
\centering
\includegraphics[width=6.8cm,height=5.2cm,angle=0]{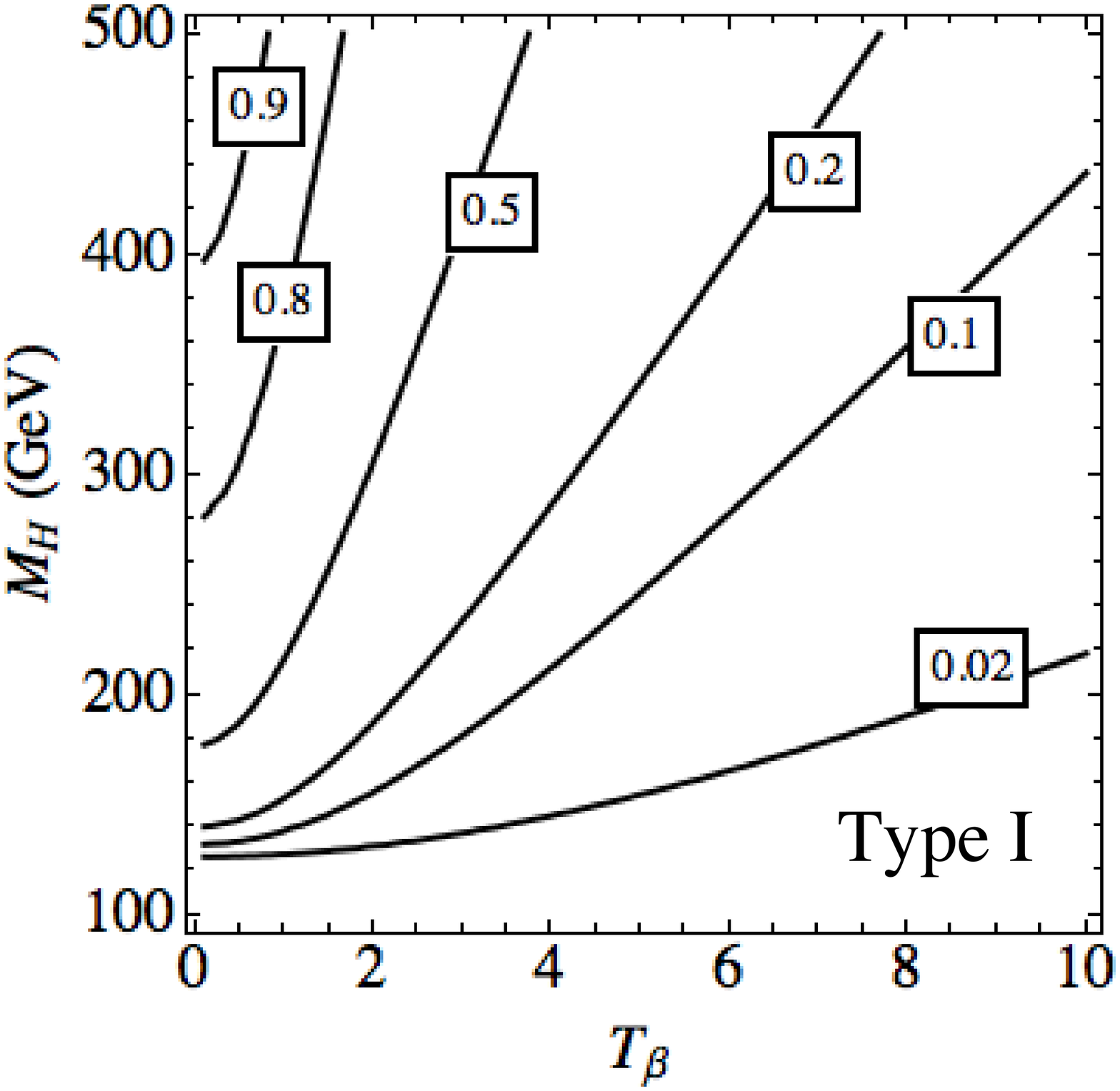}\hspace{-1.5cm}
\includegraphics[width=6.8cm,height=5.2cm,angle=0]{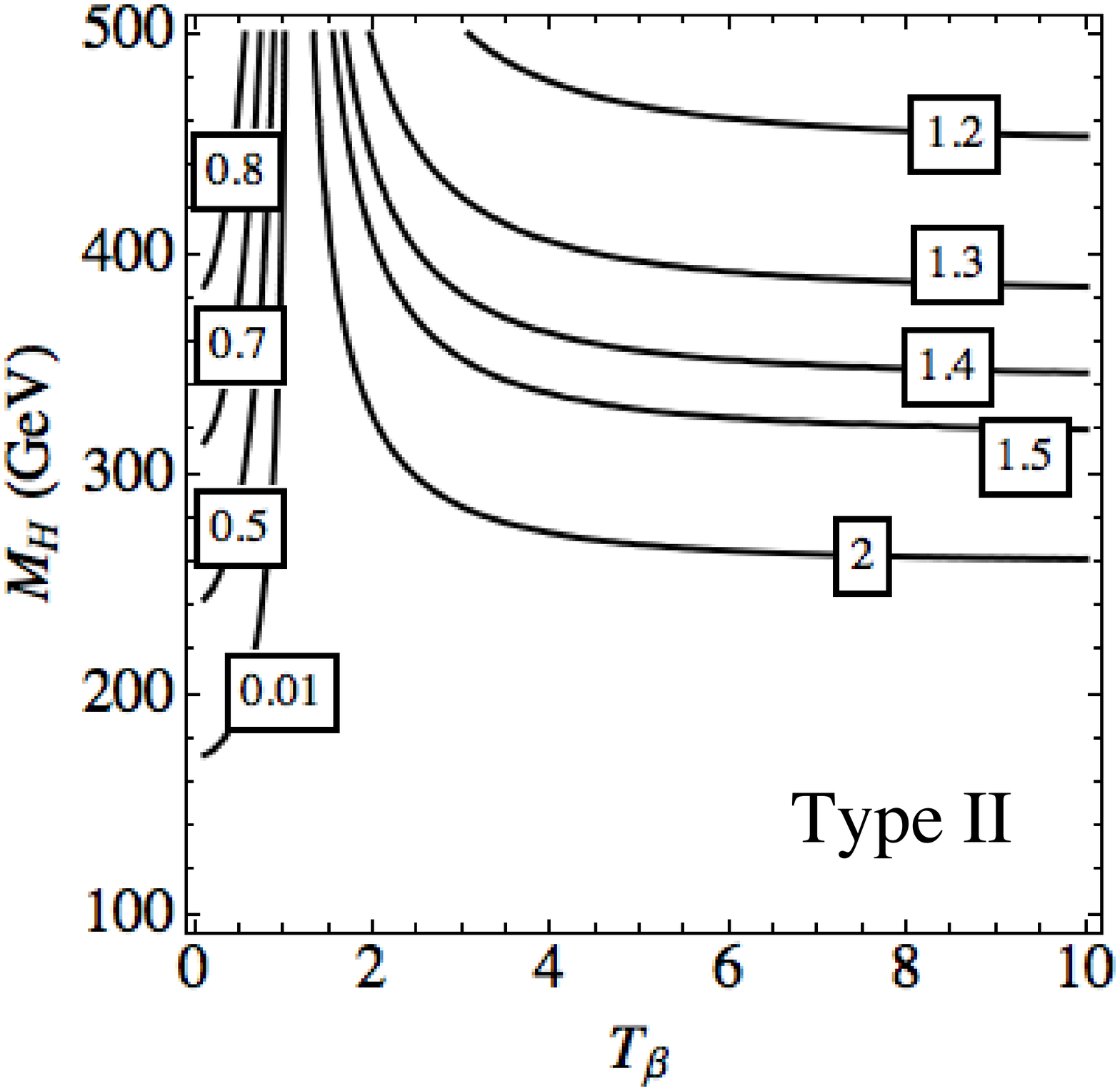}\vspace{-0.5cm}
\caption{\small Vanishing destructive interference between $h$ and $H$ exchange. We display some contour plots for $\lambda _r$ in the plane $T_{\beta}-M_H$.}
\label{fig5}
\end{figure}

\begin{figure}[t]
\centering
\includegraphics[width=7cm,height=4.4cm,angle=0]{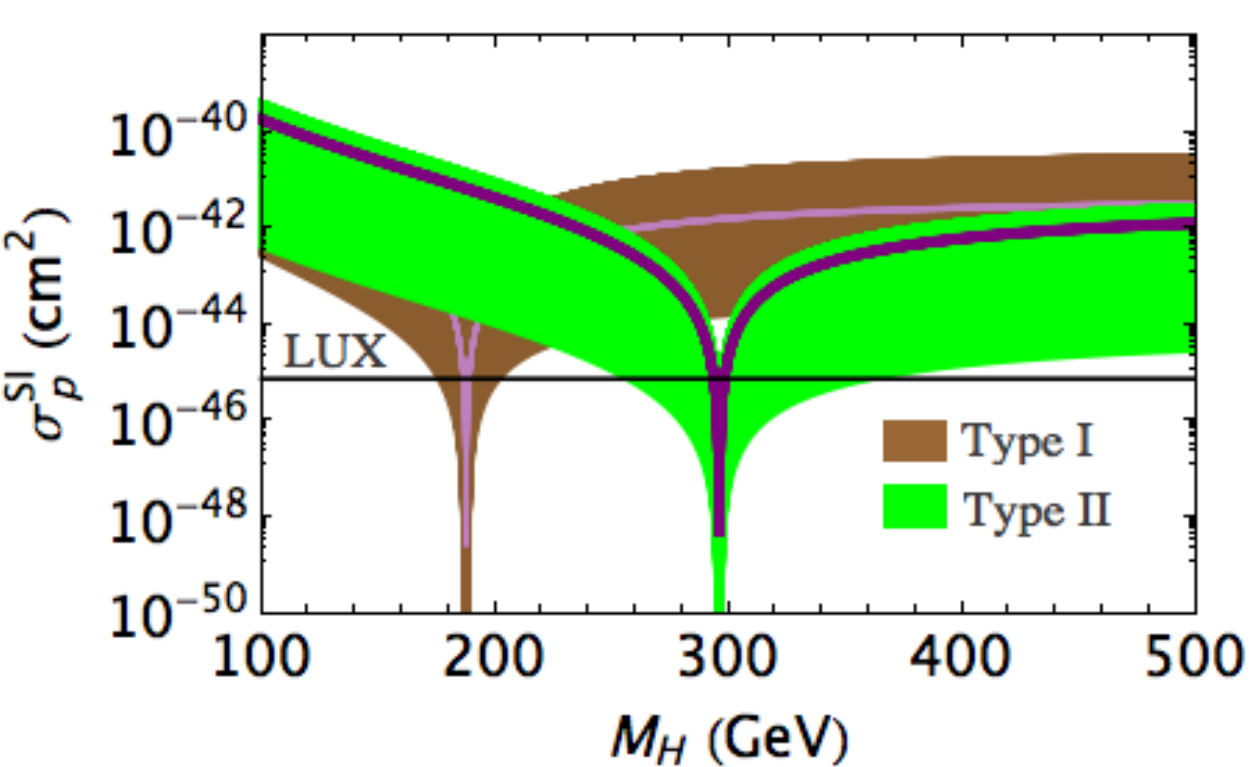}\vspace{-0.2cm}
\caption{\small WIMP-proton cross section for $\lambda _r=0.5$ and $T_{\beta}=0.5$ in the range $\lambda '_6=[0.1,3]$, where the interference peaks agree with the $0.5$ contours from the plots of figure \ref{fig5}. The horizontal line corresponds to the lowest limit measure by LUX at $M_{\sigma}=33$ GeV \cite{lux}. The purple lines correspond to the ranges $\lambda '_6=[0.97,1.02]$ (type I) and $\lambda '_6=[2.04,2.12]$ (type II), compatible with the measured DM relic density.}
\label{fig6}
\end{figure}

To explore specific values of the parameters more deeply for vanishing destructive interference, we search for solutions that lead us to $f_p=0$  in Eq. (\ref{effective-scalar-coup}), i.e., solutions that accomplish the condition $S_{h}F_p^{h}=-S_{H}F_p^{H}$, which does not depend on the values of $\lambda' _6$ and $M_{\sigma}$. The plots in figure \ref{fig5} show contour plots for some values of $\lambda _r$ in the plane $M_H$-$T_{\beta}$ and  in type I and II models. To illustrate how the interferences can affect the SI cross section of the WIMP-nucleon scattering, in figure \ref{fig6} we show the solution $\lambda _r=0.5$ at $T_{\beta}=0.5$ and $M_{\sigma}=33$ GeV, which according to the corresponding contour from \ref{fig5}, gives destructive interference near $M_H=190$ and $300$ GeV for type I and II models, respectively. From this figure, we also see that although there are broad bands due to the $\lambda' _6$ scan, they reduce to narrow peaks in the interference, which confirm that cancellations do not depend on this parameter. We compare the cross section with the lowest limit for WIMP scattering measured by the LUX collaboration at $M_{\sigma}=33$ GeV \cite{lux}. We also include the regions allowed by the measured DM relic abundance, which exhibits thin bands in the ranges $\lambda '_6=[0.97,1.02]$ and $[2.04,2.12]$ for type I and II models, respectively.

\subsection{Interactions with vector boson exchange}

\begin{figure}[tb]
\centering
\includegraphics[width=6.2cm,height=4cm,angle=0]{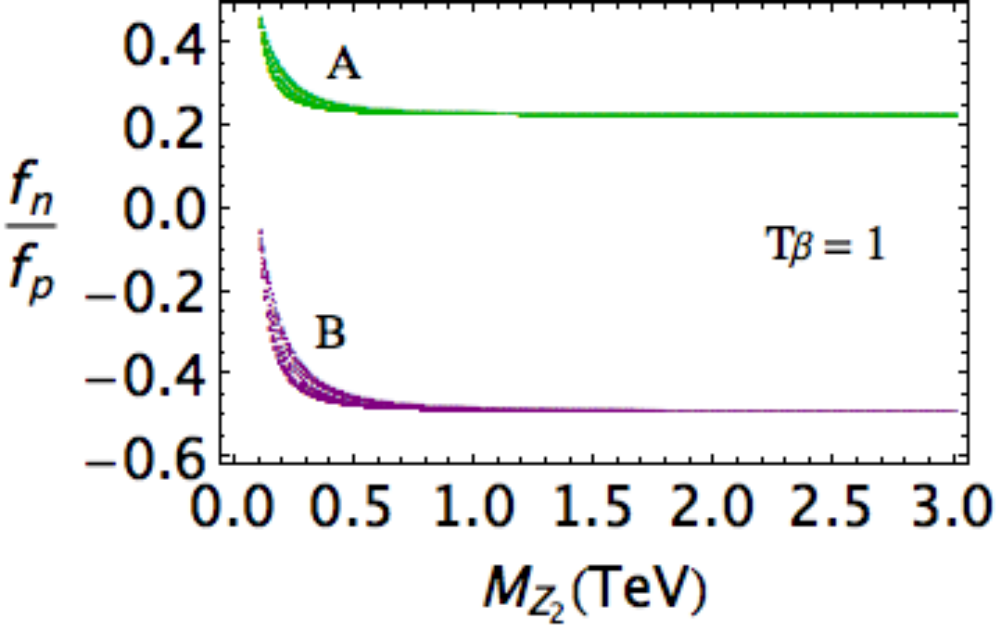}
\includegraphics[width=6.2cm,height=4cm,angle=0]{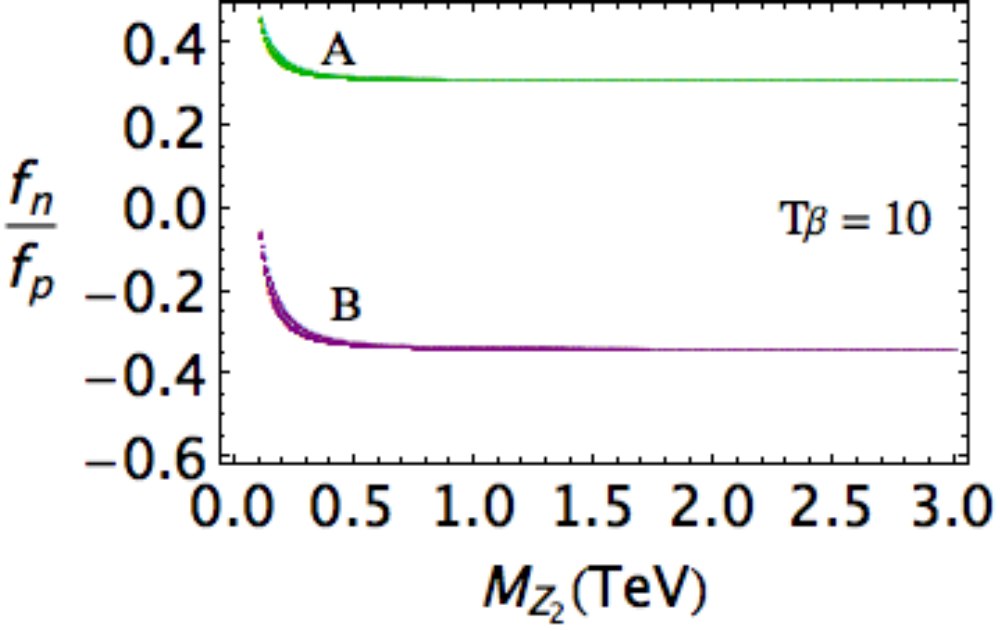}\vspace{-0.3cm}
\caption{\small Ratio between WIMP-neutron and -proton effective coupling by gauge boson exchange in models with family structures A (Green) and B (Purple) and for two values of $T_{\beta}$}
\label{fig7}
\end{figure}

If we turn off the Higgs interaction by making $\lambda '_6=\lambda _r=0$, and permit the gauge interaction, we obtain:

\begin{eqnarray}
f_N=\sum _{\mathcal{Z}=Z_1,Z_2}G_{\mathcal{Z}}V_N^{\mathcal{Z}},
\label{effective-vector-coup}
\end{eqnarray}
with $G_{\mathcal{Z}}$ and $V_N^{\mathcal{Z}}$ given by Eqs. (\ref{effective-couplings}) and (\ref{vector-formfactor}), respectively. The space of parameters in this case is reduced to $(g_X,$ $T_{\beta},$ $M_{Z_{2}})$. Figure \ref{fig7} shows the ratio $f_n/f_p$, but now we compare the two family structures A and B from table \ref{tab:family-matching} as function of the $Z_2$ mass. Since $V_N^{\mathcal{Z}}$ depends on the quark flavour, the isospin violation appears in all the mass regions. We include low mass regions to illustrate how the isospin asymmetry changes with $M_{Z_{2}}$. We find that model $A$ (green scan) exhibits positive asymmetries while $B$ (purple) is negative. This happens because in structure $B$, the $u$ and $d$ quarks have the vector couplings of the representation $q^{1}_L$, shown in table \ref{tab:SM-espectro}, leading to negative neutron coupling $f_n< 0$ and positive proton coupling $f_p>0$, while in $A$, these quarks change their couplings according to the representation $q^{2}_L$, producing positive ratios $f_n/f_p>0$. However, if we would like to fit specific values on $f_n/f_p$, there are few solutions due to the narrowness of the regions. For example, the claimed value of $-0.7$ is obtained only as an asymptotic approximation for small values of $T_{\beta}$ in the family structure B, while for A, the allowed regions are near $0$, i.e. $f_n\approx 0$ for model A at $T_{\beta} \ll 1$. In general, due to the decoupling between the quark flavour $d=D^1$ and the gauge boson $Z_2$ exhibited by the structure $A$ (i.e., $v_Q^{NSM}=0$ as shown in table \ref{tab:vector-axial-couplings}), the interaction in this case becomes nearly neutron-phobic, and the scattering is dominated by the proton structure of the nucleus.

\begin{figure}[t]
\centering
\includegraphics[width=8cm,height=5.3cm,angle=0]{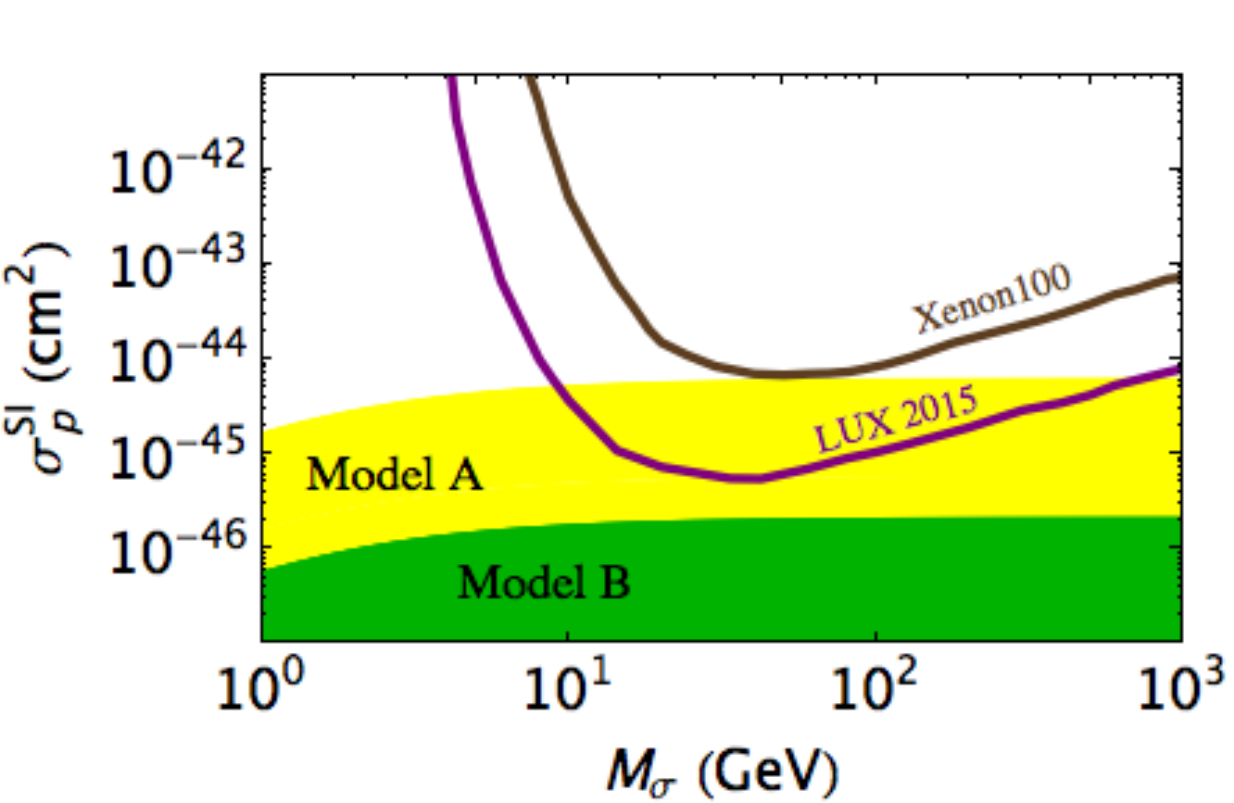}\vspace{-0.2cm}
\caption{\small  Experimental limits and theoretical regions for WIMP-nucleus scattering by considering only gauge bosons exchange. We set $M_{Z_{2}}=3000$ GeV and $g_X \leq 0.4$. A and B denotes the two family structures of the model.}
\label{fig8}
\end{figure}

On the other hand, to have proton-phobic scenarios, we would require divergent ratios $f_n/f_p\rightarrow \infty$, which do not appear as solutions in figure \ref{fig7}, at least not for finite $M_{Z_{2}}$ values. As a consequence, real solutions for destructive interference between gauge bosons, i.e., solutions for $G_{Z_1}V_p^{Z_1}=-G_{Z_2}V_p^{Z_2}$ in (\ref{effective-vector-coup}) do not exist with the values of the fundamental parameters of the model. In spite of this, it is interesting to evaluate the one-proton cross section to compare with the experimental limits. For this, we set $M_{Z_{2}}=3000$ GeV and scan over $g_X$ and $T_{\beta}$. Figure \ref{fig8} shows the limits from Xenon-based experiments (LUX and Xenon100). The shaded regions on the bottom shows the allowed points for each family structure, A and B, where we scan $g_X$ and $T_{\beta} $ in the range $0-0.4$ and $0.1-10$, respectively, and the normalization from (\ref{normalization}) was taken into account.  We observe that due to the large value of $M_{Z_2}$ (which produces small $Z-Z'$ mixing angle values) and the small limits for $g_X$, the cross section is below the experimental limits for most of the values of the parameters. Only for large values of $g_X$ (at $0.4$), the theoretical region for model A is excluded by LUX in the range $9 \leq M_{\sigma} \leq 800$ GeV, and passed the lowest limit from Xenon100 at $50$ GeV. %Model B, pass the limits from both experiments.   

\begin{figure}[t]
\centering
\includegraphics[width=5.5cm,height=4cm,angle=0]{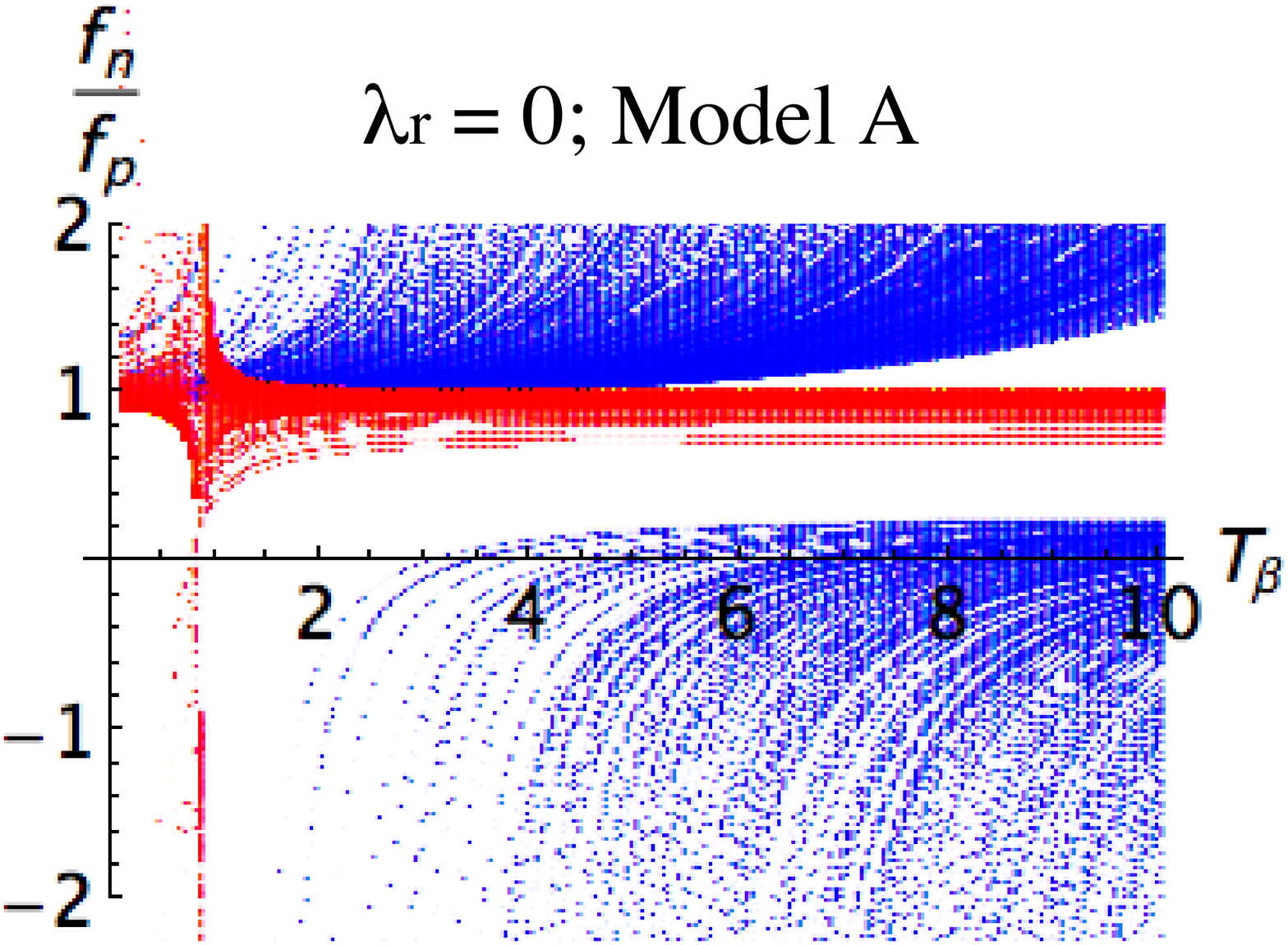}
\includegraphics[width=5.5cm,height=4cm,angle=0]{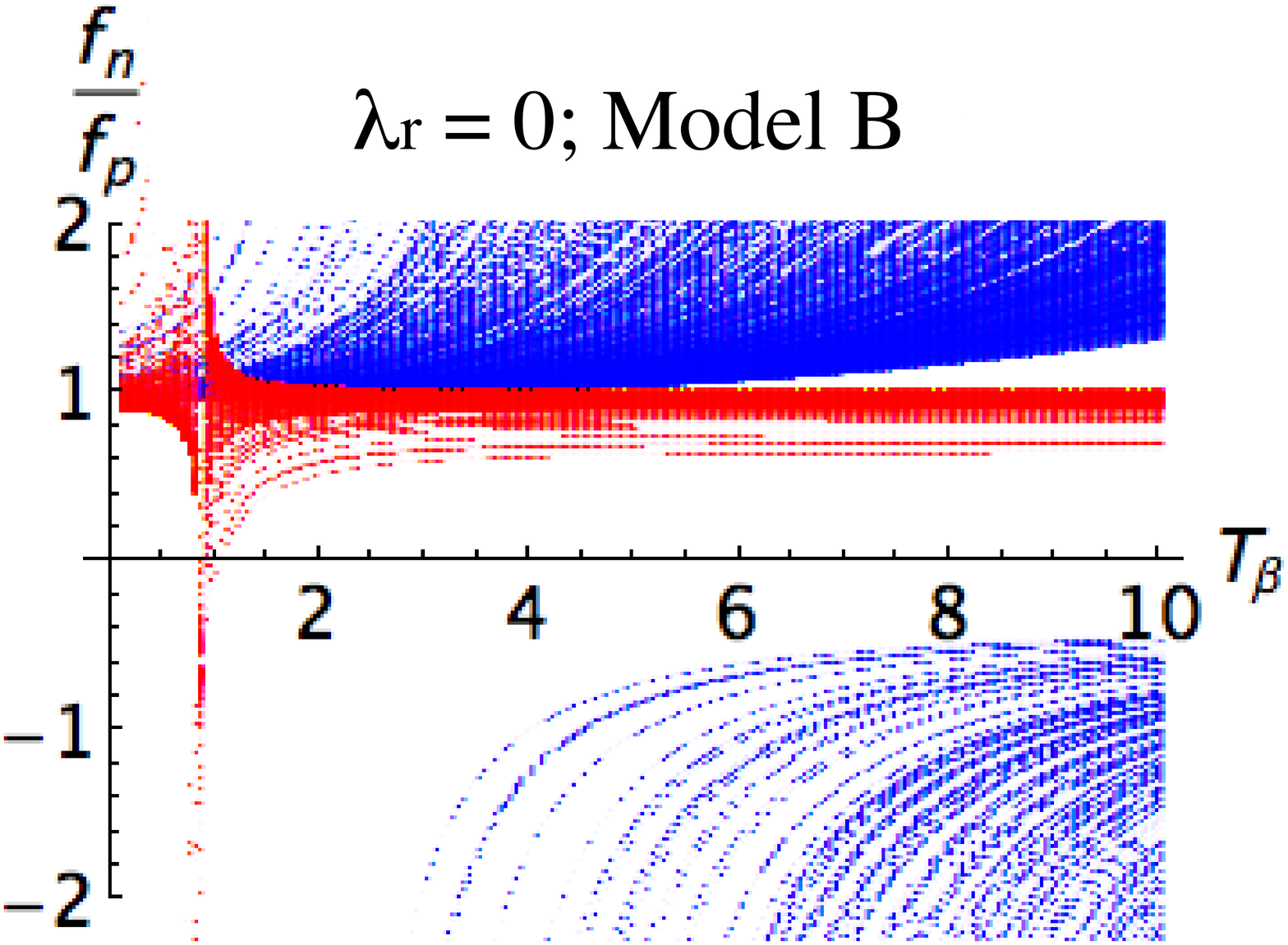}
\includegraphics[width=5.5cm,height=4cm,angle=0]{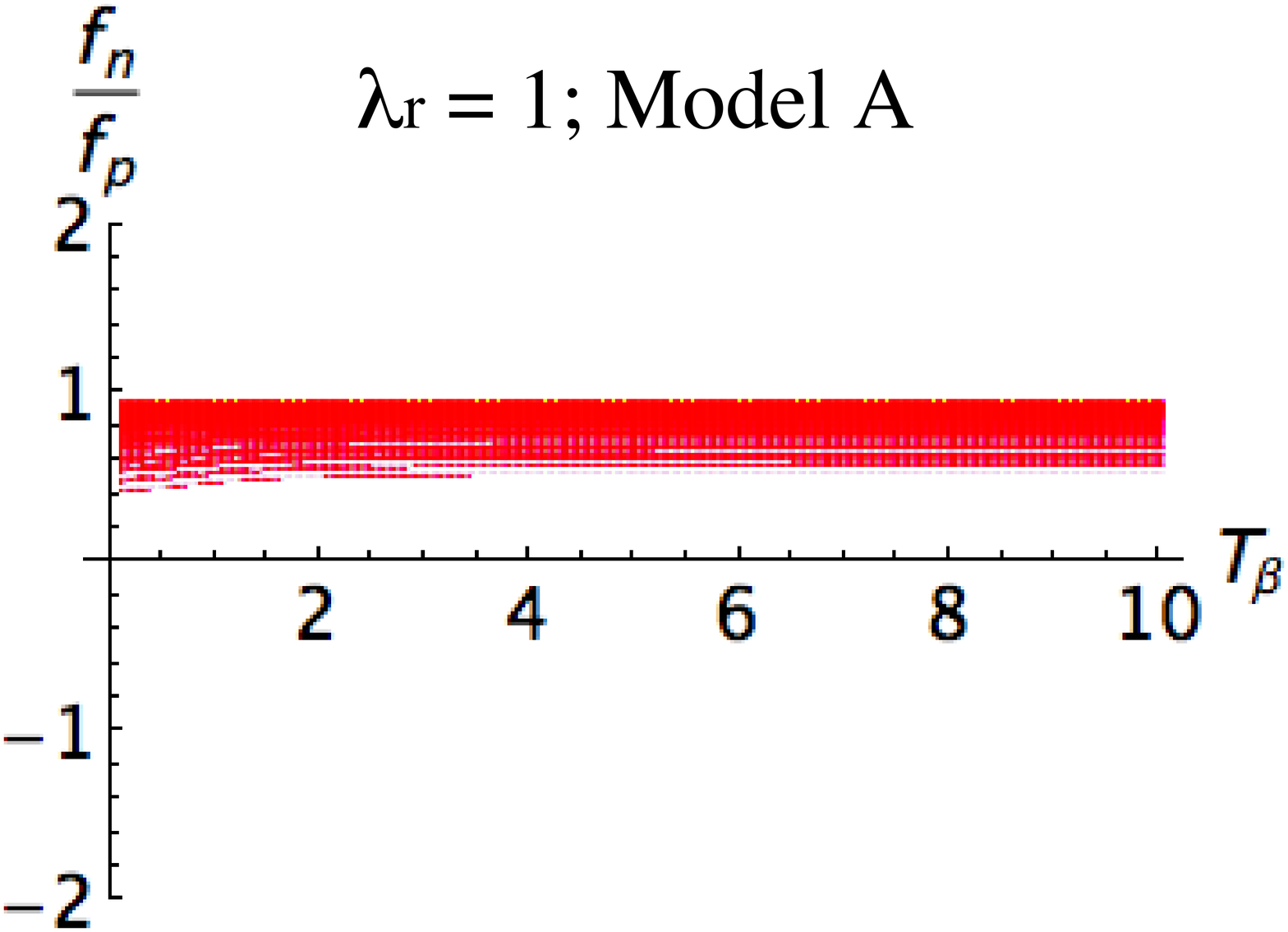}
\includegraphics[width=5.5cm,height=4cm,angle=0]{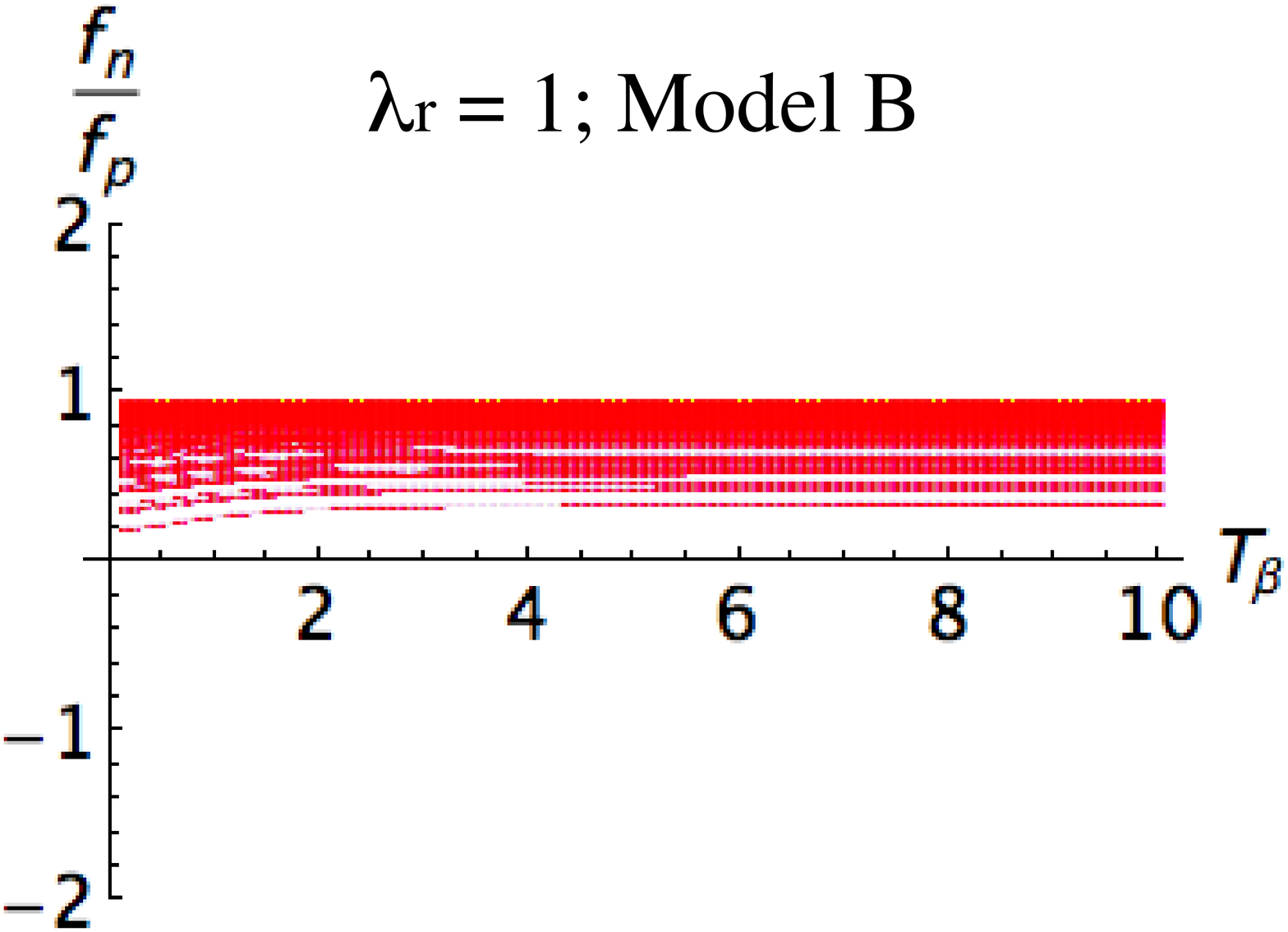}
\includegraphics[width=5.5cm,height=4cm,angle=0]{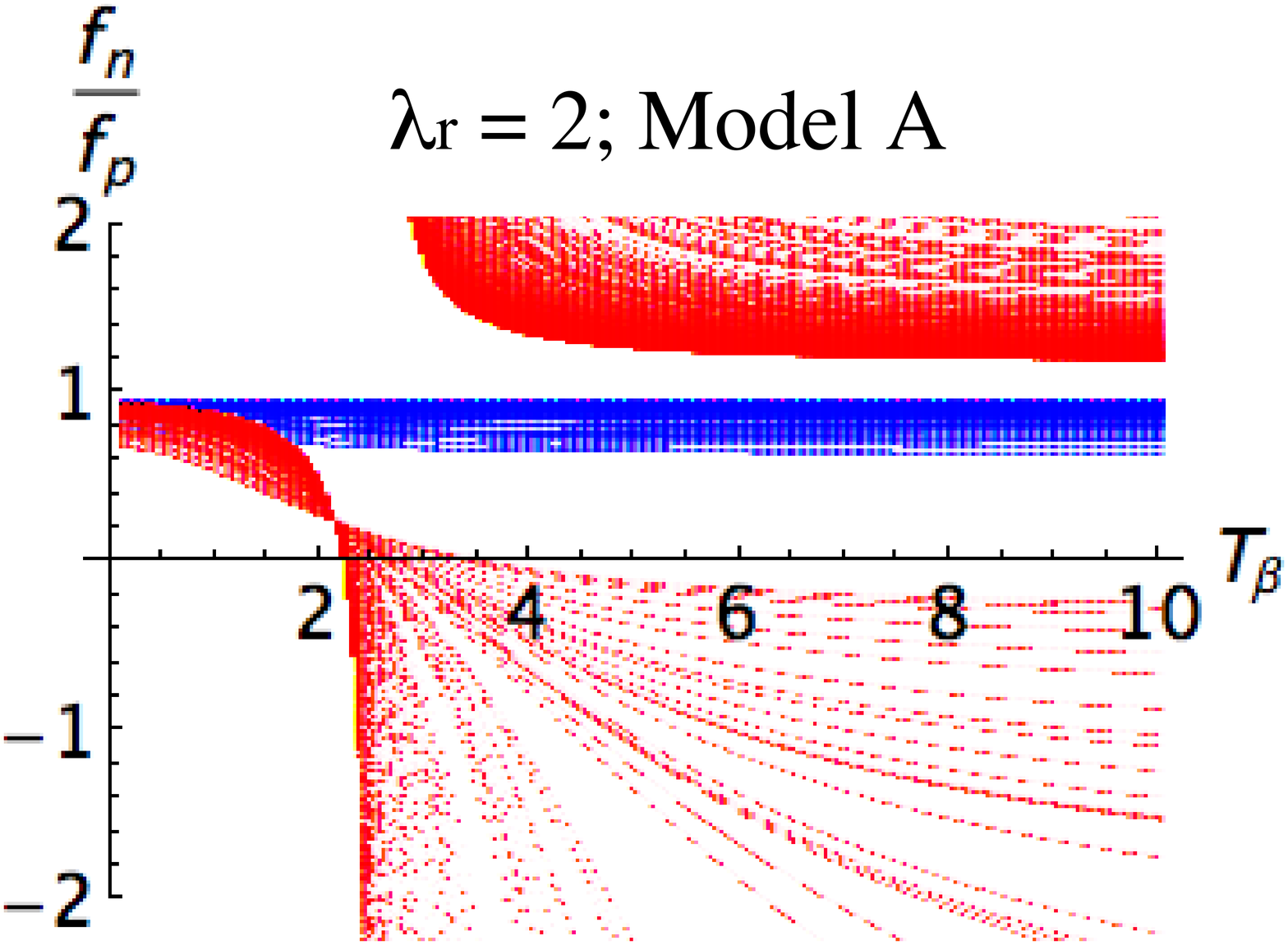}
\includegraphics[width=5.5cm,height=4cm,angle=0]{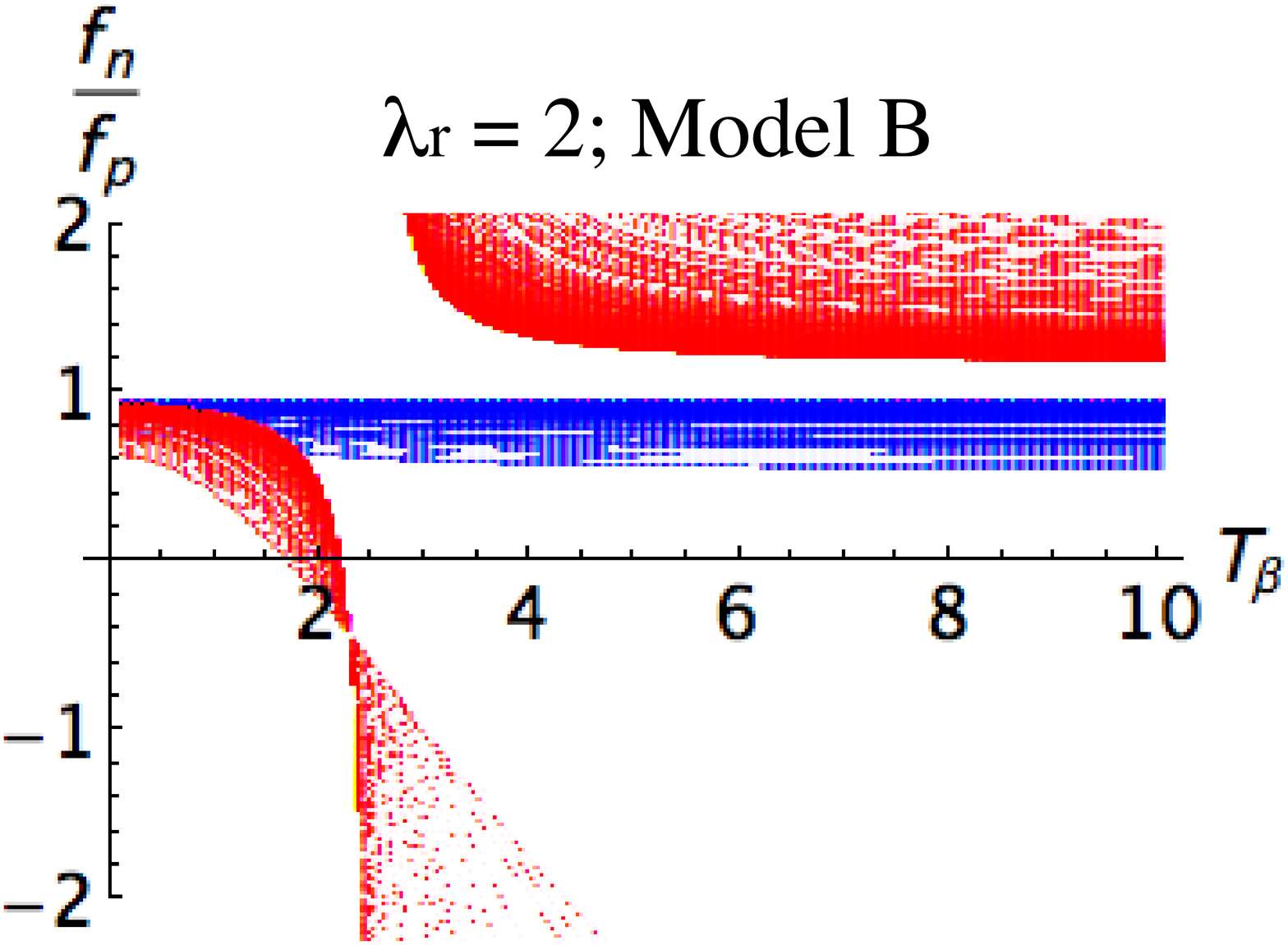}
\caption{\small Ratio between WIMP-neutron and -proton effective coupling by Higgs and gauge boson exchange in models type I (Blue) and II (Red) with family structures A and B. We set $M_H=300$ and $M_{Z_{2}}=3000$ GeV. }
\label{fig9}
\end{figure}

\begin{figure}[t]
\centering
\includegraphics[width=6.3cm,height=4.7cm,angle=0]{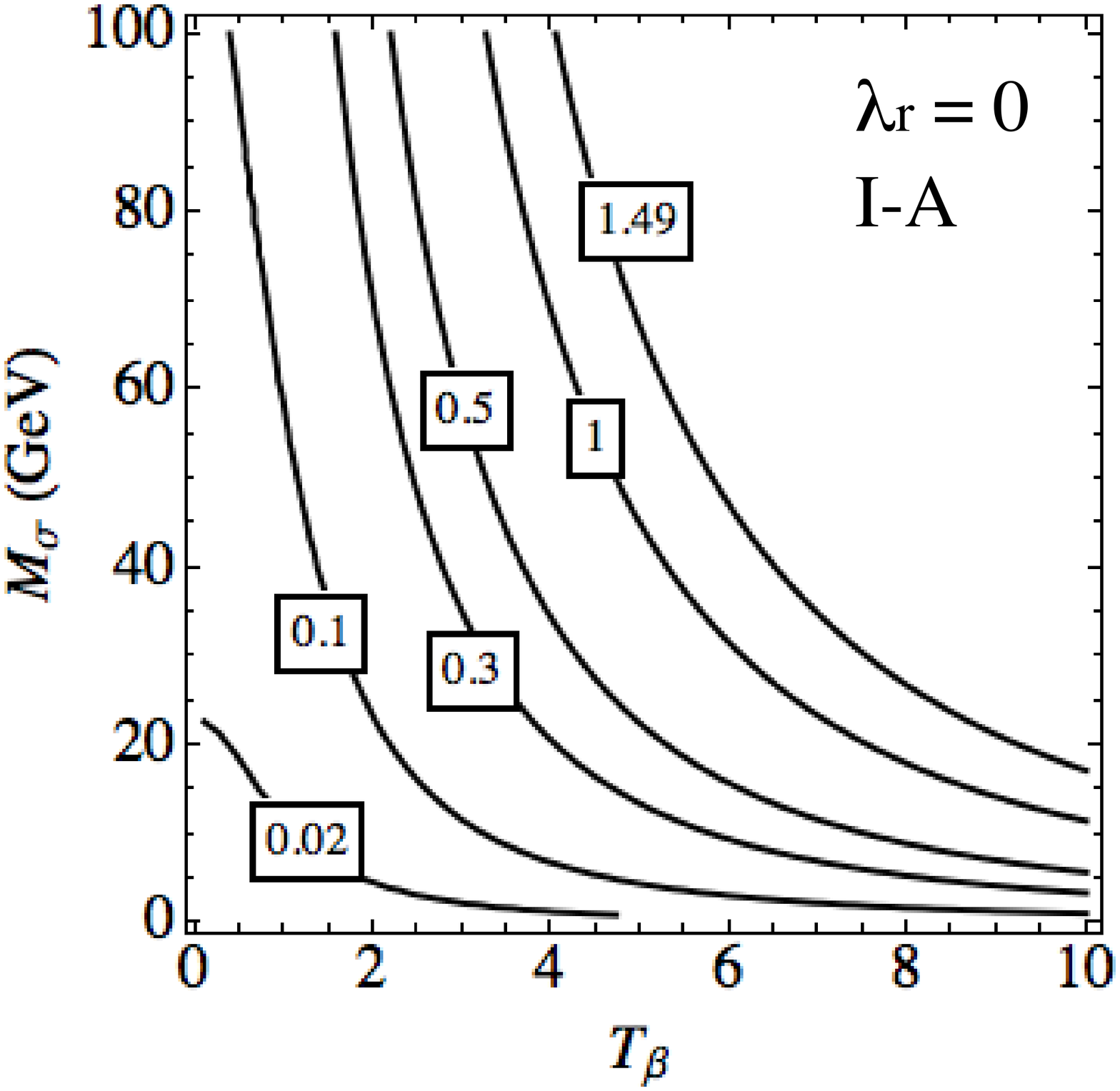}\hspace{-1.5cm}
\includegraphics[width=6.3cm,height=4.7cm,angle=0]{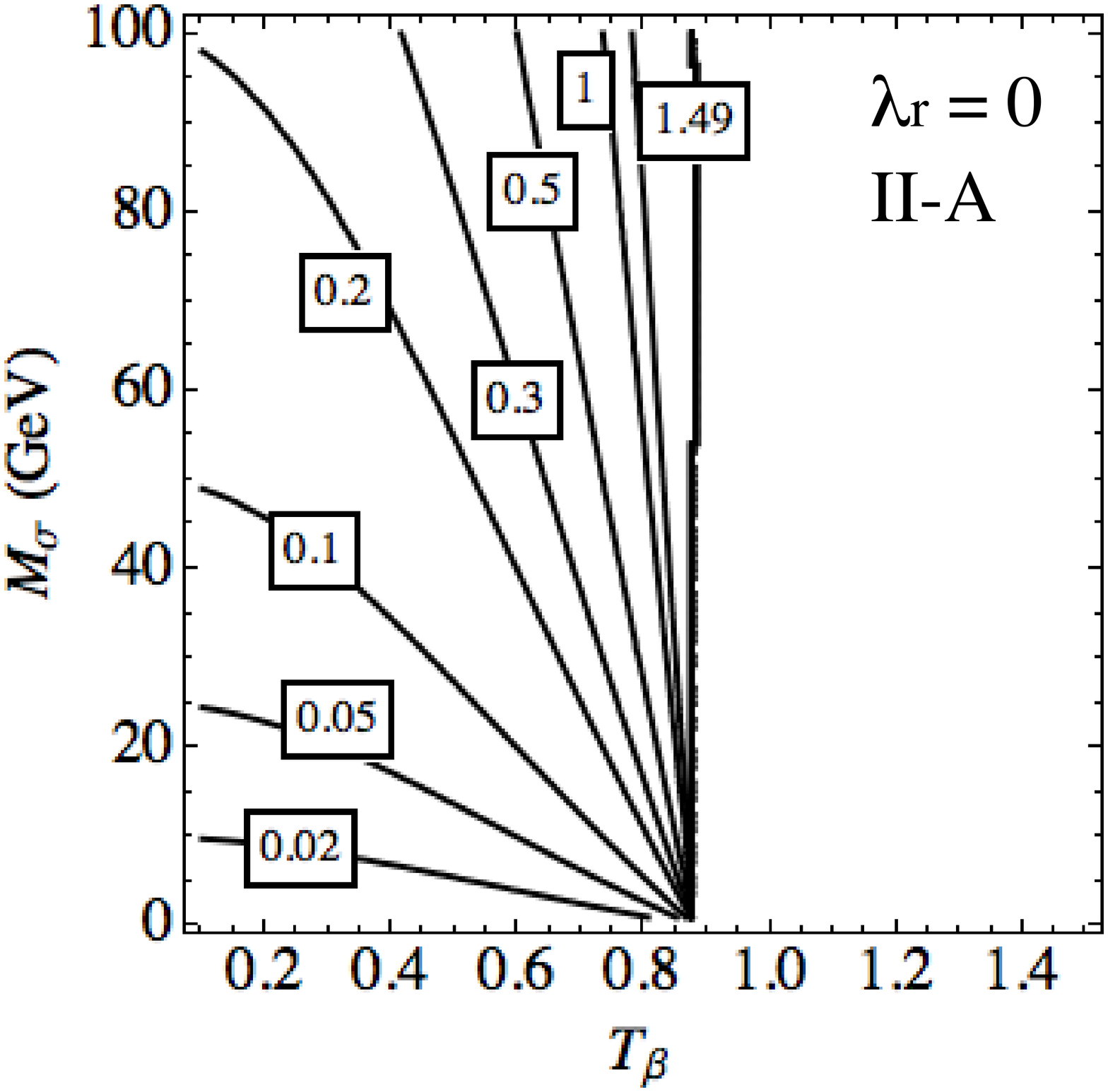}\hspace{-1.5cm}
\includegraphics[width=6.3cm,height=4.7cm,angle=0]{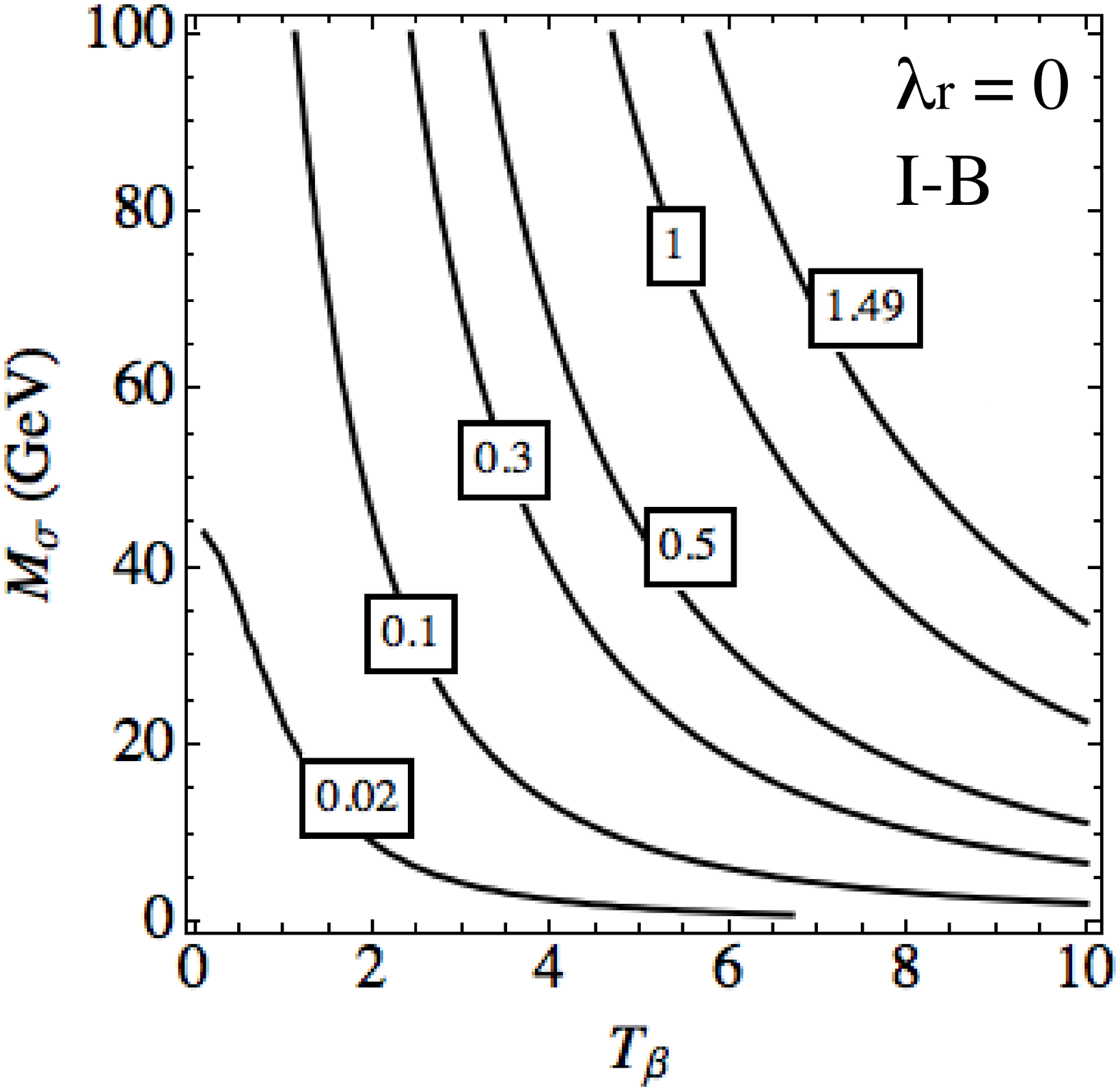}\hspace{-1.5cm}
\includegraphics[width=6.3cm,height=4.7cm,angle=0]{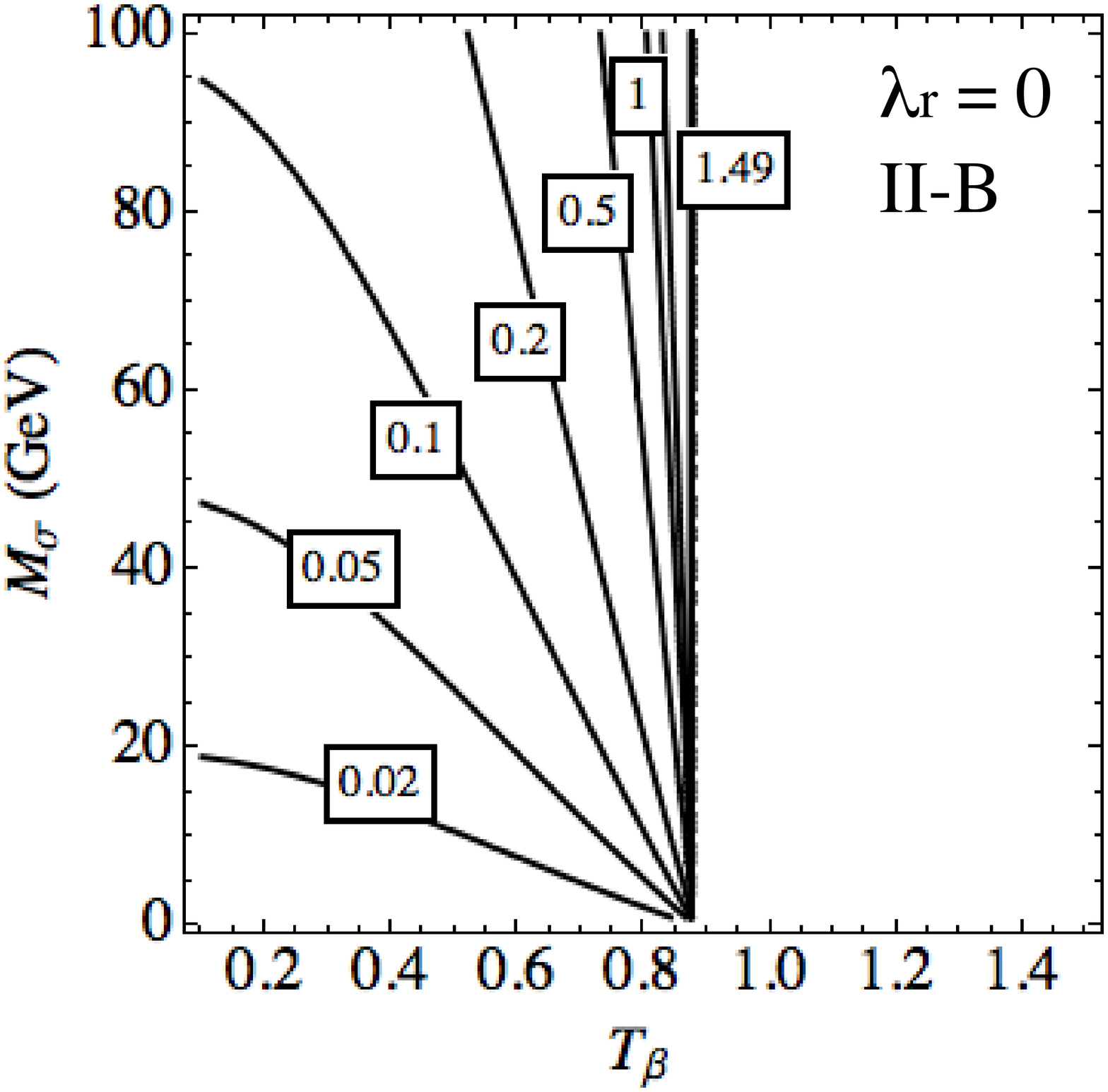}\hspace{-1.5cm}
\includegraphics[width=6.3cm,height=4.7cm,angle=0]{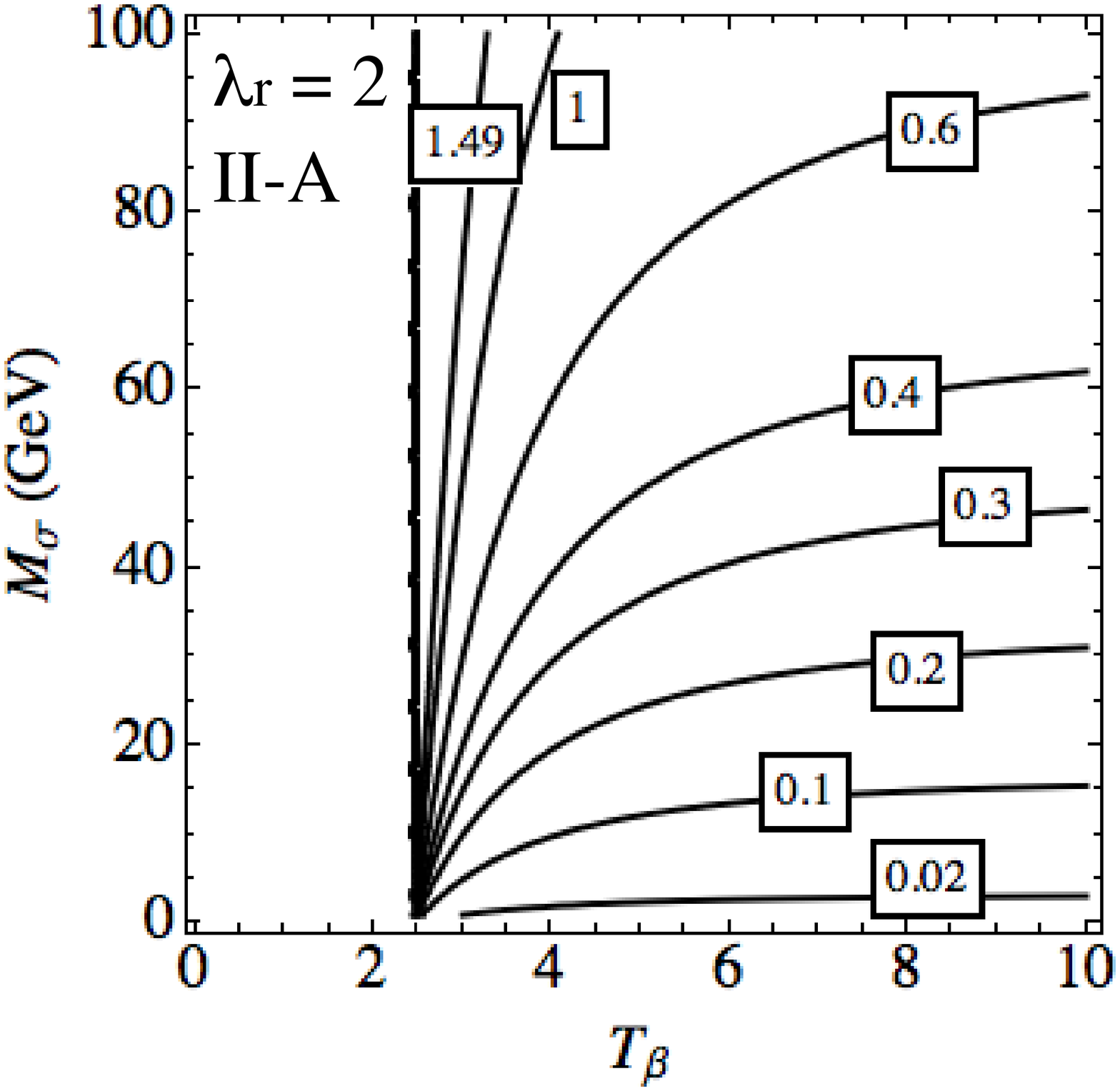}\hspace{-1.5cm}
\includegraphics[width=6.3cm,height=4.7cm,angle=0]{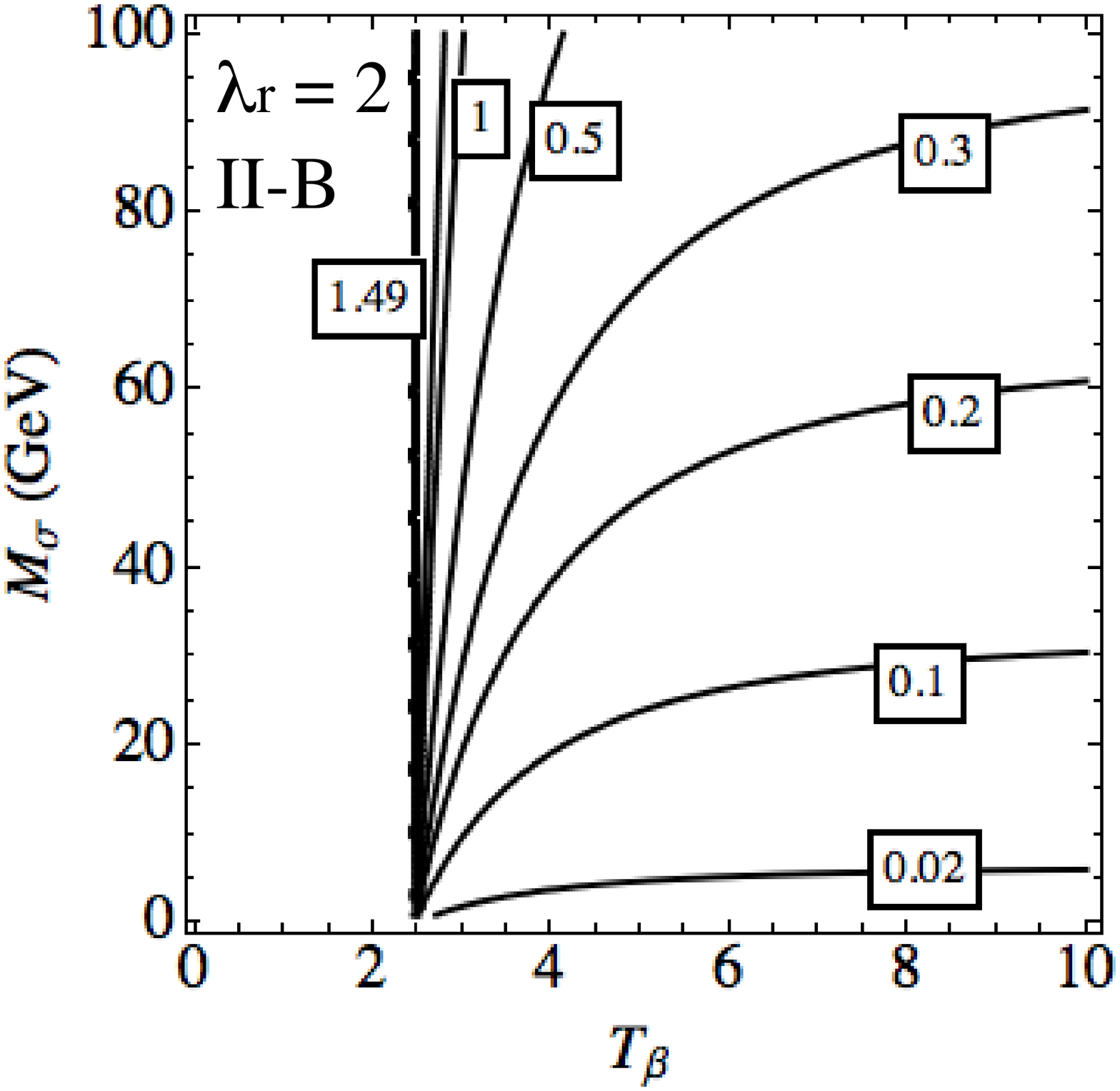}
\caption{\small  Contourplots of $\lambda '_6$ in the plane $T_{\beta}-M_{\sigma}$ for $f_p=0$ with Higgs and vector bosons exchange.}
\label{fig10}
\end{figure}

\begin{figure}[t]
\centering
\includegraphics[width=6.3cm,height=4.7cm,angle=0]{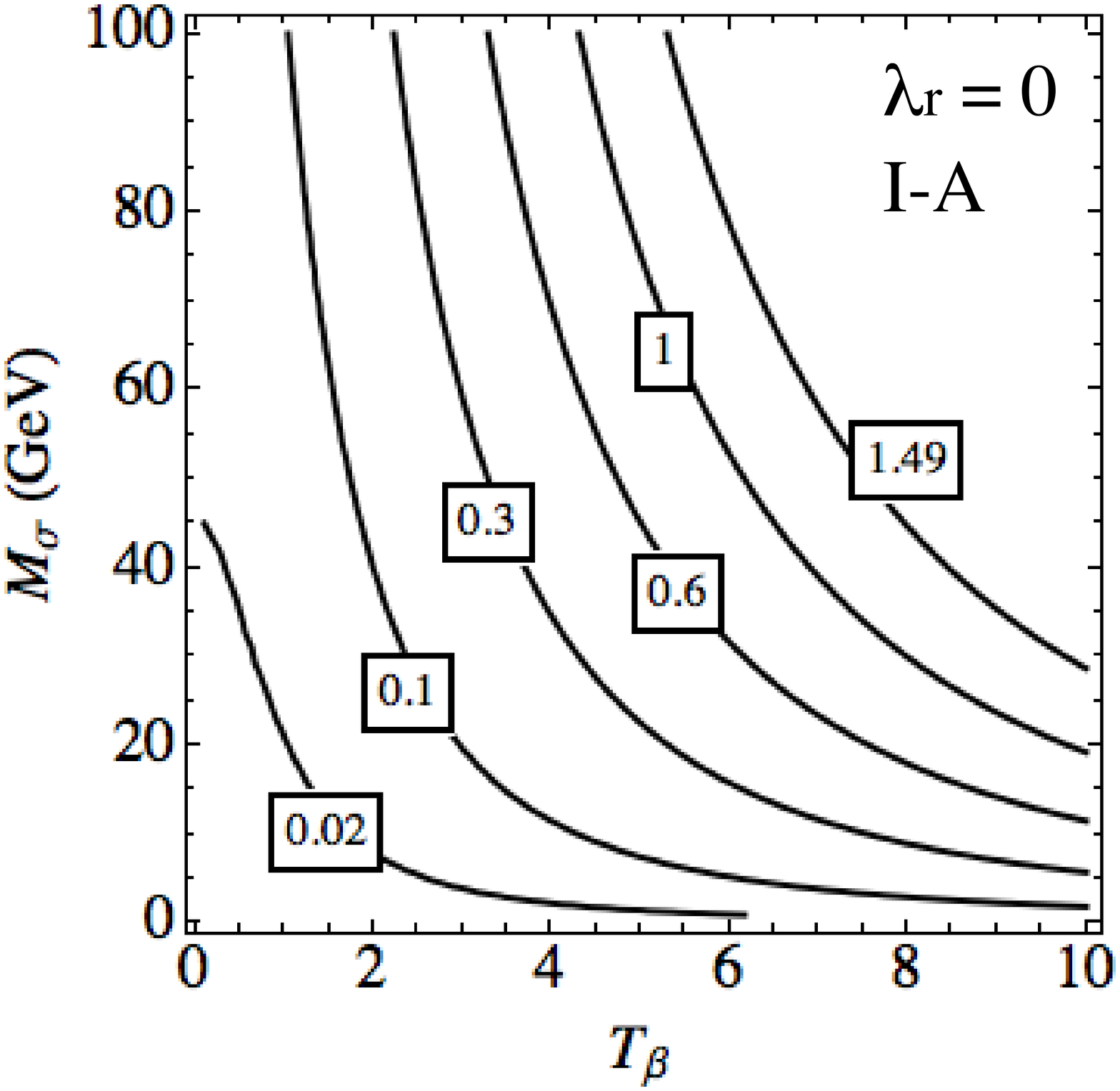}\hspace{-1.5cm}
\includegraphics[width=6.3cm,height=4.7cm,angle=0]{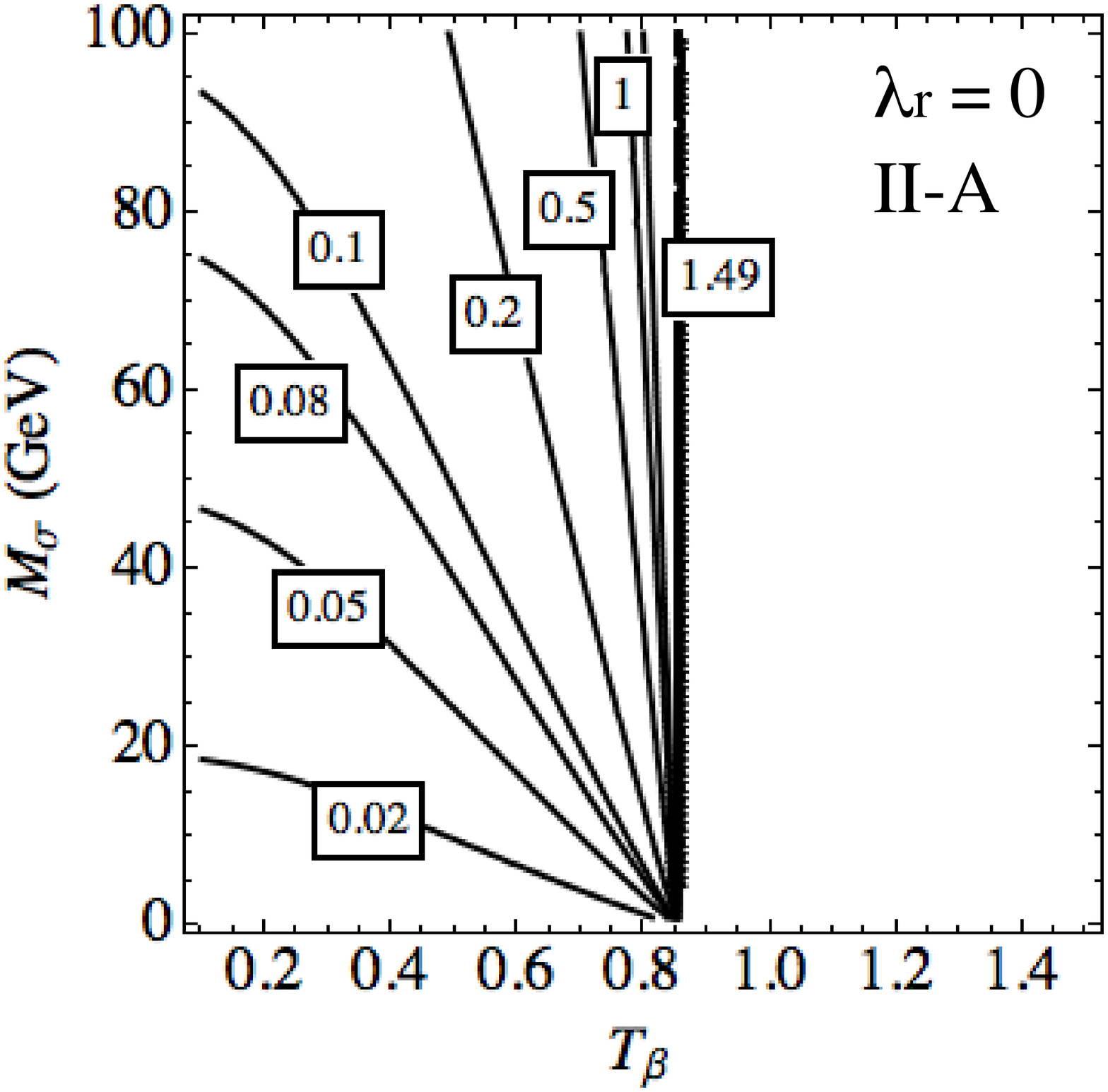}\hspace{-1.5cm}
\includegraphics[width=6.3cm,height=4.7cm,angle=0]{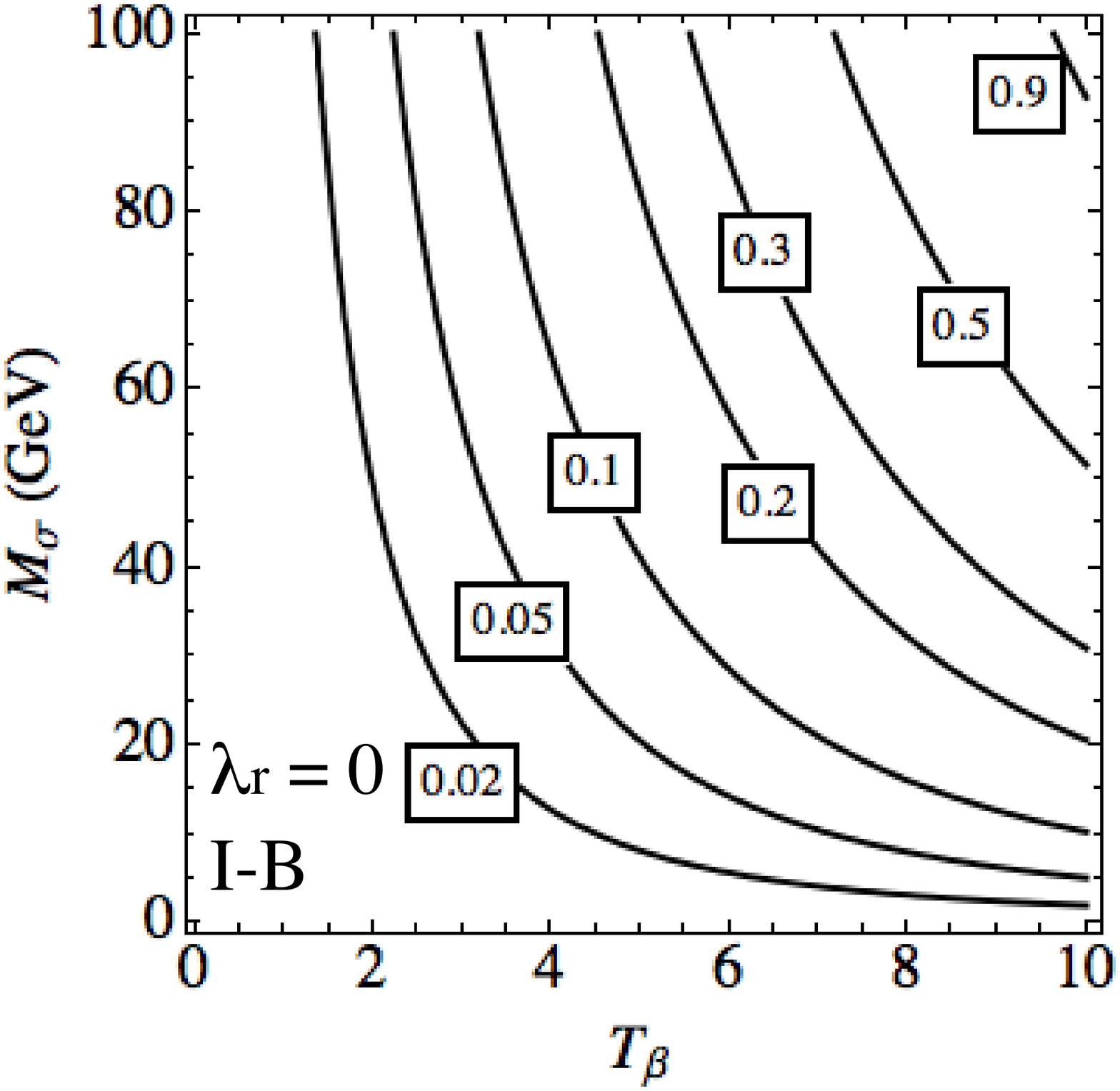}\hspace{-1.5cm}
\includegraphics[width=6.3cm,height=4.7cm,angle=0]{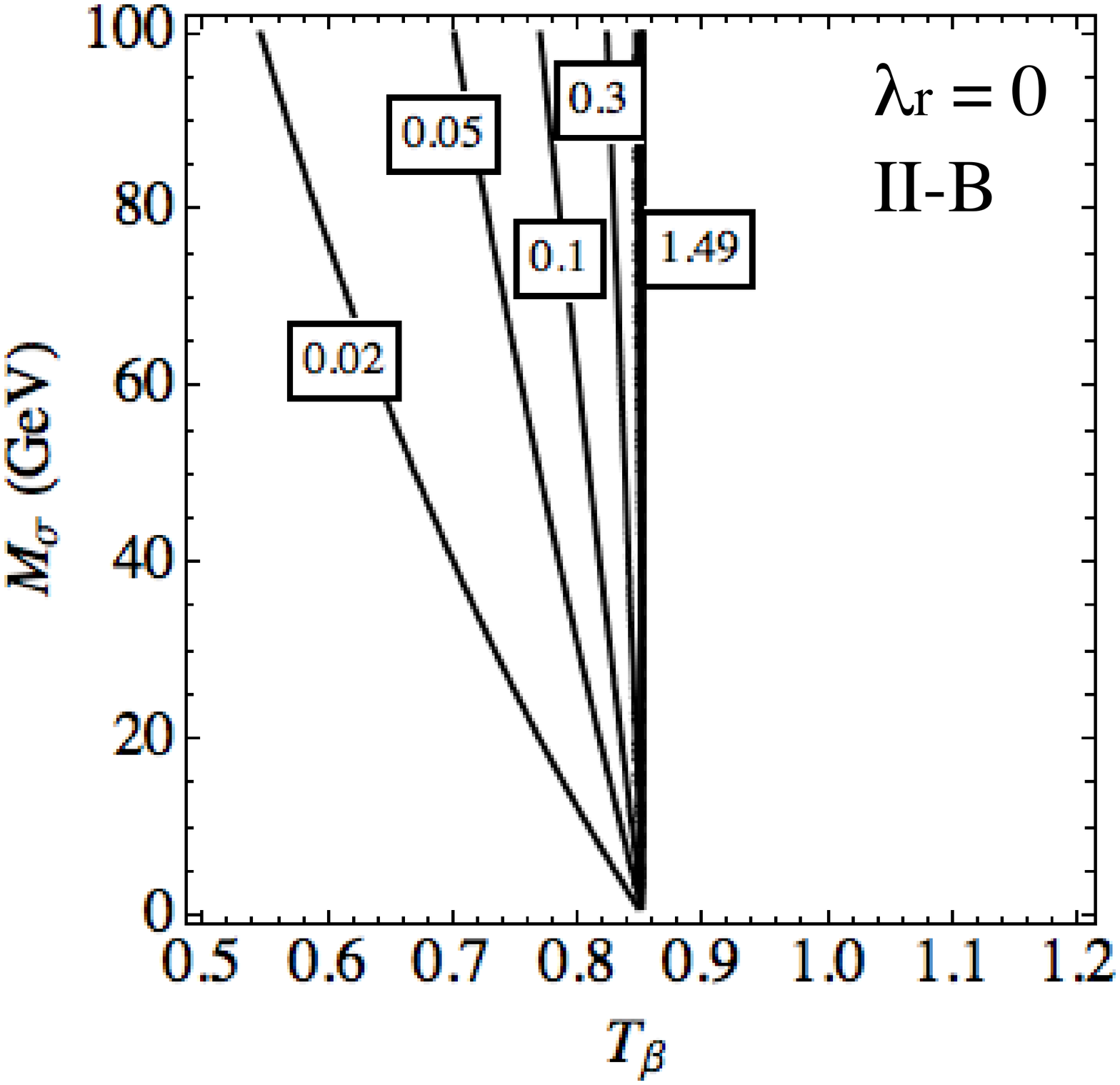}\hspace{-0.8cm}
\includegraphics[width=4.8cm,height=4.7cm,angle=0]{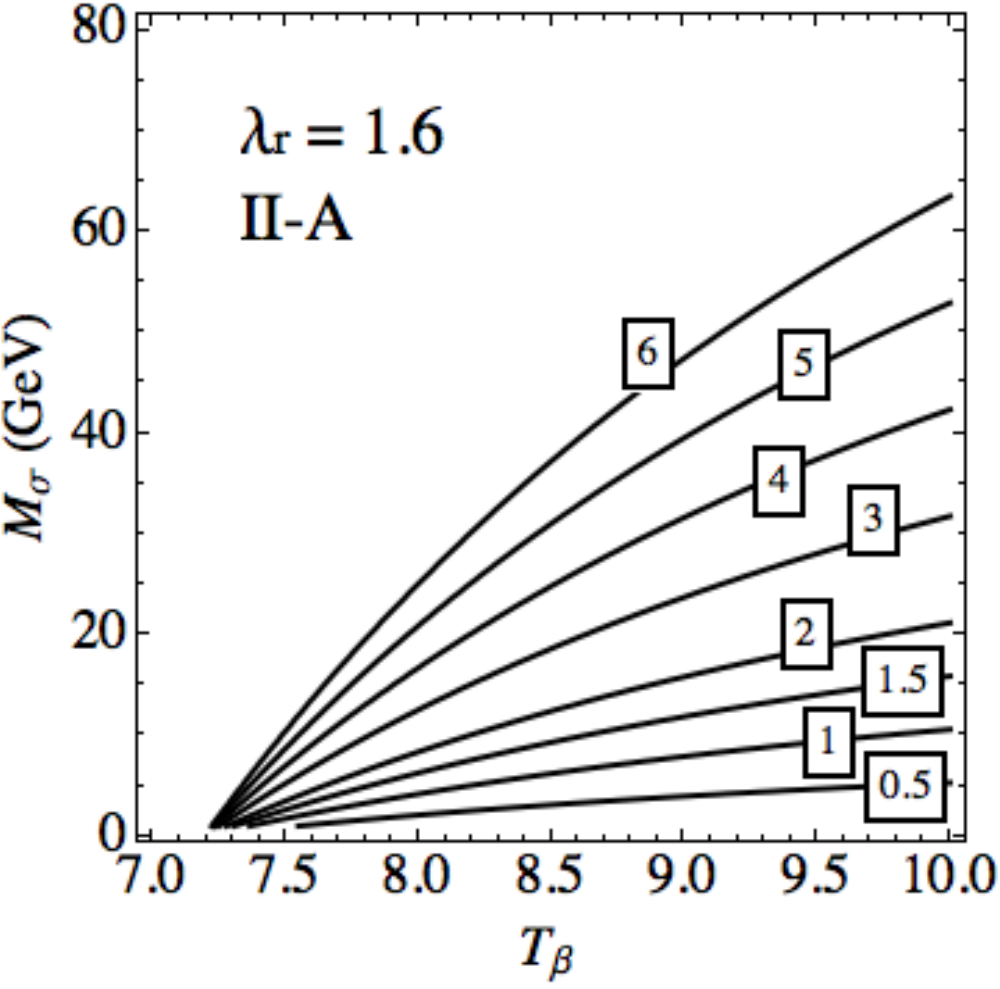}\hspace{0cm}
\includegraphics[width=4.8cm,height=4.7cm,angle=0]{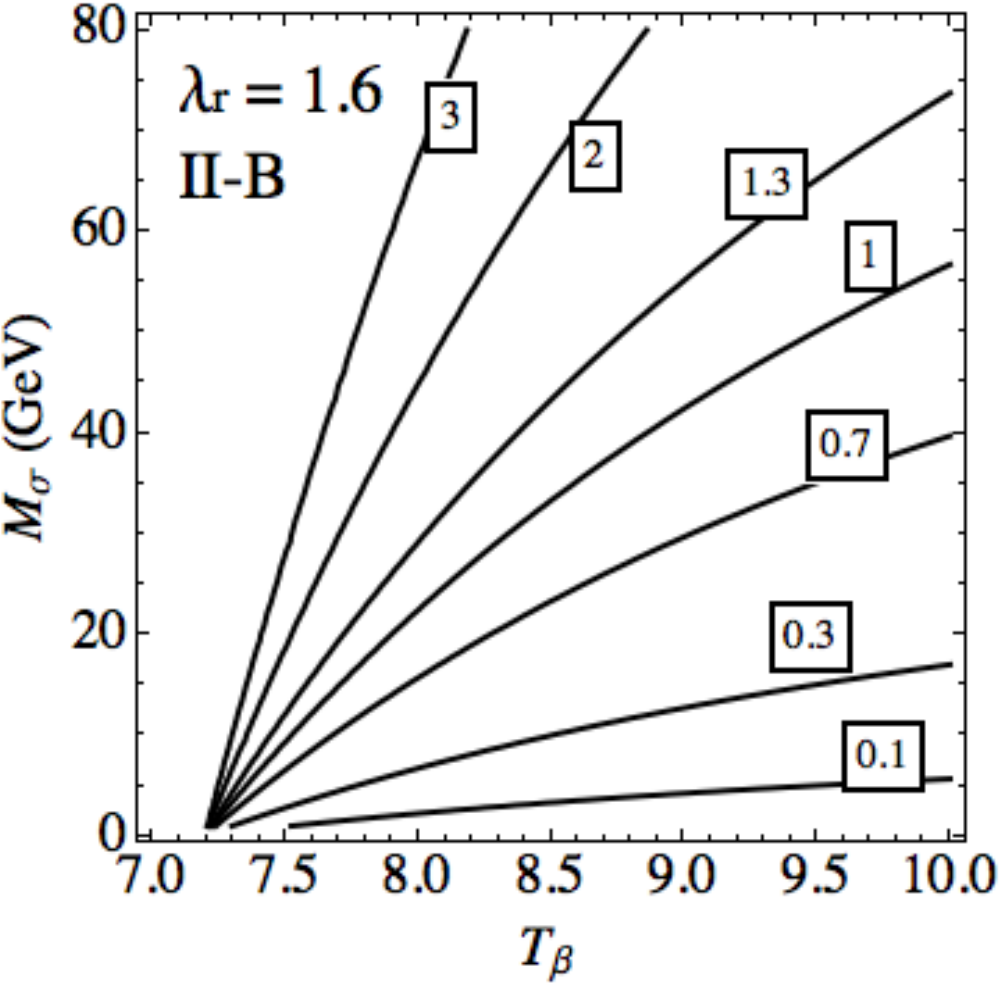}
\caption{\small  Contourplots of $\lambda '_6$ in the plane $T_{\beta}-M_{\sigma}$ for $f_n/f_p=-0.7$ with Higgs and vector bosons exchange, which cancel the scattering with $^{131}Xe$ isotopes.}
\label{fig11}
\end{figure}

\begin{figure}[t]
\centering
\includegraphics[width=8cm,height=6cm,angle=0]{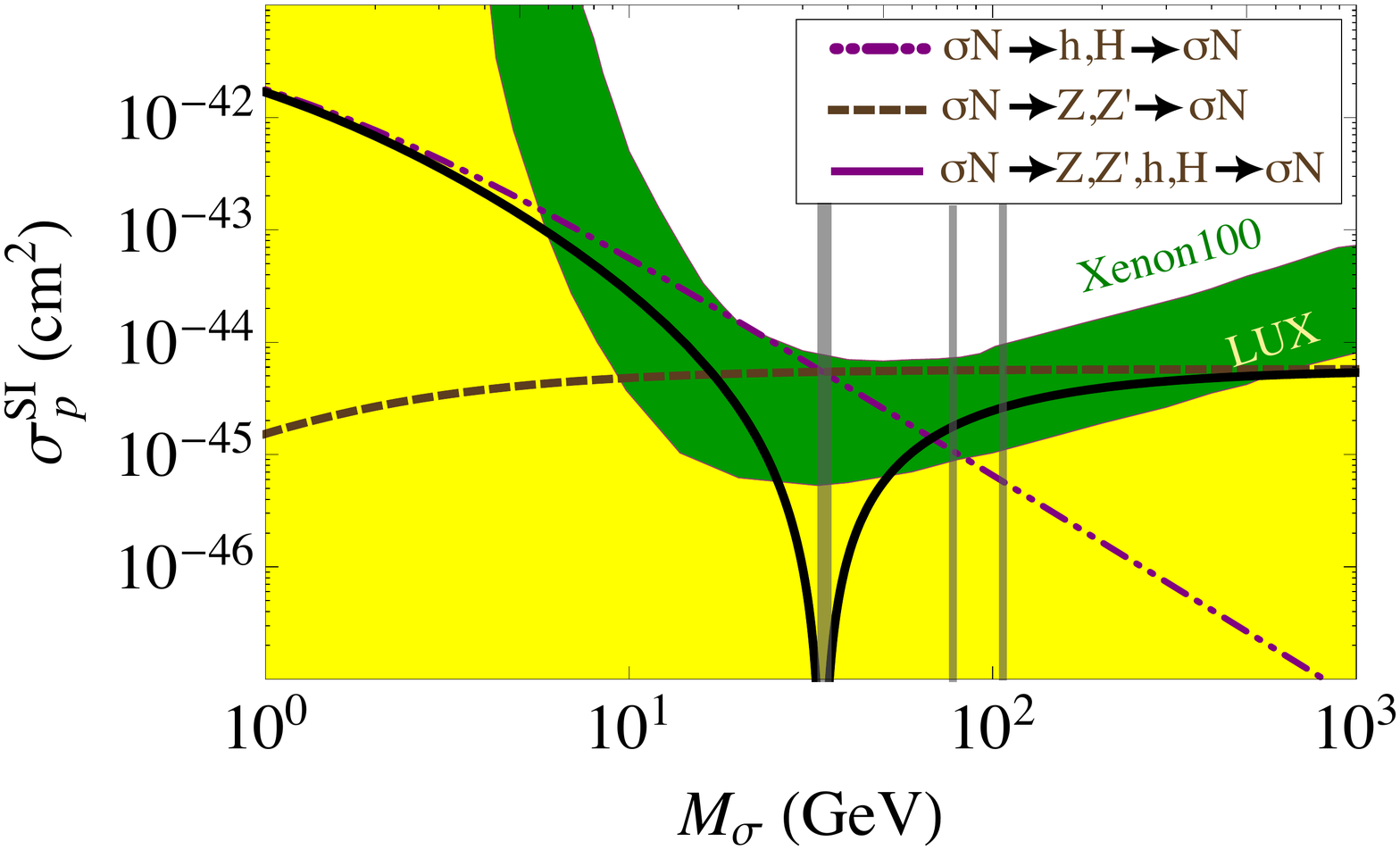}\vspace{-0.7cm}
\caption{\small WIMP-proton cross sections with only Higgs exchange (purple dotdashed), only gauge boson exchange (brown dashed) and combined exchange (black). Vanishing destructive interference between scalar and vector interactions arise at $M_{\sigma}=35$ GeV. The vertical shaded fringes are the limits compatible with the measured DM relic abundance.}
\label{fig12}
\end{figure}

\subsection{Interactions with Higgs and vector boson exchange combined}

We find different results if we take into account both the scalar and vector interactions simultaneously. Since the space of parameters in this general case has many variables, we set the values $M_H=300$ GeV, $M_{Z_{2}}=3000$ GeV and $g_X=0.4$ as inputs, which are compatible with electroweak observables and collider researches. Again, we evaluate the ratio between the effective nucleon couplings for each theoretical combination (Model I-A, I-B, II-A and II-B) and three values of $\lambda _r$: $0$, $1$ and $2$. The other two parameters, $M_{\sigma }$ and $\lambda ' _6$, are scanned in the ranges $[6,12]$ GeV and $[0.02,1.5]$, respectively. The range of $M_{\sigma}$ is choosen according to the region of interest where a positive WIMP signal is claimed by the CoGeNT experiment \cite{cogent, cogent2}. Figure \ref{fig9} shows the ratio as function of $T_{\beta}$ for each case and both types of Yukawa couplings, i.e. type I (blue points) and II (red points). We see, that due to the simultaneous contribution from Higgs and vector bosons, the bands expand to a wider region. For example, in figure \ref{fig4} the type I model is reduced to a single line near $1$ for all $T_{\beta }$ value, now, the same model exhibits density points in a larger range of the space of parameters, especially when $\lambda _r=0$. It is also noteworthy that while figure \ref{fig7} indicates that there are not total interferences, in figure \ref{fig9} we obtain points with $f_n=0$ and/or $f_p=0$. For example, when $\lambda _r=0$, model A has points where $f_n/f_p=0$ in the region $T_{\beta}>3$ for type I, while model B does not exhibit a cancelation of this ratio. By contrast, type II exhibits a narrow band near $T_{\beta}=1$ that extend to $\pm \infty$. This ``anomalous''  peak arise due to the interference between both Higgs channels (observe that when $\lambda _r=0$ and $T_{\beta}=1$, the scalar couplings become opposite, i.e., $\lambda _h=-\lambda _H$ in Eq. (\ref{higgs-coupling-2})).

On the other hand, if we compare regions for different scalar couplings, we see that $\lambda _r=0$ exhibits opposite solutions in relation to  $\lambda _r=2$. The former displays narrow and constant bands for type II model (red bands) except for the anomaly at $T_{\beta}\approx 1$, and broad regions in almost all the planes for type I (blue regions), while in the latter the situation is inverted, with broad regions for type II and narrow bands for type I. This inversion is the same as in figure \ref{fig4}, due to the change of sign of the Higgs coupling $\lambda _h$. Finally, the points for $\lambda _r=1$ do not exhibit cancellations of the nucleon couplings. 
 
Since we obtain scenarios with interference, we explore solutions that accomplish

\begin{eqnarray}
f_p=\frac{M_p}{2M_{\sigma}}\left[S_{h}F_p^{h}+S_{H}F_p^{H}\right]+G_{Z_1}V_p^{Z_1}+G_{Z_2}V_p^{Z_2}=0,
\label{effective-coup-2}
\end{eqnarray}
which occur in the cases $\lambda _r=0$ and $2$. Figure \ref{fig10} shows the contour plots of the parameter $\lambda '_6$ in the plane $T_{\beta}-M_{\sigma}$ that holds the condition (\ref{effective-coup-2}) for each theoretical model and both alternatives for $\lambda _r$. There are no solutions in the particular cases of models I-A and I-B when $\lambda _r=2$, which is compatible with the corresponding regions in figure \ref{fig9}. We also find that the contours correspond to the ranges of $T_{\beta}$ where $f_n/f_p$ diverges in figure \ref{fig9}. For example, the contour lines for $\lambda _r=2$ in models II-A are defined above $T_{\beta}>2.5$, which match with the aymptote of the borderline for the corresponding case in figure \ref{fig9}.

It is also interesting to explore solutions where $f_n/f_p=-0.7$, which will cancel the cross section for the $^{131}Xe$ isotopes, as a possible explanation for the negative results of the current data from Xenon-based experiments. Figure \ref{fig11} displays the corresponding contour plots of $\lambda '_6$ that lead to this hypotetical cancellation. To study the significance to consider both types of intermediary interactions, Higgs and gauge bosons, we compare the cross sections in three scenarios: with only Higgs exchange ($g_X=0$), only gauge boson exchange ($\lambda '_6=0$) and scattering with both contributions ($g_X$ and $\lambda' _6$ different from zero). For the purpose of illustration, we choose $\lambda _r=0$ in the framework of models I-A. According to the first plot in figure \ref{fig11}, isospin interference in $Xe$ nucleus arise at, for example, $T_{\beta}=7.5$ and $M_{\sigma}=35$ if $\lambda '_6=1$. The plot in figure \ref{fig12} shows the cross section for WIMP-proton scattering in three scenarios of particle exchange. For reference, we include the limits from Xenon100 and LUX experiments \cite{xe100, lux}, where the shaded areas are allowed regions. First, we see that the case with only Higgs exchange (purple dotdashed line) drops sharply with $M_{\sigma }$ but does not exhibit interference for intermediate values of masses. By contrast, the interaction through only gauge bosons (brown dashed line) shows a constant contribution, which corresponds to the same region from figure \ref{fig8}. Finally, taking into account Higgs and gauge boson exchange simultaneously (black line), we find the expected interference peak at $M_{\sigma}=35$ GeV, which confirms that Higgs and gauge boson exchange combines to produce the cancellation of the scattering for specific values of $M_{\sigma}$. In fact, we see that the interference peak coincides with the intersection of each contribution, which indicates that the Higgs interaction has the same strength as the vector interaction but with an opposite sign. Furthermore, if we take into account the stringent limits from DM relic abundance, we obtain the three vertical shaded fringes in the ranges $M_{\sigma}=[33,35.11], [77,78],$ and $ [104,104.5]$ GeV, that match the vanishing destructive interference.

\begin{figure}[t]
\centering
\includegraphics[width=8cm,height=5.2cm,angle=0]{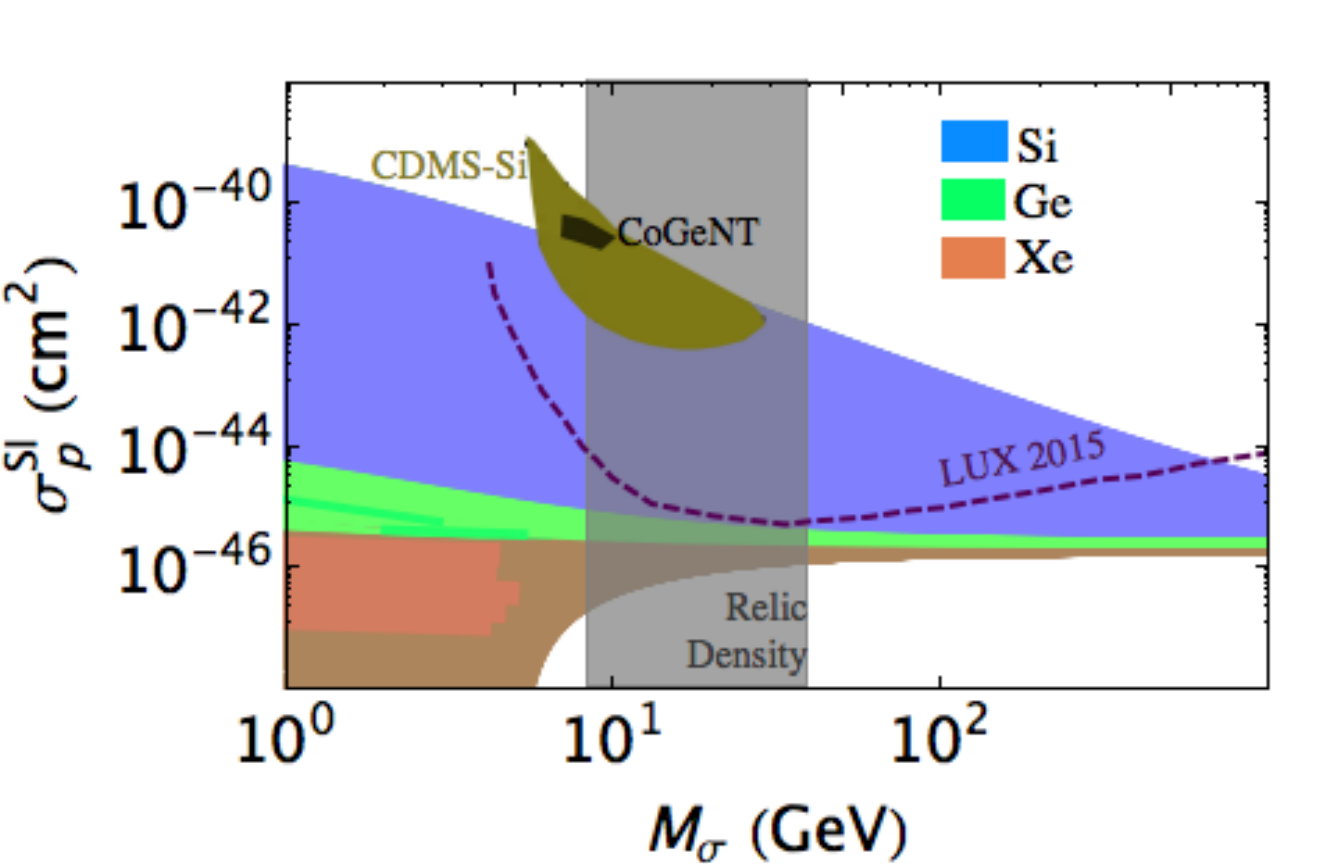}\vspace{-0.2cm}
\caption{\small Theoretical WIMP cross section regions (normalized to a single proton) for $Si$ (blue), $Ge$ (green) and $Xe$ (brown) isotopes. The points contained into the shaded rectangular region fullfill the relic abundance limits.}
\label{fig13}
\end{figure}

To examine isospin-violating effects, we consider three types of isotopes: $^{29}Si$, $^{73}Ge$ and $^{131}Xe$. For the purpose of illustration, we choose the model type II-B and the following values of the space of parameters: $(g_X,\lambda '_6,\lambda _r, M_H, M_{Z_{2}})=(0.4, 1, 1.6, 300, 3000)$. According to the last plot from figure \ref{fig11}, the interference contours converges in the limit $T_{\beta}=7.2$ in the low WIMP mass region. Below this value, there are no solutions for interferences in the context of models II-B. Figure \ref{fig13} shows regions of the theoretical cross sections for each nucleus, where we scan values of $T_{\beta}$ in the range $1$-$7.3$. We also include the experimental data from three detectors: LUX (Xenon-based), CDMS-Si (Silicon-based) and CoGeNT (Germanium-based), where LUX displays a lower limit, while CDMS and CoGeNT exhibit regions of interest where positive signals may exist. First, we see that there is a band where the regions from the three isotopes ovelap at large cross sections, above the LUX limits. In particular, we observe that the theoretical regions for Silicon and Germanium can take values up to the experimental region of interest from CoGeNT and CDMS-Si. Second, there are solutions where the interactions with the three isotopes separate at low cross sections, below the LUX limit. However, only interactions with Xenon nucleus exhibits solutions that suppress the cross section, which occur around $T_{\beta}=7.3$ and for light WIMPs ($M_{\sigma}<6$ GeV), where LUX excludes WIMP-Xe cross sections above $8.5\times 10^{-44}$ cm$^2$. Finally, with the same $T_{\beta}$ scan, we obtain the WIMP mass limits $M_{\sigma}=[8.3,38.5]$ GeV allowed by the observed DM relic density, shown as the shadow rectangular region in the plot. In particular, we see that the relic density region overlaps the CDMS-SI and CoGeNT regions in the range $M_{\sigma}=[8.3,10]$ GeV. We observe a few points into the overlapped regions for $Si$ and $Ge$ nucleus, compatible with scattering signals and relic abundance simultaneously, while $Xe$ nucleus exhibits cross sections below the LUX limits.    

\section{Spin-dependent Interaction}

Although the interaction between scalar WIMPs and the spin of the nucleus is neglegible in the extreme non-relativistic limit, this interaction could become the only source of scattering in case of cancellation of the SI interaction. In particular, we can evaluate the ratio between SD and SI events for different isotopes and explore the effects on this parameter when cancellation of the SI interaction due to interferences takes place.

\subsection{Spectrum with discrete speeds}

\begin{figure}[t]
\centering
\includegraphics[width=7cm,height=5.2cm,angle=0]{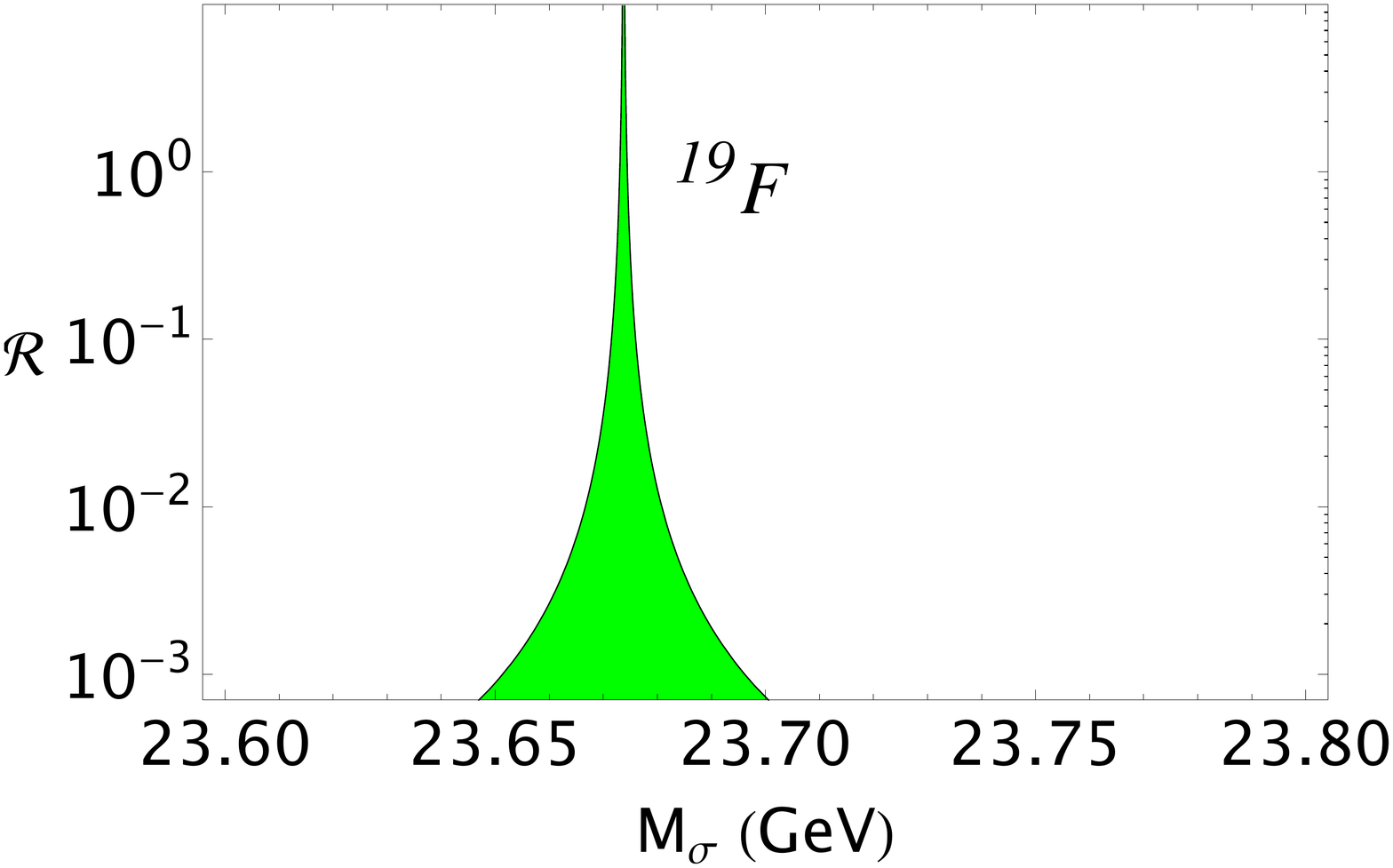}
\includegraphics[width=7.5cm,height=5.9cm,angle=0]{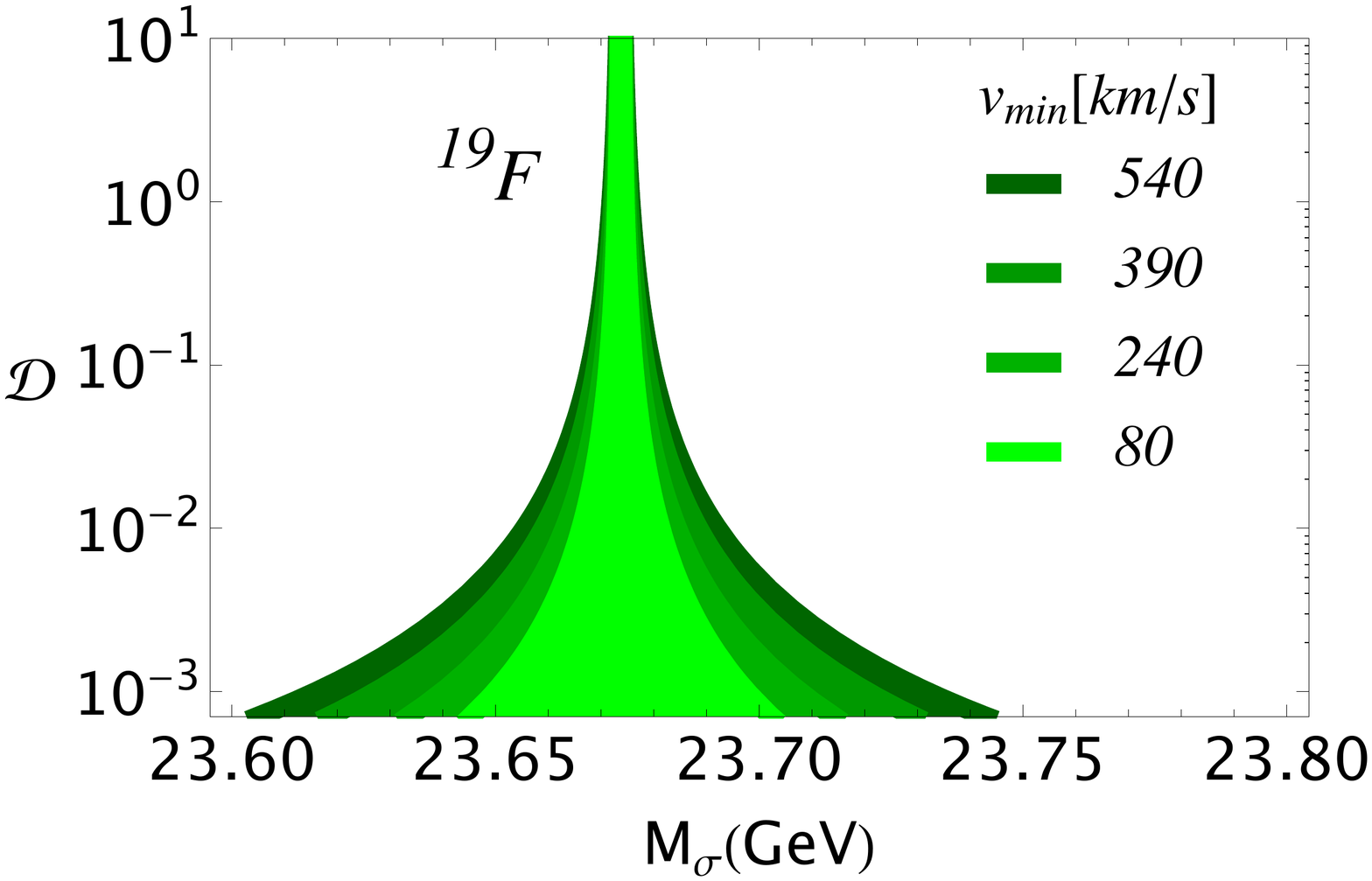}\vspace{-1.5cm}
\includegraphics[width=7cm,height=5.2cm,angle=0]{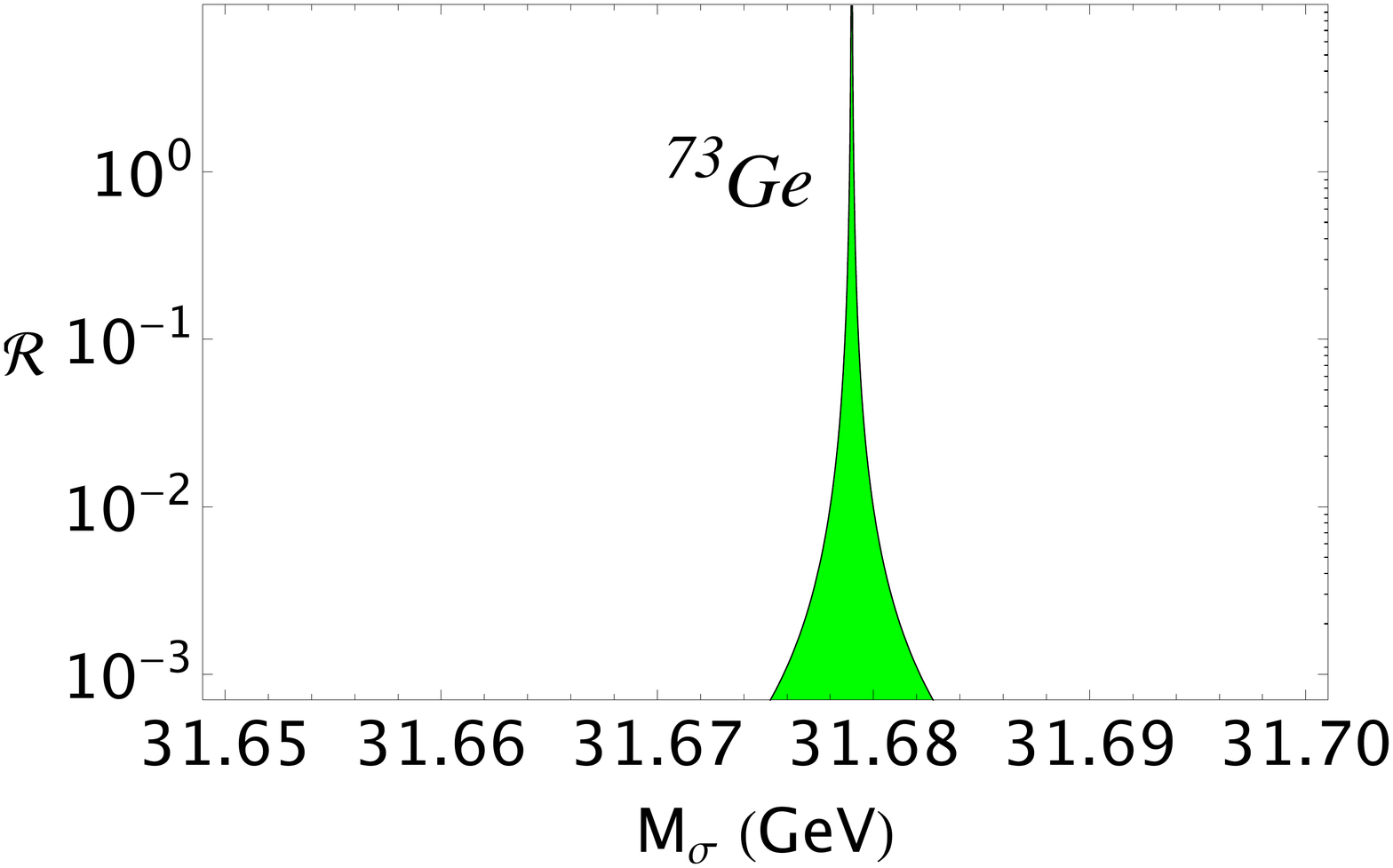}
\includegraphics[width=7.5cm,height=5.9cm,angle=0]{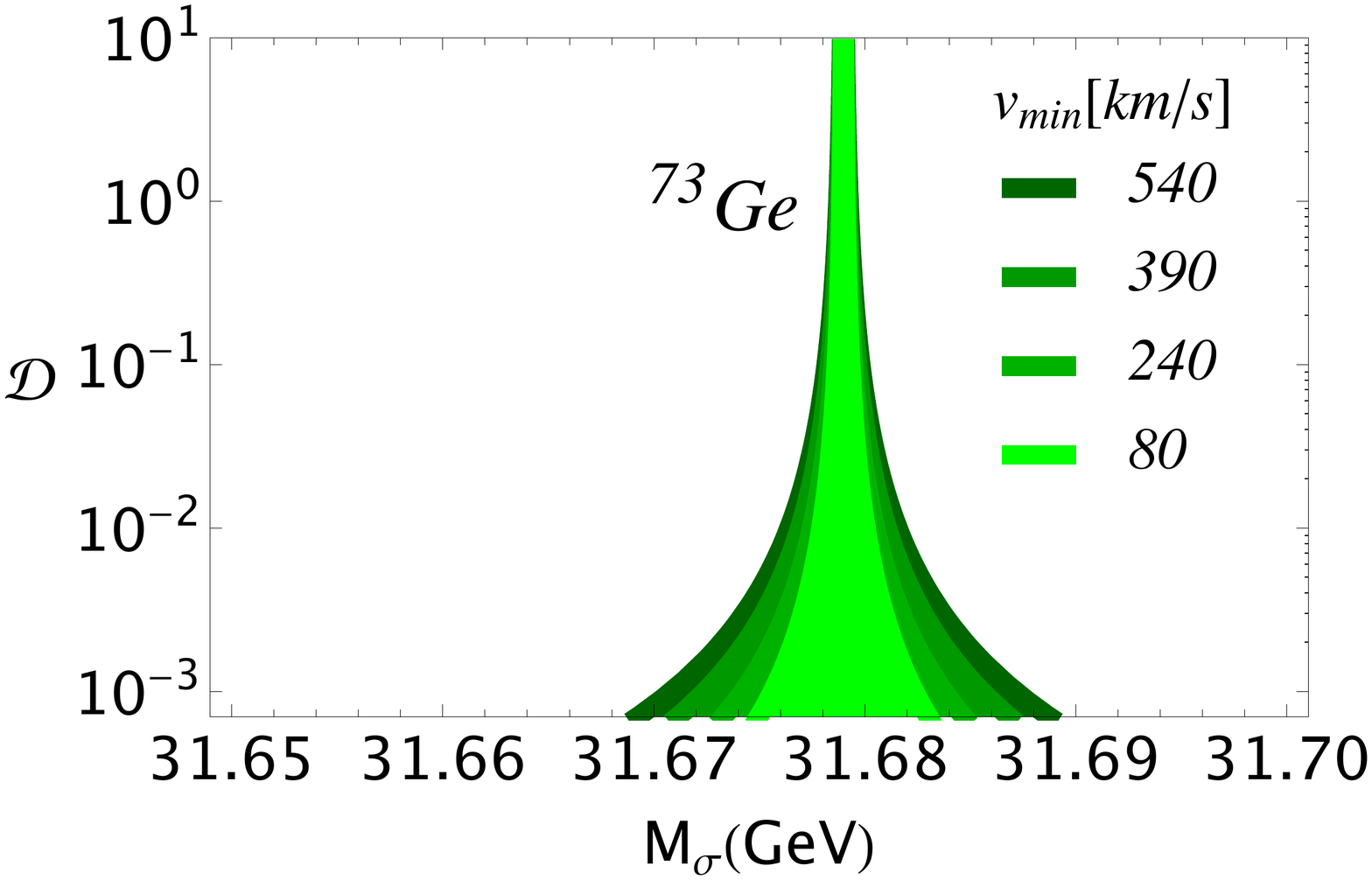}\vspace{-1.5cm}
\includegraphics[width=7cm,height=5.2cm,angle=0]{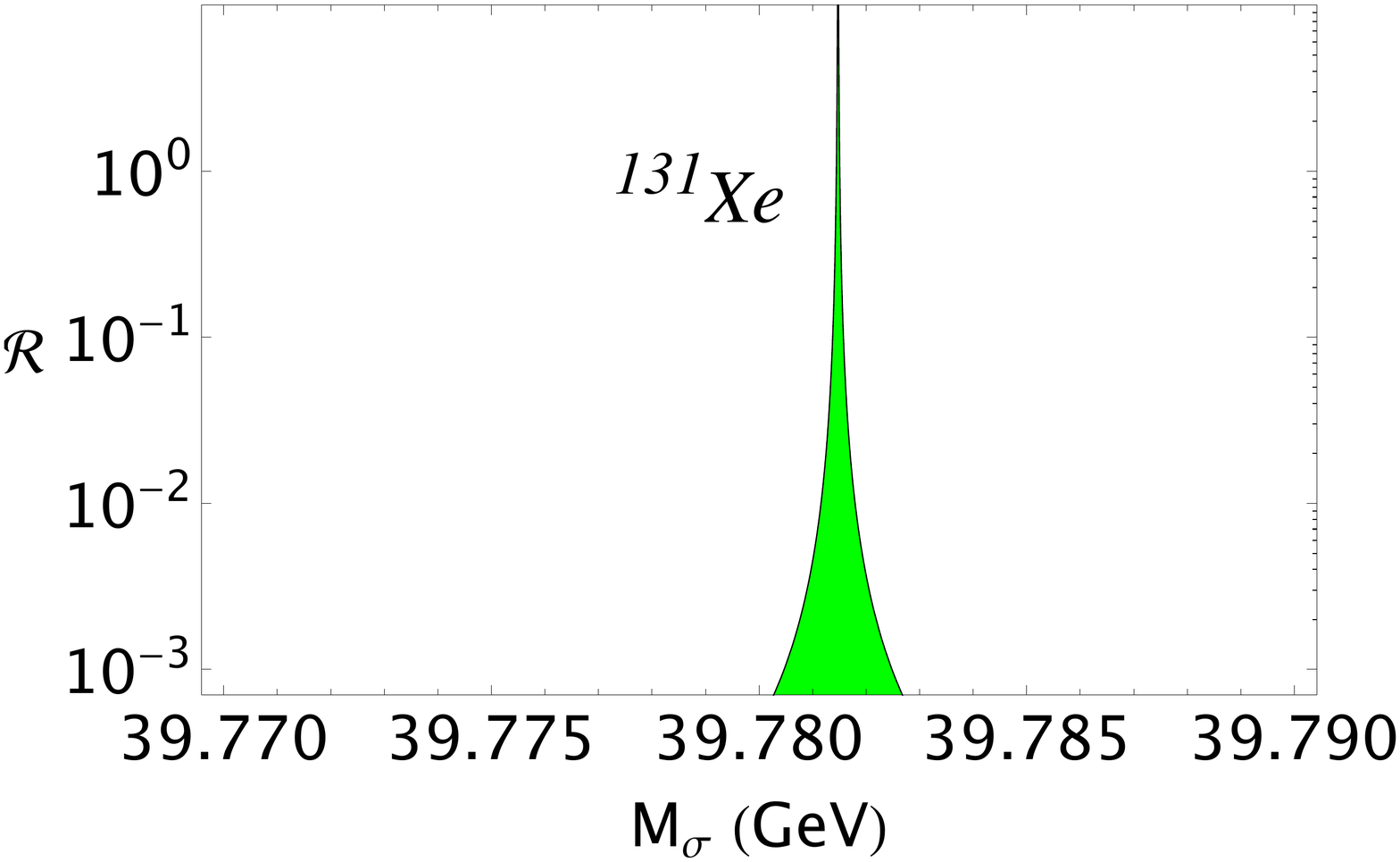}
\includegraphics[width=7.5cm,height=5.9cm,angle=0]{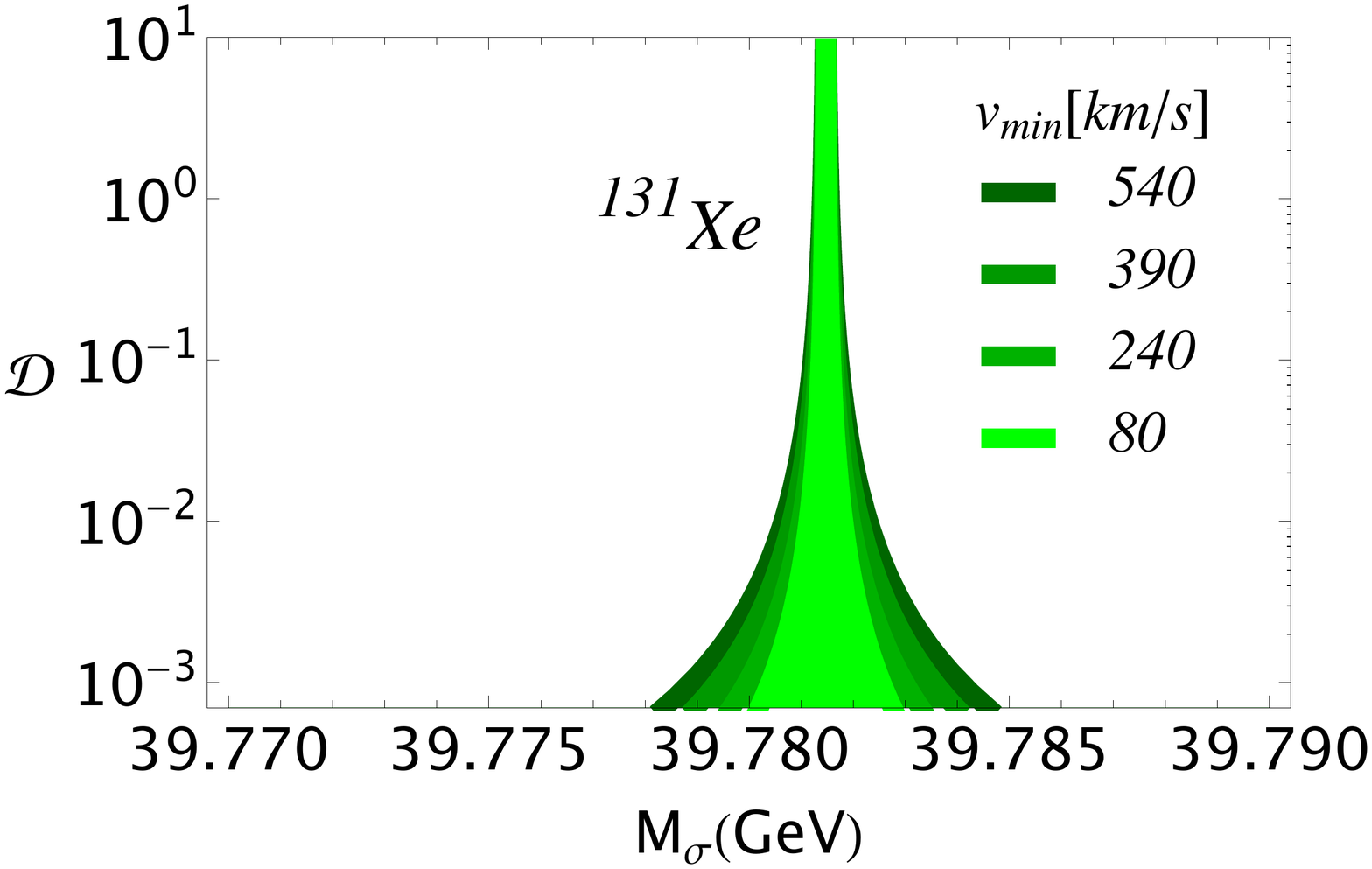}\vspace{-0.5cm}
\caption{\small SD and SI cross section ratios at zero momentum transfer (left column) and rate events at finite momentum transfer with Maxwellian distributions of velocities (right column). We choose $(\lambda _r, \lambda '_6, T_{\beta})=(1.6, 0.7, 10)$ in models II-B.}
\label{fig14}
\end{figure}

First, we consider the ideal limit when the collision speed of WIMPs is perfectly known. We choose, as an approximate value, the circular speed of the Sun around the galactic center, about $220$ km s$^{-1}$, which gives $\beta = 7 \times 10^{-4}$. From (\ref{SI-SD-cross-sections}), the ratio between the SD and SI cross section at zero momentum transfer is:

\begin{eqnarray}
\mathcal{R}=\frac{\sigma _0^{SD}}{\sigma _0^{SI}}=\frac{\left|\pmb{\beta }\right|^2}{3f_p^{2}}j(j+1)\left|\frac{\sum _{\mathcal{Z}}G_{\mathcal{Z}}\Lambda _{\mathcal{Z}}}{Z+\frac{f_n}{f_p}(A-Z)}\right|^2,
\label{SD-SI-ratio-discrete}
\end{eqnarray}
where the SD factor is mediated only by the gauge bosons. Let us take a specific case in models II-B (the general form of the ratio does not change signficantly for other cases). For example, according to figure \ref{fig11}, if $\lambda _r=1.6$ and $\lambda '_6=0.7$, the interaction with $Xe$ nucleus in this model disappears at $(T_{\beta},M_{\sigma})=(10,40$ GeV$)$. This type of cancellation will also occur for other isotopes with the same parameters, but at different WIMP masses. The plots in the left column in figure \ref{fig14} show the ratio (\ref{SD-SI-ratio-discrete}) as function of $M_{\sigma}$ for three different isotopes. Specifically, we choose $^{19}F$, $^{73}Ge$ and $^{131}Xe$, which are typical targets used in experiments for SD couplings. In particular, for Xenon, we see a very narrow peak at the expected region of interference, more precisely at $M_{\sigma}=39.7815$ GeV. Similarly, a narrow peak is found for Germanium at   $M_{\sigma}=31.679$ GeV and Fluor at $M_{\sigma}=23.675$ GeV. Since Fluor-based detectors are more sensitive to SD interactions, its line shows a larger width, and raises to values above $\mathcal{R}>1$, i.e., to values with $\sigma _0^{SD}>\sigma _0^{SI}$. %However, this dominance of the SD interaction occurs in a very thin range of the space of parameters, which requires a fine tune of the parameters. Nevertheless, this fine tuning arise in the case of an unique colision with a perfectly known value of the speed, which is not real in an experiment. 
More accurately, we must compare event rates with some distribution of velocities, as shown below.

 \subsection{Spectrum with speed distributions}
 
The predicted WIMP event rate per unit detector mass is \cite{wimpsC}

\begin{eqnarray}
dR=\frac{\rho _0}{M_{\sigma }M_A}\frac{d\sigma}{d|\mathbf{q}|^2}vf(v)dvd|\mathbf{q}|^2,
\label{diferential-rate}
\end{eqnarray}
where $\rho _0\approx 0.3$ GeV/cm$^3$ is the estimated local DM density, $f(v)$ is a distribution function of WIMP velocities and $d\sigma/d|\mathbf{q}|^2$ is the WIMP-nucleus differential cross section at finite momentum transfer. This momentum transfer can be obtained for non-relativistic elastic collisions from the nuclear recoil energy through the classical relation

\begin{eqnarray}
q_0=\frac{|\mathbf{q}|^2}{2M_A}.
\label{recoil-energy}
\end{eqnarray} 

The differential cross section can be parameterized in terms of the zero-momentum cross section as

\begin{eqnarray}
\frac{d\sigma}{d|\mathbf{q}|^2}=\frac{\sigma _0}{4m_r^2v^2}F^2(q),
\label{dif-cros-sec}
\end{eqnarray}
where $F^2(q)$ is a normalized ($F^2(0)=1$) nuclear form factor at finite momentum transfer. In the case of SI interactions, this form factor is the Fourier transform of the nucleon density, which depends on the squared energy transfer. For SD interactions, the form factor is defined as the ratio $S(|\mathbf{q}|)/S(0)$, where $S(|\mathbf{q}|)$ is the axial structure function obtained from nuclear calculations \cite{wimpsA, wimpsB, wimpsC, wimpsD}. After separate SI and SD parts, and using the relations (\ref{recoil-energy}) and (\ref{dif-cros-sec}), the differential ratio in (\ref{diferential-rate}) becomes:

 \begin{eqnarray}
dR=\frac{\rho _0}{2M_{\sigma }m_r^2}\left[\sigma _0^{SI}F^2_{SI}(q_0)+\sigma _0^{SD}F^2_{SD}(q_0)\right]\frac{f(v)}{v}dvdq_0.
\label{diferential-rate-2}
\end{eqnarray} 
In order to obtain the total rate, we must take into account that according to Eq. (\ref{SI-SD-cross-sections}), the SD cross section at zero momentum transfer is a function of the speed, such that this term can not be factorized from the speed integration. Thus, the differential ratio per transfered energy is:

\begin{eqnarray}
\frac{dR}{dq_0}&=&\frac{\rho _0}{2M_{\sigma }m_r^2}\left[\sigma _0^{SI}F^2_{SI}(q_0)\int _{v_{min}}^{v_{max}}\frac{f(v)}{v}dv+F^2_{SD}(q_0)\int _{v_{min}}^{v_{max}}\sigma _0^{SD}\frac{f(v)}{v}dv\right] \nonumber \\
&=&\frac{dR^{SI}}{dq_0}+\frac{dR^{SD}}{dq_0}.
\label{diferential-rate-3}
\end{eqnarray}
The velocity $v_{max}$ corresponds to the local escape speed of our galaxy at $544$ km/s \cite{mon.not}. $v_{min}$ is the minimum velocity of the WIMP in order to transfer some energy $q_0$, which occurs for the WIMPs that are dispersed in the backward direction. By combining equation (\ref{recoil-energy}) with the momentum transfer $|\mathbf{q}|^2=2m_r^{2}v^2\left(1-\cos \theta \right)$ evaluated at $\theta = \pi$, we obtain:

\begin{eqnarray}
v_{min}=\sqrt{\frac{q_0M_A}{2m_r^2}}.
\label{mimimum-speed}
\end{eqnarray}
To compare the SD and SI events, we normalize each rate to the corresponding form factor, and define the ratio:

\begin{eqnarray}
\mathcal{D}=\frac{\frac{1}{F^2_{SD}(q_0)}\frac{dR^{SD}}{dq_0}}{\frac{1}{F^2_{SI}(q_0)}\frac{dR^{SI}}{dq_0}}=\frac{\int _{v_{min}}^{v_{max}}\sigma _0^{SD}\frac{f(v)}{v}dv}{\sigma _0^{SI}\int _{v_{min}}^{v_{max}}\frac{f(v)}{v}dv}
\end{eqnarray}  
As a first approximation, in order to compare the discrete case discussed in the above section, we choose the Maxwellian distribution

\begin{eqnarray}
f(v)=\frac{4v^2}{\sqrt{\pi}v_0^{3}}e^{-v^2/v_{0}^{2}},
\end{eqnarray}
where $v_0\approx 220$ km s$^{-1}$ is the galactic speed of the Sun. Taking into account the cross sections from (\ref{SI-SD-cross-sections}), and defining $\beta =v/c$, we obtain after integration that:

\begin{eqnarray}
\mathcal{D}=\frac{j(j+1)}{3f_p^{2}}\beta _0^{2}\left(1+ \frac{\beta _{min}^2}{\beta _{0}^2}\Delta \right)\left|\frac{\sum _{\mathcal{Z}}G_{\mathcal{Z}}\Lambda _{\mathcal{Z}}}{Z+\frac{f_n}{f_p}(A-Z)}\right|^2,
\label{SD-SI-ratio-distribution}
\end{eqnarray} 
where the term $\Delta $ is the function

\begin{eqnarray}
\Delta = \frac{\exp\left(-\beta _{min}^2/\beta _{0}^2\right)-\frac{\beta _{max}^2}{\beta _{min}^2}\exp\left(-\beta _{max}^2/\beta _{0}^2\right)}{\exp\left(-\beta _{min}^2/\beta _{0}^2\right)-\exp\left(-\beta _{max}^2/\beta _{0}^2\right)}.
\end{eqnarray}
The ratio in (\ref{SD-SI-ratio-distribution}) is the same as the ratio at zero momentum transfer in Eq. (\ref{SD-SI-ratio-discrete}) but changing $\beta ^2$ by $\beta _0^{2}\left(1+\frac{\beta_{min}^2}{\beta _0^{2}}\Delta\right)$. The plots in the right column in figure \ref{fig14} displays the ratio from (\ref{SD-SI-ratio-distribution}) for the same isotopes and parameters used previously for the left plots. Taking into account that the recoil energy $q_0$ depends on the WIMP mass through equation (\ref{mimimum-speed}), we display plots for different minimal speeds. It is evident that the result indicates a distribution of velocities. First, the lines around the SI interference encompass a broader region where the SD interaction is comparable to the SI part. Second, the width of the lines increses as the minimum speed increases. However, below $v_{min}<80$ km/s we did not find variations in relation to the zero momentum transfer case in the left plots. The effect is stronger for fluor-based detectors.
%We use in addition the typical value $q_0 \approx 100$ keV for the recoil energy. It is evident from the plots that as a consequence to consider a distribution of velocities, the lines around the SI interference get a broader region where the SD interaction is comparable to the SI part. This effect is stronger for fluor-based detectors.  

\section{Conclusions}

In the context of a nonuniversal $U(1)'$ extension of the SM with scalar DM, we studied scenarios for destructive interference of the spin-independent interactions in WIMP-nucleus scattering. The model contains specialized 2HDM types I and II, and an extended gauge sector with a new neutral weak boson $Z'$. The chosen theoretical model allows two cases of interference. First, for pure scalar couplings, there are two sources of scattering: through the SM-like Higgs boson $h$ and an extra CP-even neutral Higgs $H$. By matching the SM-like Higgs to the observed $125$ GeV scalar boson, we found a set of values of the space of parameters where both channels of scattering mutually interferes, which cancel the WIMP-nuclear cross section for SI interactions for all WIMP mass range. Solutions for cancellation are found at $M_{H}>125$ and $170$ GeV for type I and II, respectively. Second, due to the extra gauge content of the model, it is possible to find scenarios where the Higgs exchanges cancel the $Z_1$ and $Z_2$ exchanges. These interferences are calculated by assuming two family structures, A and B, where the $U(1)'$ quantum number depends on the quark flavour. Combined scenarios of models type I and II with the two family structures were evaluated, showing several options for total interference of the SI scattering, which depends on the WIMP mass. In particular, the type II model exhibits regions free from interferences according to the value of the relative scalar coupling $\lambda _r$. For example, if $\lambda _r=0$, interferences occur only when $T_{\beta}<0.9$, while for $\lambda _r=2$, this occurs only when $T_{\beta}>2.5$, as shown in figure \ref{fig10}. There are no restrictions for type I models. If both scalars doublets have the same coupling constants (i.e. $\lambda _r=1$) there are no solutions for total interference. We also examine the case of pure vector couplings, where the gauge bosons $Z_1$ and $Z_2$ are the only source of scattering between WIMPs and nucleus. We did not find any interference effects between both gauge bosons. In this case, the cross sections exhibit different allowed regions according to the family structure. In general, one structure (B in table \ref{tab:family-matching}) exhibits smaller cross sections than the other structure (A), such that B gives values below the LUX limits, while A passes the smallest limit from XENON100 experiment. In paticular, LUX excludes the family structure A in the range $9<M_{\sigma}<800$ GeV.

On the other hand, we studied scenarios where the effective WIMP-nucleon coupling depends on the type of nucleon, i.e., couplings with isospin asymmetry. For pure scalar interactions, these scenarios arise with Yukawa couplings type II and for specific values of the space of parameters. However, the effects of interferences with isospin asymmetry affect all the WIMP mass ranges in the same amount, and is overlaped by the quantum interferences between both Higgs exchanges, producing indistinguishable cross sections among different isotopes. For that, we found that the interferences between Higgs and gauge vector bosons exhibits observable interferences that depends on the content of protons and neutrons of the target nucleus. In particular, there are solutions where the interaction with Xenon-based detectors is suppresed by Higgs-vector interferences in the low WIMP mass region (below $M_{\sigma}<6$ GeV), while the interactions with other isotopes (Germanium and Silicon) enhance their interactions in the same region. We found solutions that fit the parameters according to the experimental regions from the Si-, Ge- and Xe-based detector experiments, and simultaneously are compatible with the DM relic abundance observations in the range $M_{\sigma}=[8.3, 10]$ GeV.

Finally, we examine regions where the SD interactions can participate with the same strenght as the SI interactions into the WIMP-nucleus scattering. Although the SD cross section is suppressed in the non-relativistic limit by a factor $\beta ^2 \sim 5 \times 10^{-7}$, in case of interference of the SI contribution, the SD coupling could become in the dominant source of WIMPs scattering. In the ideal case of an unique collision with a perfectly known speed, this dominance of the spin interaction exhibits stringent limits of the WIMP mass around the interference point. By considering a more realistic scenario, where the collisions are governed by a statistical speed distribution, the regions encompass broader ranges  around the interference, where the SD interaction is comparable to the SI interaction. This effect is more significant for fluor-based detectors.

\section*{Acknowledgment}

This work is supported by El Patrimonio Aut\'onomo Fondo Nacional de Financiamiento para la Ciencia, la Tecnolog\'{i}a y la Innovaci\'on Fransisco Jos\'e de Caldas programme of COLCIENCIAS in Colombia. F.O, thanks to the Theory Unit of the Physics Department of CERN, where part of this work was developed.
 
\section*{Appendix}

\appendix

\section{Form Factors\label{App:Form factors}}

In this section, we obtain the form factors of the three matrix elements shown in Eq. (\ref{nucleon-amplitudes}). Each amplitude corresponds to scalar, vector, and axial-vector interactions, respectively. 

\subsection{Scalar amplitudes}

The trace of the QCD energy-momentum tensor is:

\begin{eqnarray}
\theta ^{\mu} _{\mu }=\sum _Q m_Q\overline{Q}Q+\frac{\beta (\alpha _s)}{2\alpha _s}Tr(G^{\mu \nu }G_{\mu \nu}),
\label{QCD-tensor}
\end{eqnarray}
where all 6 flavour quarks ($Q=u,d,s,c,b,t$) contributes, and $\beta (\alpha _s)$ is the QCD beta function. At leading order for $n=6$

\begin{eqnarray}
\beta (\alpha _s) \approx -\left(\frac{33-2n}{12\pi}\right)\alpha _s^{2}=\frac{-7}{4\pi }\alpha _s^{2}.
\label{QCD-beta}
\end{eqnarray}
Then, the matrix element of the trace in (\ref{QCD-tensor}) is:
%The matrix element of the trace is related with the nucleon mass by:

%\begin{eqnarray}
%\langle N|\theta ^{\mu} _{\mu } |N\rangle = M_N\overline{u^{s}}(k_N)u^{s}(k_N).
%\label{QCD-element}
%\end{eqnarray}
%By combining (\ref{QCD-element}) and (\ref{QCD-beta}) in (\ref{QCD-tensor}), we obtain:

\begin{eqnarray}
\langle N'|\theta ^{\mu} _{\mu } |N\rangle=\sum _Q \langle N'|m_Q\overline{Q}Q|N\rangle-\frac{7\alpha _s}{8\pi }\langle N'|Tr(G^{\mu \nu }G_{\mu \nu})|N\rangle .
\label{total-element}
\end{eqnarray}

It is convenient to separate the light sector of quarks ($Q_{l}=u,d,s$) from the heavy sector ($Q_{h}=c,b,t$), such that each light quark contribute to the total matrix element in a fraction:

\begin{eqnarray}
f_{TQ_{l}}^{(N)}=\frac{\langle N'|m_{Q_l}\overline{Q_{l}}Q_{l}|N\rangle}{\langle N'|\theta ^{\mu} _{\mu } |N\rangle},
\label{light-formfactor}
\end{eqnarray}
while the heavy sector contributes through anomaly corrections with gluons as \cite{shifman}:

\begin{eqnarray}
\sum _{Q_h} \langle N'|m_{Q_h}\overline{Q_h}Q_h|N\rangle = \frac{-\alpha _s}{4\pi }\langle N'|Tr(G^{\mu \nu }G_{\mu \nu})|N\rangle.
\label{heavy-element}
\end{eqnarray}
Thus, after combining (\ref{heavy-element}) and (\ref{light-formfactor}) into (\ref{total-element}), we obtain the following relation between the light and heavy matrix elements:

\begin{eqnarray}
\sum _{Q_h} \langle N'|m_{Q_h}\overline{Q_h}Q_h|N\rangle&=&\frac{2}{9}\left(\langle N'|\theta ^{\mu} _{\mu } |N\rangle-\sum _{Q_l}\langle N'|m_{Q_l}\overline{Q_{l}}Q_{l}|N\rangle \right) \nonumber \\
&=&\frac{2}{9}\langle N'|\theta ^{\mu} _{\mu } |N\rangle\left(1-\sum _{Q_l} f_{TQ_{l}}^{(N)}\right).
\label{light-heavy-relation}
\end{eqnarray}
Defining the gluon fraction as:

\begin{eqnarray}
 f_{TG}^{(N)}= 1-\sum _{Q_l}f_{TQ_{l}}^{(N)},
\end{eqnarray}
then, from (\ref{light-heavy-relation}), each heavy quark contributes as:

\begin{eqnarray}
\langle N'|m_{Q_h}\overline{Q_h}Q_h|N\rangle=\frac{2}{27}\langle N'|\theta ^{\mu} _{\mu } |N\rangle f_{TG}^{(N)}. 
\label{heavy-formfactor}
\end{eqnarray}
The matrix element of the trace is related with the nucleon mass by:

\begin{eqnarray}
\langle N'|\theta ^{\mu} _{\mu } |N\rangle = M_N\overline{u^{s'}}(k'_N)u^{s}(k_N).
\label{QCD-element}
\end{eqnarray}

By using the definition from Eq. (\ref{light-formfactor}), the result from (\ref{heavy-formfactor}), and the relation (\ref{QCD-element}), the amplitude of the first equation of (\ref{nucleon-amplitudes}) can be demostrated as follows:

\begin{eqnarray}
\sum _Q\langle N' |c_Q^{\mathcal{H}}m_Q\overline{Q}Q| N \rangle &=&\sum _{Q_l}\langle N' |c_{Q_l}^{\mathcal{H}}m_{Q_l}\overline{Q_l}Q_l| N \rangle+\sum _{Q_h}\langle N' |c_{Q_h}^{\mathcal{H}}m_{Q_h}\overline{Q_h}Q_h| N \rangle \nonumber \\
&=&M_N\overline{u^{s'}}(k'_N)u^{s}(k_N)\left(\sum _{Q_l}c_{Q_l}^{\mathcal{H}}f_{TQ_{l}}^{(N)}+\frac{2}{27}f_{TG}^{(N)}\sum _{Q_h}c_{Q_h}^{\mathcal{H}}\right).
\end{eqnarray}
The form factor is defined as:

\begin{eqnarray}
F_N^{\mathcal{H}}=\sum _{Q_l}c_{Q_l}^{\mathcal{H}}f_{TQ_{l}}^{(N)}+\frac{2}{27}f_{TG}^{(N)}\sum _{Q_h}c_{Q_h}^{\mathcal{H}},
\label{scalar-formfactor}
\end{eqnarray}
obtaining the final results at (\ref{nucleon-amplitudes}). The individual form factors $f_{TQ_{l}}^{(N)}$ for each light quarks are listed in table \ref{tab:nucleon-param}.

\begin{table}[tbp]
\begin{center}
%\caption{\small Scalar and axial form factors at nucleons for light quark sector \cite{micromegas}. } \vspace{-0.5cm}
%\label{tab:nucleon-param}
\begin{equation*}
\begin{tabular}{c|cc|cc}
\hline\hline
& \multicolumn{2}{c|}{Proton} & \multicolumn{2}{c}{Neutron}  \\ \hline
Q & $f_{TQ_{l}}^{(p)}$ & $\Delta Q^{p}$ & $f_{TQ_{l}}^{(n)}$ &  $\Delta Q^{n}$ \\ \hline
$u$& $0.0153$  & $0.842$  & $0.011$ &  $-0.427$  \\
$d$ & $0.0191$ & $-0.427$  & $0.0273$  &  $0.842$  \\ 
$s$ & $0.0447$ & $-0.085$ & $0.0447$ & $-0.085$ \\ \hline
\end{tabular}
\end{equation*}%
\end{center}
\vspace{-0.5cm}
\caption{\small Scalar and axial form factors at nucleons for light quark sector \cite{micromegas}. } 
\label{tab:nucleon-param}
\end{table}

\subsection{Vector amplitudes}

The contribution for vector-like interactions comes from the valence quarks of the nucleon (the contributions from the virtual sea cancels quarks with antiquarks). For protons and neutrons, they are $u$ and $d$ quarks. Thus, the amplitude add coherently over the 3 valence quarks, $(2u,d)$ for protons and $(u,2d)$ for neutrons. Then, the amplitude of the second equation in (\ref{nucleon-amplitudes}) is:

\begin{eqnarray}
\sum _Q\langle N' | \overline{Q}\gamma _{\mu }\overline{v}_{Q}^{(\mathcal{Z})}Q| N \rangle &=&\sum _{Q_V} \langle N' | \overline{Q_V}\gamma _{\mu }\overline{v}_{Q_V}^{(\mathcal{Z})}Q_V| N \rangle \nonumber \\
&=&\sum _{Q_V} \overline{u^{s'}}(k'_N)\gamma _{\mu } \overline{v}_{Q_V}^{(\mathcal{Z})}u^{s}(k_N) \nonumber \\
&=& \overline{u^{s'}}(k'_N)\gamma _{\mu } V_N^{\mathcal{Z}}u^{s}(k_N),
\end{eqnarray}  
where the nucleon vector coupling is defined as:

\begin{eqnarray}
V_N^{\mathcal{Z}}=
\begin{cases}
V_p^{\mathcal{Z}}=2\overline{v}_{u}^{(\mathcal{Z})}+\overline{v}_{d}^{(\mathcal{Z})} & \text{for protons} \\
V_n^{\mathcal{Z}}=\overline{v}_{u}^{(\mathcal{Z})}+2\overline{v}_{d}^{(\mathcal{Z})} & \text{for neutrons}.
\end{cases}
\label{vector-formfactor}
\end{eqnarray}

\subsection{Axial amplitudes}

The matrix element of the axial-vector current is parameterized in terms of two form factors that are functions of the invariant $q^2$, with $q$ the transfered momentum:

\begin{eqnarray}
\sum _Q\langle N' | \overline{Q}\gamma _{\mu }\gamma _5\overline{a}_{Q}^{(\mathcal{Z})}Q| N \rangle =\frac{1}{2}\overline{u^{s'}}(k'_N)\left[\gamma _{\mu}\gamma _{5}f_N^{\mathcal{Z}}(q^2)+q_{\mu }\gamma _{5}g_N^{\mathcal{Z}}(q^2)\right]u^{s}(k_N).
\label{axial-amplitude}
\end{eqnarray}
The first form factor is induced by the spin of the quarks. The fractional spin of the nucleon $\Delta Q^{N}$ carried by the quark $Q$ is defined such that:

\begin{eqnarray}
f_N^{\mathcal{Z}}(0)=\sum _{Q}2\overline{a}_{Q}^{(\mathcal{Z})}\Delta Q^{N}.
\label{spin-formfactor-zero}
\end{eqnarray}
The values of $\Delta Q^{N}$ are listed in table \ref{tab:nucleon-param} for each quark flavor, which are dominant only for the light sector. The second form factor is calculated from partially conserved axial current approximation induced by the exchange of virtual mesons, which can be written from the first form factor for protons and neutrons as\footnote{see reference \cite{wimpsB}}:

\begin{eqnarray}
g_N^{\mathcal{Z}}(q^2)\approx \frac{M_N\left[f_p^{\mathcal{Z}}(q^2)-f_n^{\mathcal{Z}}(q^2)\right]}{|\mathbf{q}|^2+m_{\pi }^2}.
\end{eqnarray}
However, this term contributes significantly only if $q^2\gtrsim m_{\pi}^2$. Since we are considering small momentum transfer, we just ignore this form factor in Eq. (\ref{axial-amplitude}). This equation is written in a compact form if we define the effective axial-vector coupling:

\begin{eqnarray}
\Gamma _{\mu ,5}^{(\mathcal{Z})}=\frac{1}{2}\gamma _{\mu}\gamma _{5}f_N^{\mathcal{Z}}(q^2),
\label{axial-formfactor}
\end{eqnarray}
obtaining the third amplitude in (\ref{nucleon-amplitudes}). For the explicit calculations, we will use the Dirac representation of the gamma matrices, which we write as:

\begin{eqnarray}
\gamma _{\mu}=
\begin{pmatrix}
a_{\mu } & \sigma _{\mu } \\
- \sigma _{\mu } & -a_{\mu }
\end{pmatrix}, \ \ \ \ \ \ \
\gamma _5=
\begin{pmatrix}
0 & 1 \\
1 & 0
\end{pmatrix},
\label{gamma-matrices}
\end{eqnarray}
where we have defined the following vectors:

\begin{eqnarray}
a_{\mu} =(1, \mathbf{0}), \ \ \ \ \ \ \  \sigma _{\mu }=(0, \pmb{\sigma }).
\label{c-pauli-matrices}
\end{eqnarray}
Thus, the axial-vector form factor in (\ref{axial-formfactor}) is:

\begin{eqnarray}
\Gamma _{\mu ,5}^{(\mathcal{Z})}=
\frac{1}{2}f_N^{\mathcal{Z}}(q^2)\begin{pmatrix}
\sigma _{\mu } & a_{\mu } \\
- a_{\mu } & -\sigma _{\mu }
\end{pmatrix}.
\label{axial-formfactor-2}
\end{eqnarray}

\section{Nuclear Amplitudes\label{App:nuclear-amplit}}

The nuclear amplitudes for each nucleon factor are:

\begin{eqnarray}
\langle A, m'_j |\overline{\chi _N}M_N^{2}F_N^{\mathcal{H}}\chi _N| A, m_j \rangle&=& M_N^{2}F_N^{\mathcal{H}}\left(\xi ^{j'}_{A}\right)^{\dagger}\xi ^j_{A} \nonumber \\
 \langle A, m'_j |\overline{\chi _N}V_N^{\mathcal{Z}}M_N\chi _N| A, m_j \rangle&=&V_N^{\mathcal{Z}}M_N\left(\xi ^{j'}_{A}\right)^{\dagger}\xi ^j_{A} \nonumber \\
\langle A, m'_j |\overline{\chi _N}f_N^{\mathcal{Z}}(0)M_N\left(\mathbf{p} \cdot \pmb{\sigma }\right)\chi _N| A, m_j \rangle &=& f_N^{\mathcal{Z}}(0)M_N \left(\xi ^{j'}_{A}\right)^{\dagger}\mathbf{p} \cdot\pmb{\sigma }\xi ^j_{A},
\end{eqnarray}
where $\xi ^j_{A}$ is the angular wave function of the nucleus. This wave function obey

\begin{eqnarray}
\left(\xi ^{j'}_{A}\right)^{\dagger}\xi ^j_{A}&=&A\delta _{m'_jm_j}, \nonumber \\
 \left(\xi ^{j'}_{A}\right)^{\dagger}\frac{\pmb{\sigma }}{2}\xi ^j_{A}&=&\langle A, m'_j|\mathbf{S}_A^{N}|A, m_j \rangle,
 \label{nuclear-spinors}
\end{eqnarray}
where $\mathbf{S}_A^{N}$ is the spin operator of the nucleon $N$ in the direction of the angular momentum of the nucleus $A$. This direction is defined by the unit vector $\mathbf{\hat{J}}=\mathbf{J}/j$. Then, the matrix element of the spin can be written as \cite{micromegas}:

\begin{eqnarray}
\left(\xi ^{j'}_{A}\right)^{\dagger}\frac{\pmb{\sigma }}{2}\xi ^j_{A}&=&\langle A, m'_j|S_A^{N}\mathbf{\hat{J}}|A, m_j \rangle = \frac{1}{j}\langle S_A^{N} \rangle\langle m'_j|\mathbf{J}|m_j \rangle,
\label{nuclear-spinors}
\end{eqnarray}
where $\langle S_A^{N} \rangle=\langle A|S_A^{N}|A \rangle$ is interpreted as the expectation value for a nucleon to has its spin in the direction of the total angular momentum of the nucleus.

\section{Cross Section\label{App.cross section}}

From (\ref{nuclear-amplitude-final}), the squared nuclear amplitude is:

\begin{eqnarray}
\left|\mathcal{M}_{0}^{A}\right|^2=\left|\mathcal{M}_{0}^{SI}\right|^2+\left|\mathcal{M}_{0}^{SD}\right|^2+\mathcal{M}_{0}^{SI\dagger}\mathcal{M}_{0}^{SD}+\mathcal{M}_{0}^{SI}\mathcal{M}_{0}^{SD\dagger}.
\label{amplitud-squared}
\end{eqnarray}
Taking into account the amplitudes from Eqs. (\ref{SI-amplitude2}) and (\ref{SD-amplitude2}), the sum of the spin states from the incoming and outcoming nucleus, for each term, is

\begin{eqnarray}
\sum _{m_{j}}\sum _{m'_{j}} \left|\mathcal{M}_{0}^{SI}\right|^2&=&16M_A^{2}M_{\sigma}^{2}f_p^{2}\left|Z+\frac{f_n}{f_p}(A-Z)\right|^2\sum _{m_{j}}\sum _{m'_{j}}\delta _{m'_{j}m_{j}} \nonumber \\
&=&16M_A^{2}M_{\sigma}^{2}f_p^{2}\left|Z+\frac{f_n}{f_p}(A-Z)\right|^2(2j+1),
\label{spin-sum-SI}
\end{eqnarray}

\begin{eqnarray}
\sum _{m_{j}}\sum _{m'_{j}} \left|\mathcal{M}_{0}^{SD}\right|^2&=&4M_A^{2}\left|\sum _{\mathcal{Z}}G_{\mathcal{Z}}\Lambda _{\mathcal{Z}}\right|^2\sum _{m_{j}}\sum _{m'_{j}}\left(\mathbf{p} \cdot \langle m'_j|\mathbf{J}|m_j \rangle\right)\left(\mathbf{p} \cdot \langle m_j|\mathbf{J}^{\dagger}|m_j' \rangle\right) \nonumber \\
&=&4M_A^{2}\left|\sum _{\mathcal{Z}}G_{\mathcal{Z}}\Lambda _{\mathcal{Z}}\right|^2p^{k}p^{l}\sum _{m'_{j}}\langle m'_j|J_kJ_l^{\dagger}|m_j' \rangle,
\label{spinsumSD}
\end{eqnarray}
where we use the completeness relation for the last term. From the usual set of quantum relations

\begin{eqnarray}
\langle m'_j|J_{1(2)}^{2}|m_j' \rangle&=&\frac{1}{2}\left[j(j+1)-(m'_{j})^{2}\right], \nonumber \\
\langle m'_j|J_3^{2}|m_j' \rangle&=&(m'_j)^{2} \nonumber \\
\langle m'_j|J_1J_2^{\dagger}|m_j' \rangle&=&-\frac{i}{2}m'_j \nonumber \\
\langle m'_j|J_{1(2)}J_3|m_j' \rangle&=&0,
\end{eqnarray}
the tensor part gives:

\begin{eqnarray}
p^{k}p^{l}\sum _{m'_{j}}\langle m'_j|J_kJ_l^{\dagger}|m_j' \rangle&=&\frac{1}{2}\left(p_1^2+p_2^2\right)\sum _{m'_{j}}\left[j(j+1)-(m'_{j})^{2}\right]\nonumber \\
&&+\ \ p_3^2\sum _{m'_{j}}(m'_j)^{2}-ip_1p_2\sum _{m'_{j}} m'_j \nonumber \\
&=&\frac{1}{2}\left(p_1^2+p_2^2\right)\left[\frac{2}{3}j(j+1)(2j+1)\right] \nonumber \\
&&+\ \ p_3^2\left[\frac{1}{3}j(j+1)(2j+1)\right] \nonumber \\
&=&\frac{1}{3}\left|\mathbf{p}\right|^2j(j+1)(2j+1),
\end{eqnarray}
obtaining in (\ref{spinsumSD}):

\begin{eqnarray}
\sum _{m_{j}}\sum _{m'_{j}} \left|\mathcal{M}_{0}^{SD}\right|^2&=&\frac{4}{3}M_A^{2}\left|\mathbf{p}\right|^2j(j+1)(2j+1)\left|\sum _{\mathcal{Z}}G_{\mathcal{Z}}\Lambda _{\mathcal{Z}}\right|^2.
\label{spin-sum-SD}
\end{eqnarray}

With a similar procedure, it can be shown that the mixing terms in (\ref{amplitud-squared}) cancel after add the spin states. Then, the sum of the spin states lead us:

\begin{eqnarray}
\overline{\left|\mathcal{M}_{0}^{A}\right|^2}=\overline{\left|\mathcal{M}_{0}^{SI}\right|^2}+\overline{\left|\mathcal{M}_{0}^{SD}\right|^2},
\label{spin-sum-total}
\end{eqnarray}
with each SI and SD term given by (\ref{spin-sum-SI}) and (\ref{spin-sum-SD}). By replacing Eq. (\ref{spin-sum-total}) into (\ref{unpol-cross-section}), we obtain Eqs. (\ref{unpol-cross-section2}) and (\ref{SI-SD-cross-sections}), where the known relation $\left|\pmb{\beta }\right|=\left|\mathbf{p}\right|/E_{\sigma }\approx \left|\mathbf{p}\right|/M_{\sigma }$ was applied in the SD component.

\end{document}